\documentclass[reqno]{amsart}
\usepackage{graphicx}

\theoremstyle{definition}

\theoremstyle{remark}

\numberwithin{equation}{section}

\begin{document}
%
\newpage
\noindent SFI Working paper 07-08-029    \hfill \mbox{ITFA-2007-37}  \\

\title[]{The Physics of Information}
\author{F. Alexander Bais and J. Doyne Farmer}
\email{}
\thanks{\\Santa Fe Institute, 1399
Hyde Park Road, Santa Fe, NM 87501\\ Permanent address F.A.B.: Institute for Theoretical Physics, University of Amsterdam. \\ This article is a contribution to  \textit{The Handbook on the philosophy of information}, edited by J. van Benthem and P. Adriaans, to be published by Elsevier.\\ [2mm]Email: bais@science.uva.nl,
\;jdf@santafe.edu}
\address{}


\maketitle
\tableofcontents

\section{The Physics of Information}
\begin{quote}
{\footnotesize Why cannot we write the entire 24 volumes of the
Encyclopedia Brittanica on the head of a pin?
\\ \mbox{}\hfill R.P.~Feynman}
\end{quote}

Information is carried, stored, retrieved and processed by
machines, whether they be electronic computers or living
organisms. All information, which in an abstract sense one may think of  as a string of zeros and ones, has to be carried by a physical substrate, be it paper, silicon chips or holograms, and the handling of this information is physical, so information is ultimately constrained by the fundamental laws of physics.  It is therefore not surprising that physics and information share a rich interface. 

The notion of information as used by Shannon is a generalization of the notion of entropy, which first appeared in thermodynamics.  In thermodynamics entropy is an abstract quantity depending on heat and temperature whose interpretation is not obvious.  This changed with the theory of statistical mechanics, which explains and generalizes thermodynamics.  Statistical mechanics exploits a decomposition of a system into microscopic units such as atoms to explain macroscopic phenomena such as temperature and pressure in terms of the statistical properties of the microscopic units.  Statistical mechanics makes it clear that entropy can be regarded as a measure of microscopic disorder.  The entropy $S$ can be written as $S = -\sum p_i \log p_i$, where $p_i$ is the probability of a particular microscopic state, for example the likelihood that a given atom will have its velocity and position within a given range.  

Shannon realized that entropy is useful to describe disorder in much more general settings, which might have nothing to do with atoms or physics.  The entropy of a probability distribution $\{ p_i \}$ is well defined as long as $p_i$ is well defined.  In this more general context he argued that measuring order and measuring disorder are essentially the same -- in a situation that is highly disordered, making a measurement gives a great deal of information, and conversely, in a situation that is highly ordered, making a measurement gives little information.   Thus for a system that can randomly be in one of several different states the entropy of its distribution is the same as the information gained by knowing which state $i$ it is in.  It turns out that the concept of entropy or equivalently information is useful in many applications that have nothing to do with physics.  

It also turns out that thinking in these more general terms is useful for physics.  For example, Shannon's work makes it clear that entropy is in some sense more fundamental than the quantities from which it was originally derived.  This led Jaynes to formulate all of statistical mechanics as a problem of maximizing entropy.  In fact, all of science can be viewed as an application of the principle of maximum entropy, which provides a means of quantifying the tradeoff between simplicity and accuracy of description.  If we want to understand how physical systems can be used to perform computations, or construct computer memories, it can be useful to define entropies that may not correspond to thermodynamic entropy.  But if we want to understand the limits to computation it is very useful to think in thermodynamic or statistical terms.  This has become particularly important in efforts to understand how to take advantage of quantum mechanics to improve computation.  These considerations have given rise to a subfield of physics that is often called the physics of information.

In this chapter we attempt to explain to a non-physicist where the idea of information came from.  We begin by describing the origin of the concept of entropy in thermodynamics, where entropy is just a macroscopic state variable related to heat flow and temperature, a rather mathematical device without a concrete physical interpretation.  We then discuss how the microscopic theory of atoms led to statistical
mechanics, which makes it possible to derive and extend
thermodynamics.   This led to the definition of entropy in terms
of probabilities on the set of accessible microscopic states of a system 
and provided the inspiration for modern
information theory starting with the seminal work of Shannon \cite{shannon1948}.   A close examination of the foundations of
statistical mechanics and the need to reconcile the probabilistic
and deterministic views of the world leads us to a discussion of
chaotic dynamics, where information plays a crucial role in
quantifying predictability.  We then discuss a variety of
fundamental issues that emerge in defining information and how one
must exercise care in discussing concepts such as order, disorder,
and incomplete knowledge.  We also discuss an alternative form of
entropy and its possible relevance for nonequilibrium
thermodynamics.  

Toward the end of the chapter we discuss how quantum mechanics gives rise to the concept of quantum information. Entirely new possibilities for information storage and computation are possible due to the massive parallel processing inherent in quantum mechanics. We also point out how entropy can be extended to apply to quantum mechanics to provide a useful measurement for quantum entanglement. Finally we make a small
excursion to the interface betweeen quantum theory and  general relativity, where one is confronted with the ``ultimate
information paradox" posed by the physics of Black Holes. 
In this review we have limited ourselves; not all relevant
topics that touch on physics and information have been
covered.  

In our quest for more and more volume and speed in storing and processing
information we are naturally led to the smallest scales we can
physically manipulate.   We began the introduction by quoting Feynman's visionary 1959 lecture ``Plenty of room at the bottom" \cite{feynman1960} where he discusses storing and manipulating information on the atomic level.  Currently commercially available processors work at scales of 60 nm (1 nm = 1 nanometer = $10^{-9}$ meter).  In 2006, IBM announced circuitry on a 30 nm scale, which indeed makes it possible to write the Encyclopedia Britannica on the head of a pin, so Feynmann's speculative remark in 1959 is now just a marker of the current scale of computation.   To make it clear how close this is to the atomic scale, a square with sides of length 30 nm contains about 1000 atoms.  Under the historical pattern of Moore's law, integrated circuitry halves in size every 2 years.  If we continue on the same trajectory of improvement, within about 20 years the components will be the size of individual atoms, and it is difficult to imagine that computers will be able to get any smaller.  Once this occurs information at the atomic scale will be directly connected to our use of information on a macroscopic scale.  There is a certain poetry to this:  Once a computer has components on a quantum scale, the motion of its atoms will no longer be random, and in a certain sense will not be described by classical statistical mechanics, at the same time that it will be used to process information on a macroscopic scale.

\section{Thermodynamics}\label{sectionthermo}

\begin{quote}
{\footnotesize The truth of the second law is, therefore, a
statistical and not a mathematical truth, for it depends on the
fact that the bodies we deal with consist of millions of molecules
and that we never can get a hold of single molecules
\\ \mbox{}\hfill J.C.~Maxwell }
\end{quote}

Thermodynamics is the study of macroscopic physical
systems\footnote{Many details of this brief expose of selected
items from thermodynamics and statistical mechanics can be found
in standard textbooks on these subjects \cite{reif1965,
kittel1966,huang1987,landau1980}.}. These systems contain a large
number of degrees of freedom, typically of the order of Avogadro's
number, i.e. $N_A \approx 10^{23}$. The three laws of
thermodynamics describe processes in which systems exchange energy
with each other or with their environment. For example, the system
may do work, or exchange heat or mass through a diffusive process.
A key idea is that of {\it equilibrium}, which in thermodynamics
is the assumption that the exchange of energy or mass between two
systems is the same in both directions; this is typically only
achieved when two systems are left alone for a long period of
time.  A process is \textit{quasistatic} if it always remains
close to equilibrium, which also implies that it is
\textit{reversible}, i.e that the process can be undone and the
system can return to its original state without any external energy inputs.  We distinguish various types of  processes, for example an \textit{isothermal} process in which the system is in thermal contact with a reservoir that keeps it at a fixed temperature.  Another example is an \textit{adiabatic} process in which the system is kept thermally isolated and the temperature is allowed to change.  A system may also go from one equilibrium state to another via a nonequilibrium process, such as the free expansion of
a gas or the mixing of two fluids, in which case it is not
reversible.  No real system is fully reversible, but it is
nonetheless a very useful concept.

The remarkable property of systems in equilibrium is that the
macro states can be characterized by only very few variables, such
as the volume $V$, pressure $P$, temperature $T$, entropy $S$,
chemical potential $\mu$ and particle number $N$.  These
state variables are in general not independent, but rather are
linked by an {\it equation of state}, which describes the constraints imposed by physics.  A familiar example is the
ideal gas law $PV=N_A kT$, where $k$ is the Boltzmann constant
relating temperature to energy ($k=1.4 \times 10^{-23}\;
joule/Kelvin$). In general the state variables come in pairs, one of
which is \textit{intensive} while the other
conjugate variable is \textit{extensive}.  Intensive variables like pressure or temperature are independent of system size, while extenstive variables like volume and entropy are proportional to system size.  

In this lightning review we will only highlight the essential
features of thermodynamics that are most relevant in connection with information
theory.

\subsection{The laws}

The first law of thermodynamics reads\footnote{The bars through
the differentials indicate that the quantities following them are
not state variables: the d-bars therefore refer to small
quantities rather then proper differentials.}
\begin{equation}\label{1stlaw}
dU = d\!{\bf\bar{\mbox{}}}\; Q - d\!{\bf\bar{\mbox{}}}\; W
\end{equation}
and amounts to the statement that heat is a form of energy and
that energy is conserved. More precisely, the change in internal
energy $dU$ equals the amount of heat $d\!{\bf\bar{\mbox{}}}\;Q$
absorbed by the system  minus the work done by the system,
$d\!{\bf\bar{\mbox{}}}\; W$.

The second law introduces the concept of entropy $S$, which is
defined as the ratio of heat flow to temperature. The law states that
the entropy for a closed system (with constant energy,
volume and number of particles) can never decrease. In
mathematical terms
\begin{equation}\label{2ndlaw}
  dS = \frac{d\!{\bf\bar{\mbox{}}}\; Q}{T},
\;\;\;\;\;\;\;\;\;\;\; \frac{dS}{dt} \geq 0.
\end{equation}
By using a gas as the canonical
example, we can rewrite the first law in proper differentials as
\begin{equation}\label{1stlaw2}
  dU = TdS - PdV,
\end{equation}
where $PdV$ is the work done by changing the volume of the container, for example by compressing the gas with a piston.
It follows from the relation between entropy, heat and temperature
that entropy differences can be measured by measuring the temperature with a thermometer and the change in heat with a calorimeter.  This illustrates that from the point of view of thermodynamics entropy is a purely macroscopic quantity.

 \hspace{1cm}
\begin{figure}[h!]
\begin{center}
\includegraphics[scale=0.4]{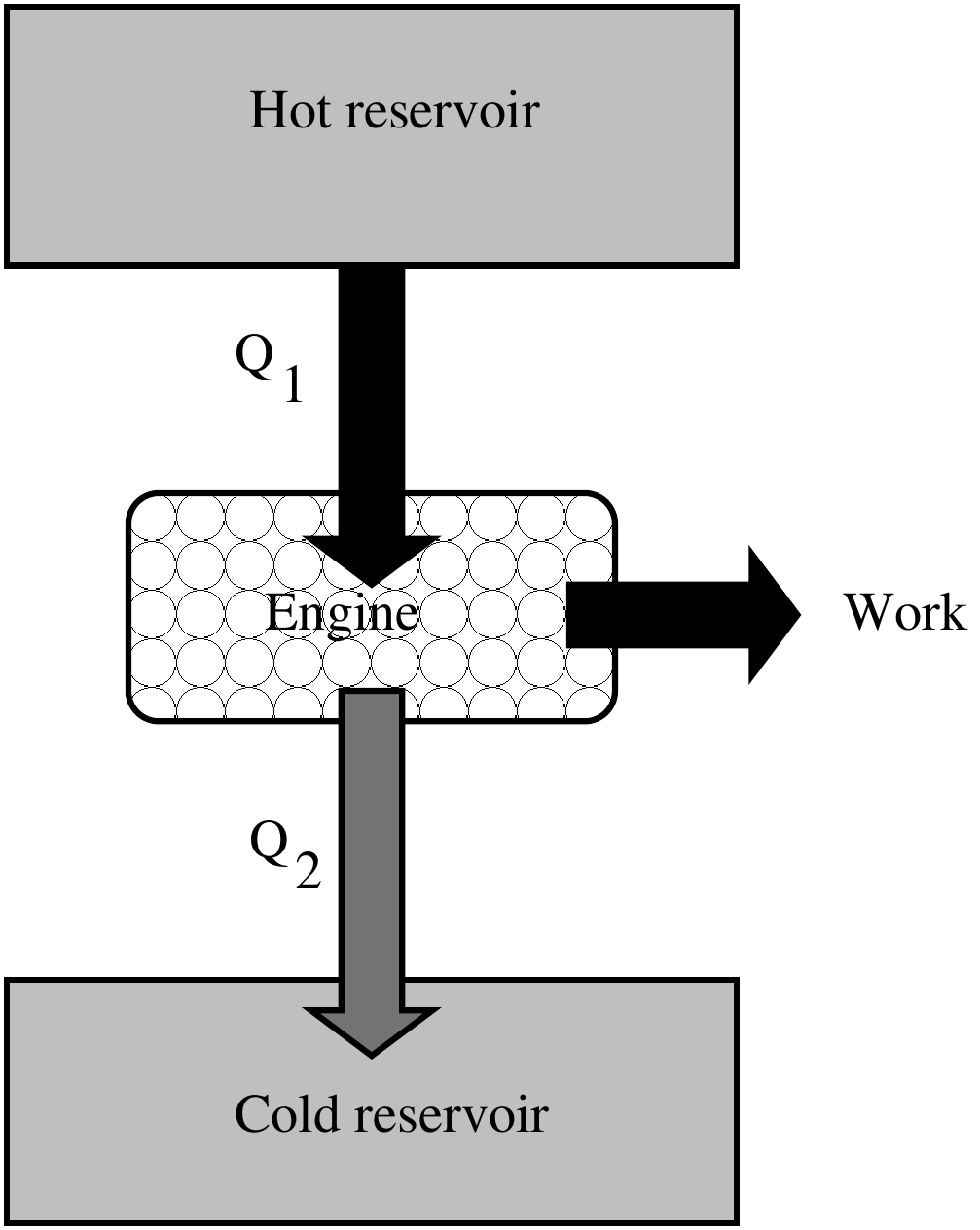}
\includegraphics[scale=0.4]{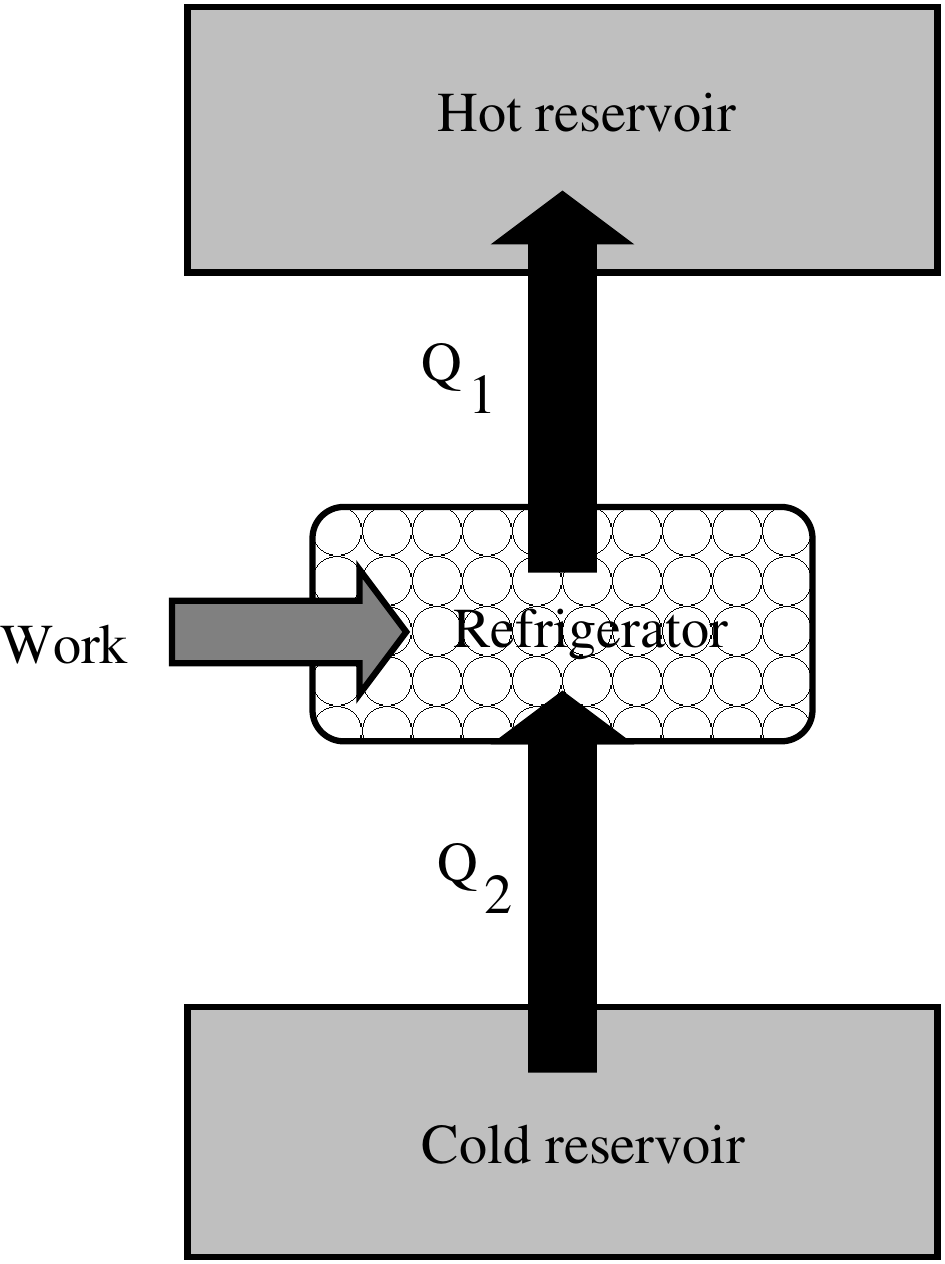}
\end{center}
\caption{The relation between heat and work illustrating the two
formulations of the second law of thermodynamics. On the left we
have the Kelvin formulation. The ideal engine corresponds to the
diagram with the black arrows only. The second law tells us that
the third, grey arrow is necessarily there. The right picture with
only the black arrows corresponds to the ideal refrigerator, and
the third, grey arrow is again required by the second
law.} \label{carnotengine}
\end{figure}

There are two different formulations of the second law. The Kelvin formulation states that it
is impossible to have a machine whose sole effect is to convert heat into work.  We can use heat to do work, but to do so we must inevitably make other alterations, e.g. letting heat flow from hot to cold and thereby bringing the system closer to equilibrium.  Clausius' formulation
says that it is impossible to have a machine that only extracts
heat from a reservoir at low temperature and delivers that same
amount of heat to a reservoir at higher temperature. Rephrasing
these formulations, Kelvin says that ideal engines cannot exist
and Clausius says that ideal refrigerators can't exist.  See figure \ref{carnotengine}.

\begin{figure}[h!]
\begin{center}
\includegraphics[scale=0.4]{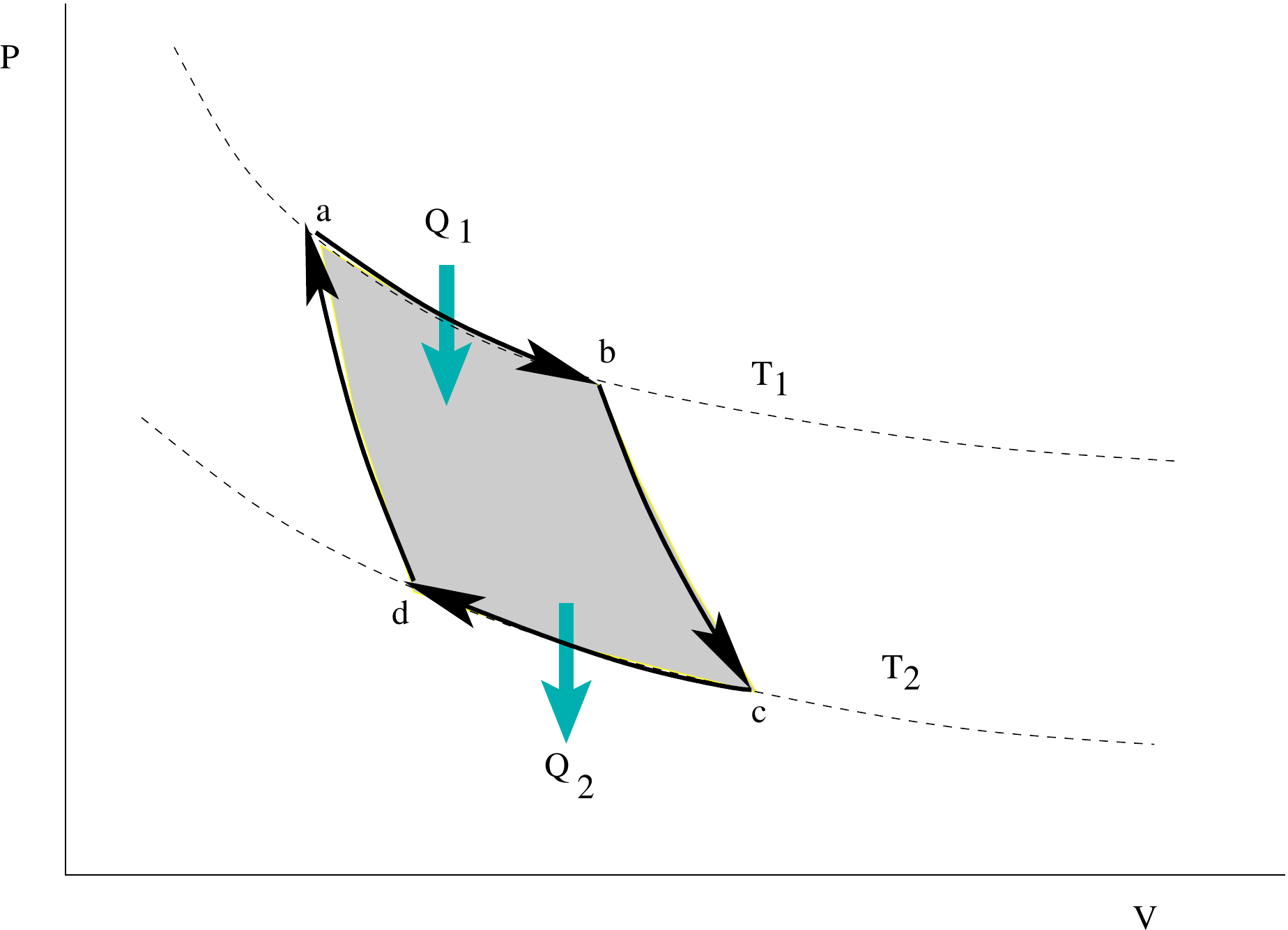} 
\end{center}
\caption{The Carnot cycle corresponding to the Kelvin formulation of the second law. The work done by the engine equals the line integral along the closed contour and is  therefore equal to the enclosed area.} \label{carnotcycle}
\end{figure}

The action of a heat engine or refrigerator machines can be pictured in a diagram in which the reversible sequence of states the system goes through are a closed curve, called a Carnot cycle.  We give an example for the Kelvin formulation in figure~\ref{carnotcycle}.  Imagine a piston in a chamber; out goal is to use the temperature differential between two reservoirs to do work.  The cycle consists of four steps:
In step $a \rightarrow b$, isothermal expansion, the system absorbs an amount $Q_1$ of heat from the reservoir at high temperature $T_1$, which causes the gas to expand and push on the piston, doing work; In step $b \rightarrow c$, adiabatic expansion, the gas continues to expand and do work, but the chamber is detached from the reservoir, so that it no longer absorbs any heat.  Now as the gas expands it cools until it reaches temperature $T_2$.  In step $c \rightarrow d$, isothermal compression, the surroundings do work on the gas, as heat flows into the cooler reservoir, giving off an amount $Q_2$ of heat; and in step $d \rightarrow a$, adiabatic compression, the surroundings continue to do work, as the gas is further compressed (without any heat transfer) and brought back up to the original temperature. The net work done by the machine is given by the line integral:
\begin{equation}
\label{cycle}
W = \oint_{cycle} PdV = \mbox{enclosed area }
\end{equation}
which by the first law should also be equal to $W=Q_1-Q_2$ because the internal energy is the same at the beginning and end of the cycle.
We also can calculate the total net change in entropy of the two reservoirs  as
\begin{equation}
\label{entropychange}
\Delta S = \frac{-Q_1}{T_1} + \frac{Q_2}{T_2} \geq 0\; ,
\end{equation}
where the last inequality has to hold because of the second law. Note that the two latter equations can have solutions with positive $W$. The efficiency of the engine $\eta$ is by definition the ratio of the work done to the heat entering the system, or
\begin{equation}
\label{efficiency }
\eta = \frac{W}{Q_1} = 1 - \frac{Q_2}{Q_1} \leq 1-\frac{T_1}{T_2}.
\end{equation}
This equals one for an ideal heat engine, but is less then one for a real engine.

A modern formulation of the second law, which in the setting
of statistical mechanics is equivalent to the statements of Kelvin
and Clausius, is the {\it Landauer principle}, which
says that there is no machine whose sole effect is the erasure of
information. There is a price to forgetting:  The principle states
that the erasure of information (which is irreversible) is
inevitably accompanied by the generation of heat.  In other words, logical irreversibility necessarily involves  thermodynamical irreversibility.  One has to
generate at least $kT\ln 2$ to get rid of one bit of information
\cite{landauer1961,landauer1991}.  We return to the Landauer principle in the  section on Statistical mechanics.  

We just showed that the second law sets fundamental limits on the possible efficiency of real machines like steam engines, refrigerators and information processing devices.  As
everybody knows, real engines give off heat and real refrigerators
and real computers need power to do their job.
The second law tells us to what extent heat can be used
to perform work. The increase of entropy as we go from one
equilibrium situation to another is related to dissipation and the
production of heat, which is intimately linked to the important
notion of \textit{irreversibility}.  A given action in a closed
system is irreversible if it makes it impossible for the system to
return to the state it was in before the action took place without external inputs.
Irreversibility is always associated with production of heat,
because heat cannot be freely converted to other forms of energy
(whereas any other form of energy can always be converted to
heat). One can decrease the entropy of a system by doing work on it, 
but in doing the work one has to increase the entropy of 
another system (or of the system's environment) by an equal or greater amount.

The theory of thermodynamics taken by itself does not connect
entropy with information. This only comes about when the results
are interpreted in terms of a microscopic theory, in which case
temperature can be interpreted as being related to uncertainty and
incoherence in the position of particles.  This requires a
discussion of statistical mechanics, as done in the next section.

There is another fundamental aspect to the second law which is
important from an operational as well as philosophical point of
view.  A profound implication of the second law is that it defines
an ``arrow of time", i.e., it allows us to distinguish the past
from the future.   This is in contrast to the fundamental microscopic laws of
physics which are time
reversal invariant (except for a few exotic interactions, that are only very rarely seen under normal
conditions as we find them on earth).  If one watches a movie of fundamental 
processes on the microscopic level it is impossible to tell 
whether it is running forwards or backwards. In contrast, if we watch a
movie of macroscopic events, it is not hard to identify
irreversible actions such as the curling of smoke, the spilling of
a glass of water, or the mixing of bread dough,  which easily
allow us to determine whether we are running in forward or
reverse.  More formally, even if we didn't know which way time
were running, we could pick out some systems at random and measure
their entropy at times $t_1,t_2, \ldots$   The direction in which
entropy increases is the one that is going forward in time.  Note
that we didn't define an a priori direction of time in formulating
the second law -- it establishes a time direction on its own,
without any reference to atomic theory or any other laws of
physics.

The second law of thermodynamics talks only about the
difference between the entropy of different macrostates. The
absolute scale for entropy is provided by the third law of
thermodynamics. This law states that when a system approaches the
absolute zero of temperature the entropy will go to zero, i.e.
\begin{equation}\label{3rdlaw}
T \rightarrow 0  \;\;\;\; \Rightarrow  \;\;\;  S \rightarrow 0.
\end{equation}
When $T = 0$ the heat is zero, corresponding classically to no atomic motion,
and the energy takes on its lowest possible value.  In quantum theory we know that
such a lowest energy ``ground" state also exists, though, if the ground 
state of the system turns out to be degenerate the entropy will 
approach a nonzero constant at zero temperature.  
We conclude by emphasizing that the laws of thermodynamics
have a wide applicability and a rich phenomenology that supports
them unequivocally.

\subsection{Free energy}

Physicists are particularly concerned with what is called the
(Helmholtz) \textit{free energy}, denoted $F$.  It is a very important
quantity because it defines the amount of energy available to do work.   
As we discuss in the next section, the free energy plays a central role in
establishing the relation between thermodynamics and statistical
mechanics, and in particular for deriving the microscopic definition of
entropy in terms of probabilities.

The free energy is defined as
\begin{equation}\label{freeenegy}
  F\equiv U-TS.
\end{equation}
This implies that in differential form we have
\begin{equation}\label{dF1}
  dF = dU-TdS-SdT,
\end{equation}
which using (\ref{1stlaw2}) can be written as
\begin{equation}\label{dF2}
  dF = -PdV - SdT.
\end{equation}
The natural independent variables to describe the free energy of a gas are
volume and temperature.

Let us briefly reflect on the meaning of the free energy. Consider
a system $A$ in thermal contact with a heat bath $A'$ kept at a
constant temperature $T_0$. Suppose the system A absorbs heat
$d\!{\bf\bar{\mbox{}}}~Q$ from the reservoir.  We may
think of the total system consisting of system plus bath as a
closed system: $A^0 = A+A'$. For $A^0$ the second law implies that
its entropy can only increase: $ dS^0= dS + dS'\ge 0$. As the
temperature of the heat bath $A'$ is constant and its absorbed
heat is $-d\!{\bf\bar{\mbox{}}}\; Q$, we may write $T_0 dS'=
-d\!{\bf\bar{\mbox{}}}\;Q $. From the first law applied to
system $A$ we obtain that  $ -d\!{\bf\bar{\mbox{}}}\;Q = -dU
-d\!{\bf\bar{\mbox{}}}\;W$, so that we can substitute the expression $T_0dS'=-dU
-d\!{\bf\bar{\mbox{}}}\;W$ in $T_0 dS + T_0 dS'\ge 0$ to get
 $-dU + T_0dS \ge d\!{\bf\bar{\mbox{}}}\;
W$.   As the system $A$ is kept at a constant temperature the left
hand side is just equal to $-dF$, demonstrating that
\begin{equation}\label{Fmin}
-dF \ge d\!{\bf\bar{\mbox{}}}\; W.
\end{equation}
The maximum work that can be done by the system in contact with a
heat reservoir is $(-dF)$. If we keep the system parameters fixed,
i.e. $d\!{\bf\bar{\mbox{}}}\; W =0$, we obtain that $dF \le 0$,
showing that for a system coupled to a heat bath the free energy
can only decrease, and consequently in a thermal equilibrium
situation the free energy reaches a minimum. This should be compared with the entropy, which reaches a maximum at equilibrium.

We can think of the second law as telling us how different kinds
of energy are converted into one another:  In an isolated system,
work can be converted into heat, but heat cannot be converted into
work.  From a microscopic point of view forms of energy that are ``more
organized", such as light, can be converted into those that are
``less organized", such as the random motion of particles, but the
opposite is not possible.

From Equation (\ref{dF2})  the pressure and
entropy of a gas can be written as partial derivatives of the free energy
\begin{equation}\label{PandS}
  P = \left(\frac{\partial F}{\partial V}\right)_T, \;\;\;\; S=\left(\frac{\partial F}{\partial T}\right)_V.
\end{equation}
So we see that for a system in thermal equilibrium the entropy is
a state variable, meaning that if we reversibly traverse a closed
path we will return to the same value (in contrast to other
quantities, such as heat, which do not satisfy this property). The
variables $P$ and $S$ are dependent variables.  This is evident from the Maxwell relation, obtained by
equating the two second derivatives
\begin{equation}\label{maxwell1}
  \frac{\partial^2 F}{\partial T \partial V}=\frac{\partial^2 F}{\partial V \partial
  T},
\end{equation}
yielding the relation
\begin{equation}\label{maxwell2}
  \left(\frac{\partial P}{\partial T}\right)_V=\left(\frac{\partial S}{\partial
  V}\right)_T.
\end{equation}

\section{ Statistical mechanics}\label{sectionstatmech}
\begin{quote}
{\footnotesize In dealing with masses of matter, while we do not
perceive the individual molecules, we are compelled to adopt what
I have described as the statistical method of calculation, and to
abandon the strict dynamical method, in which we follow every
motion by
the calculus.\\
\mbox{}\hfill J.C.~Maxwell}
\end{quote}
\begin{quote}
{\footnotesize We are forced to be contented with the more modest
aim of deducing some of the more obvious propositions relating to
the statistical branch of mechanics. Here there can be no mistake
in regard to the agreement with the facts of nature. \\ \mbox{}
\hfill J.W.~Gibbs}
\end{quote}

Statistical mechanics is the explanation of the macroscopic
behavior of physical systems using the underlying microscopic laws
of physics, even though the microscopic states, such as the
position and velocity of individual particles, are unknown.  The
key figures in the late 19th century development of statistical
mechanics were Maxwell, Boltzmann and Gibbs \cite{maxwell1872,
boltzmann1896-98, gibbs1902}. One of the outstanding questions was
to derive the laws of thermodynamics, in particular to give a
microscopic definition of the notion of entropy. Another objective
was the understanding of phenomena that cannot be computed from thermodynamics alone, such as transport phenomena.  For our purpose of highlighting the links with information theory we will give a
brief and somewhat lopsided introduction.  Our main goal is to show the origin of the famous
expression due to Gibbs for the entropy, $S=-\sum_i p_i \ln p_i$, which was
later used by Shannon to define information.

\subsection{Definitions and postulates}
\begin{quote}
{\footnotesize Considerable semantic confusion has resulted from
failure to distinguish between prediction and interpretation
problems, and attempting a single formalism to do both. \\ \mbox{}
\hfill T.S.~Jaynes}
\end{quote}

Statistical mechanics considers systems with many degrees of
freedom, such as atoms in a gas or spins on a lattice.  We can
think in terms of the {microstates} of the system which are, for
example, the positions and velocities of all the particles in a
vessel with gas.  The space of possible microstates is called the
\textit{phase space}.  For a monatomic gas with $N$ particles, the
phase space is $6N$-dimensional, corresponding to the fact that
under Newtonian mechanics there are three positions and three
velocities that must be measured for each particle in order to
determine its future evolution. A microstate of the whole system
thus corresponds to a single point in phase space.

Statistical mechanics involves the assumption that, even though we
know that the microstates exist, we are largely ignorant of their
actual values.  The only information 
we have about them comes from
macroscopic quantities, which are bulk properties such as the
total energy, the temperature, the volume, the pressure, or the
magnetization.   Because of our ignorance we have to treat the
microstates in statistical terms.  But the knowledge of the
macroscopic quantities, along with the laws of physics that the
microstates follow, constrain the microstates and allow us to
compute relations between macroscopic variables that might
otherwise not be obvious.  Once the values of the macroscopic
variables are fixed there is typically only a subset of
microscopic states that are compatible with them, which are called
the \textit{accessible states}.   The number of accessible states
is usually huge, but differences in this number can be very
important.  In this chapter we will for simplicity assume a
discrete set of microstates, but the formalism can be
straightforwardly generalized to the continuous case.

The first fundamental assumption of statistical mechanics is that in equilibrium
a closed system has an equal a priori probability to be in any of
its accessible states. For systems that are not closed, for
example because they are in thermal contact or their particle
number is not constant, the set of accessible states will be
different and their probabilities have to be calculated.
In either case we associate an
\textit{ensemble} of systems with a characteristic probability
distribution over the allowed microscopic states. Tolman
\cite{tolman1938} clearly describes the notion of an ensemble:
\begin{quote} {\small  In using ensembles for statistical
purposes, however, it is to be noted that there is no need to
maintain distinctions between individual systems since we shall be
interested merely in the number of systems at any time which would
be found in the different states that correspond to different
regions of phase space. Moreover, it is also to be noted for
statistical purposes that we shall wish to use ensembles
containing a large enough population of separate members so that
the number of systems in such different states can be regarded as
changing continuously as we pass from the states lying in one
region of the phase space to those in another. Hence, for the purpose
in view, it is evident that the condition of an ensemble at any
time can be regarded as appropriately specified by the density
{\em r} with which representative points are distributed over
phase space.}
\end{quote}
The second postulate of statistical mechanics, called
\textit{ergodicity}, says that time averages correspond to
ensemble averages.  That is, on one hand we can take the time
average by following the deterministic motion of the all the
microscopic variables of all the particles making up a system. On
the other hand, at a given instant in time we can take an average
over all possible accessible states, weighting them by their
probability of occurrence.  The ergodic hypothesis says that these
two averages are the same. We return to the restricted validity of
this hypothesis in the section on nonlinear dynamics.

\subsection{Counting microstates for a system of magnetic spins}

\label{spins} In the following example we show how it is possible
to derive the distribution of microscopic states through the
assumption of equipartition and simple counting arguments. This
also illustrates that the distribution over microstates
becomes extremely narrow in the thermodynamic (i.e. $N\rightarrow
\infty$ limit).  Consider a system of $N$ magnetic spins that can
only take two values $s_j = \pm 1$, corresponding to whether the
spin is pointing up or down (often called {\it Ising spins}). The
total number of possible configurations equals $2^N$. For
convenience assume $N$ is even, and that the spins do not
interact.  Now put these spins in an upward pointing magnetic field $H$ and ask how many configurations of spins are consistent
with each possible value of the energy.  The energy of each spin
is $e_j=\mp \mu H$, and because they do not interact, the total
energy of the system is just the sum of the energies of each spin.
For a configuration with $k$ spins pointing up and $N-k$ spins
pointing down the total energy can be written as $\varepsilon_m =
2m\mu H$ with $m\equiv (N-2k)/2$ and $-N/2\leq m\leq N/2 $.  The
value of $\varepsilon_m$ is bounded : $-N \mu H\leq \varepsilon_m
\leq N\mu H$ and the difference between two adjacent energy
levels, corresponding to the flipping of one spin, is $\Delta
\varepsilon = 2 \mu H$. The number of microscopic configurations
with energy $\varepsilon_m$ equals
\begin{equation}\label{binomial}
g(N,m)= g(N,-m)= \frac{N!}{(\frac{1}{2}N+m)!(\frac{1}{2}N-m)!}.
\end{equation}
The total number of states is $\sum_m g(N,m) = 2^N$.  For a
thermodynamic system $N$ is really large, so we can approximate
the factorials by the Stirling formula
\begin{equation}\label{stirling}
  N!\cong \sqrt{2\pi N}N^N e^{-N+ \cdots}
\end{equation}
Some elementary math gives the Gaussian approximation for
the binomial distribution for large $N$,
\begin{equation}\label{gnm}
  g(N,m)\cong 2^N \left(\frac{2}{\pi N}\right)^\frac{1}{2}
  e^{-2m^2/N}.
\end{equation}
We will return to this system later on, but at this point we
merely want to show that for large N the distribution
is sharply peaked. Roughly speaking
the width of the distribution grows with $\sqrt N$ while the peak
height grows as $2^N$, so the degeneracy of the states around $m=0$ increases very rapidly.  For
example $g(50,0)= 1.264 \times 10^{14}$,  but for $N\approx N_A$
one has $g(N_A,0)\cong 10^{10^{22}}$.  
We will return to this example in the following
section to calculate the magnetization of a spin system in thermal
equilibrium.

\subsection{The Maxwell-Boltzmann-Gibbs distribution}
Maxwell was the first to derive an expression for the
probability distribution {$p_i$} for a system in thermal
equilibrium, i.e. in thermal contact with a heat reservoir kept at
a fixed temperature $T$. This result was later generalized by Boltzmann 
and Gibbs. An equilibrium distribution
function of an ideal gas without external force applied to it should
not depend on either position or time, and thus can only depend on
the velocities of the individual particles.   In general there are
interactions between the particles that need to be taken into
account.  A simplifying assumption that is well justified by
probabilistic calculations is that processes in which two
particles interact at once are much more common than those in
which three or more particles interact. If we assume that the
velocities of two particles are independent before they interact
we can write their joint probability to have velocities $v_1$ and
$v_2$ as a product of the probability for each particle alone.
This implies $p(v_1, v_2) = p(v_1) p(v_2)$. The same holds after
they interact: $p(v_1', v_2') = p(v_1') p(v_2')$.  In
equilibrium, where nothing can depend on time,  the probability
has to be the same afterward, i.e. $p(v_1, v_2) = p(v_1', v_2')$.
How do we connect these conditions before and after the
interaction? A crucial observation is that there are conserved
quantities that are preserved during the interaction and the
equilibrium distribution function can therefore only depend on
those. Homogeneity and isotropy of the distribution function
selects the total energy of the particles as the only
function on which the distribution depends. The conservation of
energy in this situation boils down to the simple statement that
$\frac{1}{2}m v_1^2 +\frac{1}{2}m v_2^2 = \frac{1}{2}m {v'_1}^2 +
\frac{1}{2}m {v'_2}^2$. From these relations Maxwell derived the
well known thermal equilibrium velocity distribution,
\begin{equation}\label{maxwell}
  p_0(v) = n\left(\frac{m}{2\pi T}\right)^{3/2}\;e^{-mv^2/2kT}.
\end{equation}
The distribution is Gaussian.  As we saw, to derive it Maxwell
had to make a number of assumptions which were plausible 
even though they couldn't be derived from the fundamental laws 
of physics. Boltzmann generalized the result to
include the effect of an external conservative force,  leading to
the replacement of the kinetic energy in (\ref{maxwell}) by the
total conserved energy, which includes potential as well as
kinetic energy.

Boltzmann's generalization of Maxwell's result makes it clear that
the probability distribution $p_i$ for a general system in thermal
equilibrium is given by
\begin{equation}\label{psubi}
  p_i =\frac{e^{-\varepsilon_i/T}}{Z}.
\end{equation}
$Z$ is a normalization factor that ensures the conservation of
probability, i.e. $\sum_i p_i =1$.  This implies that
\begin{equation}\label{partitionfunction}
  Z \equiv \sum_i e^{-\varepsilon_i/T}\;.
\end{equation}
$Z$ is called the \textit{partition function}.  The
Boltzmann distribution describes the {\it canonical
ensemble}, that is it applies to any situation where a system is
in thermal equilibrium and exchanging energy with its environment.
This is in contrast to the {\it microcanonical ensemble}, which
applies to isolated systems where the energy is constant, or the
{\it grand canonical ensemble}, which applies to systems that are
exchanging both energy and particles with their environment\footnote{Gibbs 
extended the Boltzmann result to situations where the number of particles 
is not fixed, leading to the introduction of the \textit
{chemical potential}. Because of its complicated history, the
 exponential distribution is referred to by a variety of names, including Gibbs, Boltzmann, Boltzmann-Maxwell, and Boltzmann-Gibbs.}.
To illustrate the power of the Boltzmann distribution let us
briefly return to the example of the thermal distribution of
Ising spins on a lattice in an external magnetic field.  As we
pointed out in section (\ref{spins}), the energy of a single spin
is $\pm\mu H$. According to the Boltzmann distribution, the
probabilities of spin up or spin down are
\begin{equation}\label{spinupdown}
 p_\pm = \frac{e^{\mp \mu H/T}}{Z}.
\end{equation}
The spin antiparallel to the field has lowest energy and therefore
is favored. This leads to an average  field dependent
magnetization $m_H$ (per spin)
\begin{equation}\label{magnetization}
m_H = \langle \mu \rangle = \frac{\mu p_+ + (-\mu) p_-}{p_+ +p_-} = \mu
\tanh\frac{u H}{T}.
\end{equation}
This example shows how statistical mechanics can be used to
establish relations between macroscopic variables that cannot be
obtained using thermodynamics alone.

\subsection{Free energy revisited}

In our discussion of thermodynamics in section 2.2 we introduced 
the concept of the free energy $F$ defined by equation
\ref{freeenegy}, and argued that it plays a central role for
systems in thermal contact with a heat bath, i.e. systems kept at
a fixed temperature $T$.  In the previous section we introduced
the concept of the partition function $Z$ defined by equation
\ref{partitionfunction}.  Because all thermodynamic quantities
can be calculated from it, the importance of the partition function
$Z$ goes well beyond its role as a normalization factor.
The free energy is of particular importance,
because its functional form leads directly to the definition of
entropy in terms of probabilities. We can now directly link the 
thermodynamical quantities to the
ones defined in statistical mechanics.  This is done by postulating\footnote{Once we have identified a certain macroscopic
quantity like the free energy with a microscopic expression, then
of course the rest follows. Which expression is taken as the
starting point for the identification is quite arbitrary. The
justification is {\it a posteriori} in the sense that the well known
thermodynamical relations should be recovered.}
 the relation between the free energy and the partition
function as\footnote{Boltzmann's constant $k$ relates energy
to temperature.  Its value in conventional units is
$1.4\times10^{-23}joule/kelvin$, but we have set it equal to unity, which
amounts to choosing a convenient unit for energy or temperature.}
\begin{equation}\label{FandZ}
  F = -T\ln Z ,
\end{equation}
or alternatively $Z=e^{-F/T}$. From this definition it is possible
to calculate all thermodynamical quantities, for example
using equations (\ref{PandS}). We will now derive
the expression for the entropy in statistical mechanics in terms of probabilities.
\subsection{Gibbs entropy}
The definition of the free energy in equation
(\ref{freeenegy}) implies that
\begin{equation}\label{entropyenergy}
  S = \frac{U - F}{T}.
\end{equation}
From (\ref{FandZ}) and (\ref{psubi}) it follows that
\begin{equation}\label{Pandpsubi}
  F = \varepsilon_i + T\ln p_i.
\end{equation}
Note that even though both the terms on the right depend on $i$
the free energy $F$ is independent of $i$.  
The equilibrium value for the internal energy is by
definition
\begin{equation}\label{Uandpsubi}
  U = \langle  \varepsilon \rangle \equiv \sum_i \varepsilon_i\;p_i\;.
\end{equation}
With these expressions for $S$, $F$ and $U$, and making 
use of the fact that $F$ is independent of $i$ 
and $\sum_i p_i = 1$,  we can rewrite the entropy in
terms of the probabilities $p_i$ and arrive at the famous
expression for the entropy:
\begin{equation}\label{S-psubi}
  S=-\sum_i\; p_i \ln p_i \;.
\end{equation}
This expression is usually called the Gibbs entropy\footnote{In
quantum theory this expression is replaced by $S=-Tr \;\rho \ln
\rho$ where $\rho$ is the density matrix of the system.}.

In the special case where the total energy is fixed, the $w$
different (accessible) states all have equal a priori probability
$p_i= p= 1/w$. Substitution in the Gibbs formula yields the
expression in terms of the number of accessible states, originally
due to Boltzmann (and engraved on his tombstone):
\begin{equation}\label{logW}
 S = \ln w.
\end{equation}
We emphasize that the entropy grows logarithmically with the number of accessible states\footnote{These 
numbers can be overwhelmingly
large. Imagine two macrostates of a system which differ by $1$
millicalorie at room temperature.  The difference in entropy is
 $\Delta S = - \Delta Q/T = 10^{-3}/293 \approx
10^{-5}$.  Thus the ratio of the number of accessible states is 
$w_2/w_1 = \exp(\Delta S/k)\approx \exp (10^{18})$, a big number!}. 
Consider a system consisting of a single particle that can be in one of two states.  Assuming equipartition the entropy is $ S_1=\ln 2$.  For a system with Avogadro's number of particles $N \sim 10^{23}$, so there are $2^N$ states and if we assume independence the entropy is $S_N= \ln 2^N = N S_1$, a very large number.  The tendency of a system to maximize its entropy is a probabilistic statement:  The number of states with half of the particles in one state and half in the other is enormously larger than the number in which all the particles are in the same state, and when the system is left free it will relax to the most probable accessible state.  The state of a gas particle depends not only on its allowed position (i.e. the volume of the vessel), but also on its allowed range of velocities:  If the vessel is hot that range is larger then when the vessel is cold. So for an ideal gas one finds that the entropy increases with the logarithm of the temperature. The fact that the law is a probabilistic implies that it is not completely impossible that the system will return to a highly improbable initial state.  Poincar\'e showed that it is bound to happen and gave an estimate of  the recurrence time (which for a macroscopic system is much larger than the lifetime of the universe). 

The Gibbs entropy transcends its origins in statistical mechanics.  It can be used to describe any system with 
states $\{\psi_i\}$ and a given probability distribution
$\{p_i\}$.  Credit for realizing this is usually given to Shannon \cite{shannon1948}, although antecedents 
include Szilard, Nyquist and Hartley.  Shannon proposed that by
analogy to the entropy $S$, information can be defined as
\begin{equation}\label{infodefinition}
H \equiv -\sum_i p_i \;  \log_2 \; p_i.
\end{equation}
In information theory it is common to take
logarithms in base two and drop the Boltzmann
constant\footnote{In our convention k=1, so  $H = S/\ln 2$.}. Base
two is a natural choice of units when dealing with binary numbers and the
units of entropy in this case are called  \textit{bits}; in
contrast, when using the natural logarithm the units are called
\textit{nats}, with the conversion that  1 \em{nat} =1.443
\em{bits} . For example a memory consisting of 5 bits (which is
the same as a system of 5 Ising spins), has $N=2^5$ states.
Without further restrictions all of these states (messages) have
equal probability i.e. $p_i = 1/N$ so that the information content
is $H=-N \frac{1}{N}\;\log_2\; \frac{1}{N}=\log_2\; 2^{5}=5 \;
bits$. Similarly consider a DNA-molecule with 10 billion base pairs, each of which can be in one of four
combinations (A-T,C-G,T-A,G-C).  The molecule can a priori be in any of
$4^{10^{10}}$ configurations so the naive information content (assuming independence) is $H= 2 \times
10^{10}\; bits$. The logarithmic nature of the definition is unavoidable 
if one wants the additive property of information under the addition of bits. 
If in the previous spin
example we add another string of 3 bits then the total number of
states is $N= N_1 N_2= 2^5\times 2^3=2^8$ from which it also
follows that $H= H_1+H_2=8$. If we add extra ab initio
correlations or extra constraints
we reduce  the number of independent configurations and
consequently $H$ will be smaller. 

As we will discuss in Section \ref{sectionentropy},
this quantitative definition of information and its
applications  transcend the limited origin and scope of
conventional thermodynamics and statistical mechanics, as well as 
Shannon's original purpose of describing properties of communication channels.  See also \cite{Brillouin56}.

\section{Nonlinear dynamics}\label{sectiondynamics}
\begin{quote}
{\small The present state of the system of nature is evidently a
consequence of what it was in the preceding moment, and if we
conceive of an intelligence which at a given instant comprehends
all the relations of the entities of this universe, it could state
the respective position, motions, and general effects of all these
entities at any time in the past or future. \\
\mbox{}\hfill Pierre Simon de Laplace (1776)}
\end{quote}

\begin{quote}
{\small A very small cause which escapes our notice determines a
considerable effect that we cannot fail to see, and then we say
that the effect is due to chance.
 \\ \mbox{} \hfill Henri Poincar\'{e} (1903).}
\end{quote}
From a naive point of view statistical mechanics seems to
contradict the determinism of Newtonian mechanics.  For any initial state $x(0)$ (a
vector of positions and velocities) Newton's laws define a dynamical
system $\phi^t$ (a set of differential equations) that maps $x(0)$ into its future states
$x(t) = \phi^t(x(0))$.  This is completely deterministic.  As
Laplace so famously asserted, if mechanical objects obey Newton's
laws, why do we need to discuss perfect certainties in
statistical terms?\\
Laplace partially answered his own question:
\begin{quote}
{\small $\ldots$ But ignorance of the different causes involved in
the production of events, as well as their complexity, taken
together with the imperfection of analysis, prevent our reaching
the same certainty [as in astronomy] about the vast majority of
phenomena.  Thus there are things that are uncertain for us,
things more or less probable, and we seek to compensate for the
impossibility of knowing them by determining their different
degrees of likelihood.  So it is that we owe to the weakness of
the human mind one of the most delicate and ingenious of
mathematical theories, the science of chance or probability.}
\end{quote}
Laplace clearly understood the need for statistical descriptions, but at that point in time was not fully aware of 
the consequences of nonlinear dynamics.  As Poincar\'{e} later
showed, even without human uncertainty (or quantum mechanics),
when Newton's laws give rise to differential equations with chaotic dynamics, we inevitably
arrive at a probabilistic description of nature.   Although
Poincar\'{e} discovered this in the course of studying the three
body problem in celestial mechanics, the answer he found turns out to have relevance for the
reconciliation of the deterministic Laplacian universe with
statistical mechanics.
\subsection{The ergodic hypothesis}

As we mentioned in the previous section, one of the key
foundations in Boltzmann's formulation of statistical mechanics is
the {\it ergodic hypothesis}. Roughly speaking, it is the
hypothesis that a given trajectory will eventually find its way
through all the accessible microstates of the system, e.g. all
those that are compatible with conservation of energy.  At
equilibrium the average length of time that a trajectory spends in
a given region of the state space is proportional to the number of
accessible states the region contains.  If the ergodic hypothesis
is true, then time averages equal ensemble averages, and
equipartition is a valid assumption.

The ergodic hypothesis proved to be highly controversial for good
reason:  It is generally not true. The first numerical experiment
ever performed on a computer took place in 1947 at Los Alamos when
Fermi, Pasta, and Ulam set out to test the ergodic hypothesis.
They simulated a system of masses connected by nonlinear springs.
They perturbed one of the masses, expecting that the disturbance
would rapidly spread to all the other masses and equilibrate, so
that after a long time they would find all the masses shaking more
or less randomly.  Instead they were quite surprised to discover
that the disturbance remained well defined -- although it
propagated through the system, it kept its identity, and after a
relatively short period of time the system returned very close to
its initial state.  They had in fact rediscovered a phenomenon that
has come to be called a {\it soliton}, a localized but very stable
travelling disturbance.  There are many examples of nonlinear
systems that support solitons.  Such systems do not have equal probability
to be in all their accessible states, and so are not ergodic.

Despite these problems, there are many examples where we know that statistical
mechanics works extremely well. There are even a few cases, such as the hard sphere gas, where the 
ergodic hypothesis can actually
be proved. But more typically this is not
the case. The evidence for statistical mechanics is largely
empirical:  we know that it works, at least to a very high degree
of approximation.   Subsequent work has made it clear that
the typical situation is much more complicated than was originally imagined.
While some trajectories may wander in more
or less random fashion around much of the accessible phase space, they are
blocked from entering certain regions by what are called KAM
(Kolomogorov-Arnold-Moser) tori. Other initial conditions yield trajectories that make regular motions and lie on KAM tori trajectories.  The KAM tori are separated from each other, and have a lower
dimension than the full accessible phase space.  Such KAM tori correspond to situations in which there are other conversation laws in addition to the conservation of energy, which may depend on initial conditions as well as other parameters\footnote{Dynamical systems that conserve energy and obey Newton's laws have special properties that cause the existence of KAM tori.  Dissipative systems typically have {\it attractors}, subsets of the state space that orbits converge onto.  Energy conserving systems do not have attractors, and often have chaotic orbits tightly interwoven with regular orbits.}.  
Solitons are examples of this in which
the solutions can be interpreted as a geometrically isolated pulse. 

There have now been an enormous number of studies of ergodicity in nonlinear dynamics.  While there are no formal theorems that definitively resolve this, the accumulated lore from these studies suggests that for nonlinear systems that do not have hidden symmetries, as the number of interacting components increases and nonlinearities become stronger, the generic behavior is that chaotic behavior becomes  more and more likely -- the KAM tori shrink, fewer and fewer initial conditions are trapped on them, and the regions they exclude become smaller.  The ergodic
hypothesis becomes an increasingly better approximation, a typical
single trajectory can reach almost all accessible states, and
equipartition becomes a good assumption.  The problems occur in understanding when there are hidden symmetries that can support phenomena like solitons.  The necessary and
sufficient conditions for ergodicity to be a good assumption remains an active field of research.  

\subsection{Chaos and limits to prediction\label{chaos}}

\begin{figure}[!h]
\begin{center}
\includegraphics[scale=0.65]{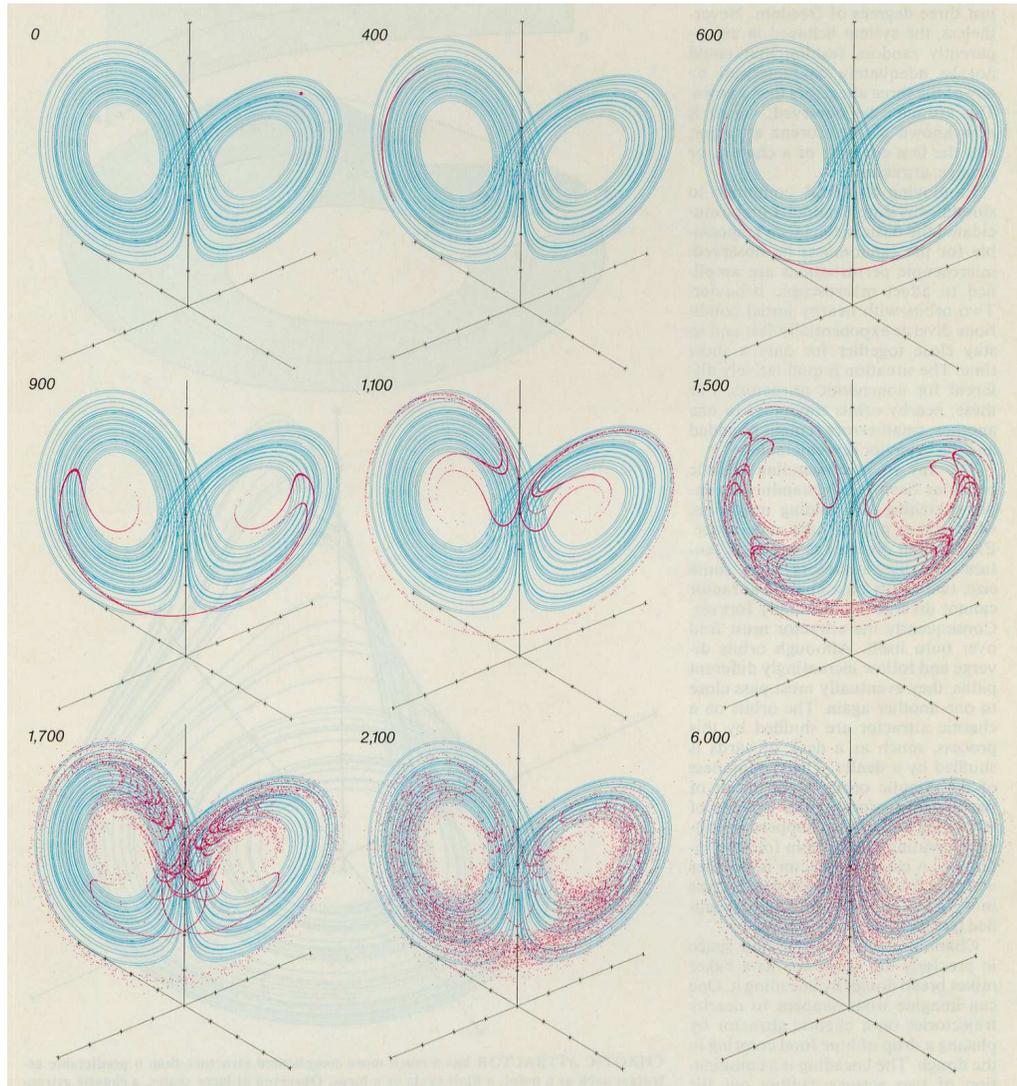}
\end{center}
\caption{ The divergence of nearby trajectories for the Lorenz equations. 
See the text for an explanation }
\label{lorenzFigure}
\end{figure}

The discovery of chaos makes it clear that Boltzmann's use of
probability is even more justified than he realized.  When motion
is chaotic, two infinitesimally nearby trajectories separate at an
exponential rate \cite{Lorenz63,Shaw81,Crutchfield86,Strogatz94}.  This is a geometric property of the 
underlying nonlinear
dynamics.  From a linear point of view the dynamics are locally
unstable.  To make this precise, consider two $N$ dimensional
initial conditions $x(0)$ and $x'(0)$ that are initially separated
by an infinitesimal vector $\delta x(0) = x(0) - x'(0)$.
Providing the dynamical system is differentiable, the separation
will grow as
\begin{equation}
\delta x(t) = D\phi^t (x(0)) \delta x(0),
\label{separationVectors}
\end{equation}
where $D\phi^t (x(0))$ is the derivative of the  dynamical system
$\phi^t$ evaluated at the initial condition $x(0)$.  For any fixed
time $t$ and initial condition $x(0)$,  $D\phi^t$ is just an $N
\times N$ matrix, and this is just a linear equation.   If the
motion is chaotic the length of the separation vector $\delta x$
will grow exponentially with $t$ in at least one direction, as
shown in Figure~\ref{lorenzFigure}.
The figure shows how the divergence of nearby trajectories is the underlying
reason chaos leads to unpredictability.  A perfect measurement
would correspond to a point in the state space, but any real
measurement is inaccurate, generating a cloud of uncertainty.  The
true state might be anywhere inside the cloud.  As shown here for
the Lorenz equations (a simple system of three coupled nonlinear
differential equations \cite{Lorenz63}), the uncertainty of the
initial measurement is represented by 10,000 red dots, initially
so close together that they are indistinguishable; a single
trajectory is shown for reference in light blue.  As each point
moves under the action of the equations, the cloud is stretched
into a long, thin thread, which then folds over onto itself many
times, until the points are mixed more or less randomly over the
entire attractor.  Prediction has now become impossible:  the
final state can be anywhere on the attractor.  For a regular
motion, in contrast, all the final states remain close together.
We can think about this in information theoretic terms; for a
chaotic motion information is initially lost at a linear rate
which eventually results in all the information being lost -- for
a regular motion the information loss is relatively small.  The
numbers above the illustration are in units of 1/200 of the natural time
units of the Lorenz equations.  (From \cite{Crutchfield86}).

Nonetheless, at the same time the motion can be globally stable,
meaning that it remains contained inside a finite volume in the
phase space.  This is achieved by stretching and folding -- the
nonlinear dynamics knead the phase space through local stretching
and global folding, just like a baker making a loaf of bread.  Two
trajectories that are initially nearby may later be quite far
apart, and still later, may be close together again.  This
property is called {\it mixing}.  More formally, the dynamics are
mixing over a given set $\Sigma$ and invariant measure\footnote{A
measure is invariant over a set $\Sigma$ with respect to the dynamics
$\phi^t$ if it satisfies the condition $\mu(A) =
\mu(\phi^{-t}(A))$, where $A$ is any subset of $\Sigma$.  There can be
many invariant measures, but the one that we have in mind
throughout is the one corresponding to time averages.} $\mu$ with
support $\Sigma$ such that for any subsets $A$ and $B$
\begin{equation}
\lim_{t \to \infty} \mu(\phi^t B \cap A) = \mu(A) \mu(B).
\end{equation}
Intuitively, this just means that B is smeared throughout $\Sigma$ by
the flow,  so that the probability of finding a point originating
in $B$ inside of $A$ is just the original probability of $B$,
weighted by the probability of $A$.    Geometrically, this happens
if and only if the future trajectory of $B$ is finely ``mixed"
throughout $\Sigma$ by the stretching and folding action of $\phi^t$.

Mixing implies ergodicity, so any dynamical system that is mixing
over $\Sigma$ will also be ergodic on $\Sigma$.   It only satisfies the
ergodic hypothesis, however, if $\Sigma$ is the set of accessible
states.  This need not be the case.  Thus, the fact that a system
has orbits with chaotic dynamics doesn't mean that it necessarily
satisfies the ergodic hypothesis -- there may be still be subsets
of finite volume in the phase space that are stuck making regular
motions, for example on KAM tori. 

Nonetheless, chaotic dynamics has strong implications for
statistical mechanics. If a dynamical system is ergodic but not
mixing\footnote{A simple example of a system that is ergodic but
not mixing is a dynamical system whose solution is the sum of two
sinusoids with irrationally related frequencies.}, by measuring
the microstates it is in principle possible to make detailed long
range predictions by measuring the position and velocity of all
its microstates, as suggested by Laplace.  In contrast, if it is
mixing then even if we know the initial values of the microstates
at a high (but finite) level of precision, all this information is
asymptotically lost, and statistical mechanics is unavoidable\footnote{An exception is that some systems display phase invariance even while they are chaotic.  The orbits move around an attractor, being chaotically scrambled transverse to their direction of motion but keeping their timing for completing a circuit of the attractor \cite{Farmer80}.}.

\subsection{Quantifying predictability}

Information theory can be used to quantify predictability \cite{Shaw81}.  To
begin the discussion, consider a measuring instrument with a
uniform scale of resolution $\epsilon$.  For a ruler, for example,
$\epsilon$ is the distance between adjacent graduations.  If such
a measuring instrument is assigned to each of the $N$ real
variables in a dynamical system, the graduations of these
instruments induce a {\it partition} $\Pi$ of the phase space,
which is a set of non-overlapping $N$ dimensional cubes, labeled
$C_i$, which we will call the outcomes of the measurement.   A
measurement determines that the state of the system is in a given
cube $C_i$.  If we let transients die out, and restrict our
attention to asymptotic motions without external perturbations,
let us assume the motion is confined to a set $\Sigma$ (which in
general depends on the initial condition).  We can then compute
the asymptotic probability of a given measurement by measuring its
frequency of occurrence $p_i$, and if the motion is ergodic on
$\Sigma$, then we know that there exists an invariant measure $\mu$
such that $p_i = \mu(C_i)$.  To someone who knows the invariant
measure $\mu$ but knows nothing else about the state of the
system, the average information that will be gained in making a
measurement is just the entropy
\begin{equation}
I(\epsilon) = - \sum_i p_i \log p_i .
\end{equation}
We are following Shannon in calling this ``information" since it
represents the element of surprise in making the measurement.  The
information is written $I(\epsilon)$ to emphasize its dependence
on the scale of resolution of the measurements. This can be used
to define a dimension for $\mu$.   This is just the asymptotic
rate of increase of the information with resolution, i.e.
\begin{equation}
D = \lim_{\epsilon \to 0} \frac{I(\epsilon)}{|\log \epsilon|}.
\end{equation}
This is called the {\it information dimension}
\cite{Farmer82}.  Note that this reduces to what is commonly
called the fractal dimension when $p_i$ is sufficiently
smooth, i.e. when $\sum_i p_i \log p_i \approx \log n$, where $n$
is the number of measurement outcomes with nonzero values of
$p_i$.

This notion of dimension can be generalized by using the R\'enyi entropy $R_\alpha$
\begin{equation}
R_\alpha = \frac{1}{1 - \alpha} \log \sum_i p_i^\alpha
\end{equation}
where $\alpha \ge 0$ and $\alpha \ne 1$.  The value for $\alpha =
1$ is defined by taking the limit as $\alpha \to 1$, which reduces
to the usual Shannon entropy.   By replacing the Shannon entropy
by the R\'enyi entropy it is possible to define a generalized
dimension $d_\alpha$.  This contains the information dimension in
the special case $\alpha = 1$.  This has proved to be very useful
in the study of multifractal phenomena (fractals whose scalings
are irregular).  We will say more about the use of such
alternative entropies in the next section.

The discussion so far has concerned the amount of information
gained by an observer in making a single, isolated measurement,
i.e. the information gained in taking a ``snapshot" of a dynamical
system.  We can alternatively ask how much new information is
obtained per unit time by an observer who is watching a movie of a
dynamical system.  In other words, what is the information
acquisition rate of an experimenter who makes a series of
measurements to monitor the behavior of a dynamical system?  For a
regular dynamical system (to be defined more precisely in a
moment) new measurements asymptotically provide no further
information in the limit $t \to \infty$.  But if the dynamical
system is chaotic, new measurements are constantly required to
update the knowledge of the observer in order to keep the
observer's knowledge of the state of the system at the same
resolution.

This can be made more precise as follows.  Consider a sequence of
$m$ measurements $(x_1, x_2, \ldots, x_m) = X_m$, where each
measurement corresponds to observing the system in a particular
$N$ dimensional cube.  Letting $p(X_m)$ be the probability of
observing the sequence $X_m$, the entropy of this sequence of
measurements is
\begin{equation}
H_m = -\sum_i p(X_m) \log p(X_m)
\end{equation}
We can then define the information acquisition rate as
\begin{equation}
h = \lim_{m \to \infty} \frac{H_m}{m \Delta t}.
\end{equation}
$\Delta t$ is the sampling rate for making the measurements.
Providing $\Delta t$ is sufficiently small and other conditions
are met, $h$ is equal to the {\it metric entropy}, also called the
{\it Kolmogorov-Sinai (KS) entropy}\footnote{In our discussion of
metric entropy we are sweeping many important mathematical
formalities under the rug.  For example, to make this definition
precise we need to take a supremum over all partitions and
sampling rates.  Also, it is not necessary to make the
measurements in $N$ dimensions -- there typically exists a one
dimensional projection that is sufficient, under an optimal
partition.}.   Note that this is not really an entropy, but an
entropy production rate, which (if logs are taken to base 2) has
units of bits/second.  If $h \rangle  0$ the motion is chaotic, and if $h
= 0$ it is regular.   Thus, when the system is chaotic, the
entropy $H_m$ contained in a sequence of measurements continues to
increase even in the limit as the sequence becomes very long.  In
contrast, for a regular motion this reaches a limiting value.

Although we have so far couched the discussion in terms of
probabilities, the metric entropy is determined by geometry.  The
average rates of expansion and contraction in a trajectory of a
dynamical system can be characterized by the spectrum of Lyapunov
exponents.  These are defined in terms of the eigenvalues of $D
\phi^t$, the derivative of the dynamical system, as  defined in
equation~\ref{separationVectors}.   For a dynamical system in $N$
dimensions, let the $N$ eigenvalues of the matrix $D\phi^t (x(0))$
be $\alpha_i (t)$.  Because $D\phi^t$ is a positive definite
matrix, the $\alpha_i$ are all positive.  The Lyapunov exponents
are defined as $\lambda_i = \lim_{t \to \infty} \log \alpha_i (t)
/ t$.  To think about this more geometrically, imagine an
infinitesimal ball that has radius $\epsilon (0)$ at time $t = 0$.
As this ball evolves under the action of the dynamical system it
will distort.  Since the ball is infinitesimal, however, it will
remain an ellipsoid as it evolves.  Let the principal axes of this
ellipsoid have length $\epsilon_i (t)$.  The spectrum of Lyapunov
exponents for a given trajectory passing through the initial ball
is
\begin{equation}
\lambda_i = \lim_{t \to \infty} \lim_{\epsilon(0) \to 0}
\frac{1}{t} \log \frac{\epsilon_i (t)}{\epsilon(0)}.
\end{equation}
For an $N$ dimensional dynamical system there are $N$ Lyapunov
exponents.  The positive Lyapunov exponents $\lambda^+$ measure
the rates of exponential divergence, and the negative ones
$\lambda^-$ the rates of convergence.  They are related to the
metric entropy by {\it Pesin's theorem}
\begin{equation}
h = \sum_i \lambda_i^+.
\end{equation}
In other words, the metric entropy is the sum of the positive
Lyapunov exponents, and it corresponds to the average exponential
rate of expansion in the phase space.

Taken together the metric entropy and information dimension can be
used to estimate the length of time that predictions remain valid.
The information dimension allows an estimate to be made of the
information contained in an initial measurement, and the metric
entropy estimates the rate at which this information decays.

As we have already seen, for a series of measurements the metric entropy tells us the
information gained with each measurement.  But if each measurement is made with the same
precision, the information gained must equal the information that
would have been lost had the measurement not been made.  Thus the
metric entropy also quantifies the initial rate at which knowledge
of the state of the system is lost after a measurement.

To make this more precise, let $p_{ij}(t)$ be the probability that
a measurement at time $t$ has outcome $j$ if a measurement at time
$0$ has outcome $i$.  In other words, given the state was
measured in partition element $C_i$ at time $0$, what is the
probability it will be in partition element $C_j$ at time $t$?.
By definition $p_{ij}(0) = 1$ if $i = j$ and $p_{ij}(0) = 0$
otherwise.  With no initial information, the information gained
from the measurement is determined solely by the asymptotic
measure $\mu$, and is $-\log \mu(C_j)$.  In contrast, if $C_i$ is
known the information gained on learning outcome $j$ is $- \log
p_{ij}(t)$.  The extra information using a prediction from the
initial data is the difference of the two or
$\log(p_{ij}(t)/\mu(C_j))$.  This must be averaged over all
possible measurements $C_j$ at time $t$, and all possible initial
measurements $C_i$.  The measurements $C_j$ are weighted by their
probability of occurrence $p_{ij}(t)$, and the initial
measurements are weighted by $\mu(C_i)$.  This gives
\begin{equation}
I(t) = \sum_{i,j} \mu(C_i) p_{ij}(t) \log( \frac{p_{ij}(t)}{\mu(C_j)}).
\end{equation}
It can easily be shown that in the limit where the initial
measurements are made arbitrarily precise, $I(t)$ will initially
decay at a linear rate, whose slope is equal to the metric
entropy.  For measurements with signal to noise ratio $s$, i.e.
with $\log s \approx |\log \epsilon |$, $I(0) \approx D_I \log s$.
Thus $I(t)$ can be approximated as $I(t) \approx D_I \log s - h
t$, and the initial data becomes useless after a characteristic
time $\tau = (D_I / h) \log s$.

To conclude, chaotic dynamics provides the link that connects deterministic dynamics with probability.  While we can discuss chaotic systems in completely deterministic terms, as soon as we address problems of measurement and long-term predictability we are forced to think in probabilistic terms.  The language we have developed above, of information dimension, Lyapunov exponents, and metric entropy, provide the link between the geometric and probabilistic views.  Chaotic dynamics can happen even in a few dimensions, but as we move to high dimensional systems, e.g. when we discuss the interactions between many particles, probability is thrust on us for two reasons:  The difficulty of keeping track of all the degrees of freedom, and the ``increased likelihood" that nonlinear interactions will give rise to chaotic dynamics.  ``Increased likelihood" is in quotations because, despite more than a century of effort, understanding the necessary and sufficient conditions for the validity of statistical mechanics remains an open problem.

\section{About Entropy}\label{sectionentropy}
In this section we will discuss various aspects of entropy, its
relation with information theory and the sometimes confusing
connotations of order, disorder, ignorance and incomplete
knowledge. This will be done by treating several well known puzzles 
and paradoxes related with the concept of entropy. A derivation 
of the second law using the procedure called \textit{coarse graining} is 
presented. The extensivity or
additivity of entropy is considered in some detail, also when we
discuss nonstandard extensions of the definition of entropy.

\subsection{Entropy and information}

The important innovation Shannon made was to show that the
relevance of the concept of entropy considered as a measure of
information was not restricted to thermodynamics, but could be
used in any context where probabilities can be defined.  He
applied it to problems in communication theory and showed that it
can be used to compute a bound on the information transmission
rate using an optimal code.

One of the most basic results that Shannon obtained was to show
that the choice of the Gibbs form of entropy to describe uncertainty is not arbitrary, even when it is used in a very general context.   Both Shannon and Khinchin
\cite{khinchin1949} proved that if one wants certain conditions to
be met by the entropy function then the functional form originally proposed by Gibbs is the unique choice. The
fundamental conditions as specified by Khinchin are:
\begin{enumerate}
  \item For a given $n$ and $\sum_{i=1}^n\; p_i=1$, the required
  function $H(p_1,...p_n)$ is maximal for
  all $p_i = 1/n$.
  \item The function should satisfy $H(p_1,...p_n,0) = H(p_1,...p_n)$.
  The inclusion of an impossible event should
  not change the value of $H$.
  \item If $A$ and $B$ are two finite sets of events, not
  necessarily independent, the entropy $H(A, B)$ for the occurrence of
  joint events $A$ and $B$ is the entropy for the set $A$ alone plus the
  weighted average of the conditional entropy $H(B|A_i)$ for
  $B$ given the occurrence of the $i^{th}$ event $A_i$ in $A$,
\begin{equation}\label{khinchin}
  H(A, B) = H(A) + \sum_i p_i H(B|A_i)
\end{equation}
where event $A_i$ occurs with probability $p_i$.
\end{enumerate}
The important result is that given these conditions the function
$H$ given in equation (\ref{infodefinition}) is the \textit{unique} solution.
Shannon's key insight was that the results of Boltzmann and Gibbs
in explaining entropy in terms of statistical mechanics had
unintended and profound side-effects, with a broader and more
fundamental meaning that transcended  their physical origin of
entropy. The importance of the abstract conditions formulated by
Shannon and Khinchin show the very general context in which the
Gibbs-Shannon function is the unique answer. Later on we will pose
the question of whether there are situations where \textit{not} all three conditions are appropriate, leading to 
alternative expressions for the entropy.

\subsection{The Landauer principle}
\label{sec:landauer}

\begin{figure}[ht]
\begin{center}
\includegraphics[scale=0.5]{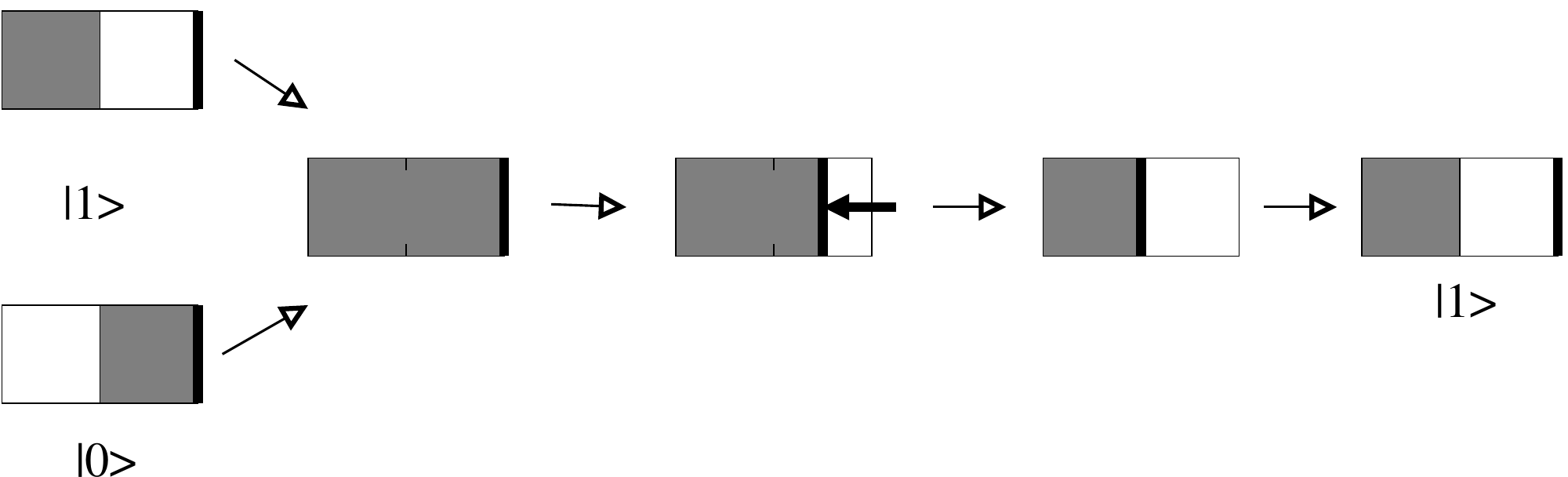}
\end{center}
\caption{An illustration of the Landauer principle using a
very simple thermodynamical system. } 
\label{landauer}
\end{figure}

Talking about the relation between information and entropy it may be illuminating to return briefly to the Landauer principle\cite{landauer1961,landauer1991}, which as we mentioned in the first section,  is a particular formulation of the second law of thermodynamics well suited for the context of information theory.  The principle expresses the fact that erasure of data in a system necessarily involves producing heat, and thereby increasing the entropy.  We have illustrated the principle in  figure \ref{landauer}. 
Consider a ``gas'' consisting of a single atom in a symmetric container with volume 2V, in contact with a heat bath.  We imagine that the position of the particle acts as a memory with one bit of information,
corresponding to whether the atom is on the left or on the right.  Erasing the information amounts
to resetting the device to the ``reference" state $1$  independent of the
initial state. Erasure corresponds therefore to reinitializing the system rather then making a measurement. It can be done by first opening a diaphragm in
the middle, then reversibly moving the piston from the right in,
and finally closing the diaphragm and moving the piston back. In
the first step the gas expands freely to the double volume. The particle
doesn't do any work, the energy is conserved,  and therefore no heat
will be absorbed from the reservoir. This is an irreversible adiabatic process 
by which the entropy $S$ of the gas increases by a factor $k\ln2V/V = k\ln2$.  (The number of states the particle can be in is just the volume; the average velocity is conserved because of the contact with the thermal bath and will not contribute to the change in entropy). 
In the second part of the erasure procedure we bring the system back to a state which
has the same entropy as the initial state. We do this through a quasistatic (i.e. reversible) isothermal process at temperature T.
During the compression the entropy decreases by $k \ln2$. 
This change of entropy is nothing but the amount of heat delivered by the gas to the reservoir divided by the temperature, i.e. $ \Delta S = \int dS= \int dQ/T =\Delta Q/T$. The heat produced  $\Delta Q$ equals the net amount of work $W$ 
that has been done in the cycle by moving the piston during the compression.  The conclusion is that during the 
erasure of one bit of information the device had to produce at least $\Delta Q=kT\ln2$ of heat.

\begin{figure}[!h]
\begin{center}
\includegraphics[scale=0.7]{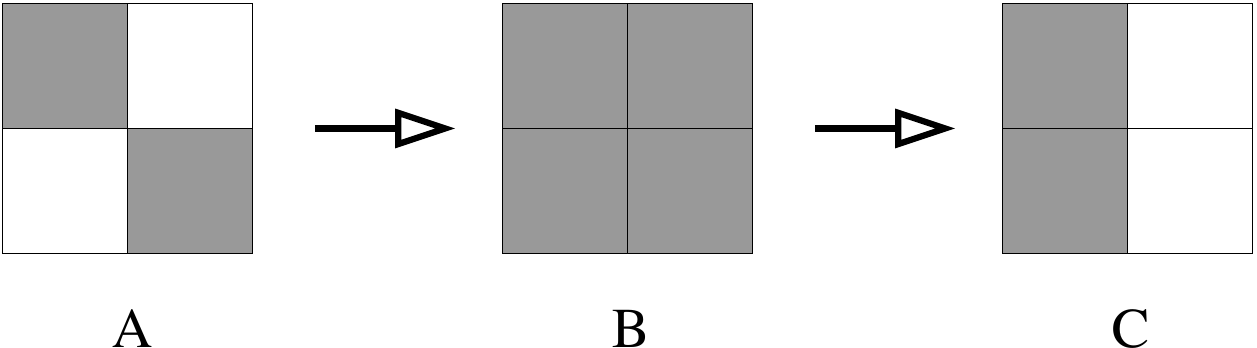}
\end{center}
\caption{A phase space picture of Landauer's principle. See text for an explanation.}
\label{erasure}
\end{figure}

We may look at the same process somewhat more abstractly, purely from the point of view of information.  We map the erasure of information for the simple  memory device on the sequence of diagrams depicted in figure \ref{erasure}.  
We choose this representation of the accessible (phase) space  to clearly mark the differences  between the situation where the particle is in the left {\em or} the right (A), the left {\em and} the right (B), and the left compartment only (C).  In part A the memory corresponds to the particle being either in the left or in the right compartment. In B the partition has  been removed and through the free expansion the  phase space has doubled and consequently the entropy increased by $\ln 2$.  In C the system is brought back to the reference state, i.e. the particle is brought in the left compartment. This is done  by moving a piston in from the right, inserting the partition, and moving the piston out again. It is in the compressing step that the phase space is reduced by a factor of two and hence entropy is reduced by $\ln 2$.  This is possible because we did work, producing a corresponding amount of heat ($\Delta Q \geq T\ln 2$). Note that in this representation one can in principle change the sizes of the partitions along the horizontal directions and the a priori probabilities along the vertical direction to model different types or aspects of memory  devices. 

\subsection{The entropy as a relative concept}

\begin{quote}
{\footnotesize Irreversibility is a consequence of the explicit
introduction of ignorance into the fundamental laws. \\ \mbox{}
\hfill M.~Born}
\end{quote}

There is a surprising amount of confusion about the interpretation
and meaning of the concept of entropy
\cite{guttman1999,denbigh1985}. One may wonder to what extent  the ``entropic
principle" just is an ``anthropocentric principle"?   That is, does
entropy depend only on our perception, or is it something more
fundamental? Is it a subjective attribute in the domain of  the
observer or is it an intrinsic property of the physical system we
study?
Let us consider the common definition of entropy as a measure of
disorder. This definition can be confusing unless we are careful
in spelling out what we mean by order or disorder. We may for instance look at the crystallization of a supercooled liquid under conditions where it is a
closed system, i.e. when no energy is exchanged with the environment.
Initially the molecules of the liquid are free to randomly move
about, but then (often through the addition of a small
perturbation that breaks the symmetry) the liquid suddenly turns into a solid 
by forming a crystal in which the molecules are pinned to the
sites of a regular lattice.  From one point of view this a
splendid example of the creation of order out of chaos.  Yet from
standard calculations in statistical mechanics we know that the
entropy increases during crystallization. This is because what
meets the eye is only part of the story.  During crystallization
entropy is generated in the form of latent heat, which is stored
in the vibrational modes of the molecules in the lattice.  Thus, even though in the
crystal the individual molecules are constrained to be roughly in
a particular location, they vibrate around their lattice
sites more energetically than when they were free to wander. From
a microscopic point of view there are more accessible states in
the crystal than there were in the liquid, and thus the entropy
increases.  The thermodynamic entropy is indifferent to whether
motions are microscopic or macroscopic -- it only counts the
number of accessible states and their probabilities.

In contrast, to measure the sense in which the crystal is more
orderly, we must measure a different set of probabilities.   To do
this we need to define probabilities that depend only on the
positions of the particles and not on their velocities.   To make
this even more clear-cut, we can also use a more macroscopic
partition, large enough so that the thermal motions of a molecule
around its lattice site tend to stay within the same partition
element. The entropy associated with this set of probabilities,
which we might call the ``spatial order entropy",  will behave
quite differently from the thermodynamic entropy.   For the
liquid, when every particle is free to move anywhere in the
container, the spatial order entropy will be high, essentially at
its largest possible value.  After the crystallization occurs, in
contrast, the spatial order entropy will drop dramatically.  Of
course, this is {\it not} the thermodynamic entropy, but rather an
entropy that we have designed to quantitatively capture the aspect
of the crystalline order that we intuitively perceive.

As we emphasized before, Shannon's great insight was that it is
possible to associate an entropy with any set of probabilities.
However, the example just given illustrates that when we use
entropy in the broader sense of Shannon we must be very careful to
specify the context of the problem.  Shannon entropy is just a
function that reduces a set of probabilities to a number,
reflecting how many nonzero possibilities there are as well as the
extent to which the set of nonzero probabilities is uniform or
concentrated. Within a fixed context, a set of probabilities that
is smaller and more concentrated can be interpreted as more
``orderly", in the sense that fewer numbers are needed to specify
the set of possibilities. Thermodynamics dictates a particular
context -- we have to measure probabilities in the full state
space. Thermodynamic entropy is a special case of Shannon entropy.
In the more general context of Shannon, in contrast, we can
define probabilities however we want, depending on what we want
to do. But to avoid confusion we must always be careful to keep
this context in mind, so that we know what our computation means.

\subsection{Maxwell's demon}
\begin{quote}
{\footnotesize The ``being" soon came to be called  Maxwell's demon, because of its far-reaching subversive effects on the natural order of things. Chief among these effects  would be to abolish the need for energy sources such as oil, uranium and sunlight. \\ \mbox{}
\hfill C.H.~Bennett}
\end{quote}
The second law of thermodynamics is statistical, deriving from the fact that the individual motions of the molecules are not observed or  controlled in any way. Would things be different if we could  intervene on a molecular scale?  This question gives rise to  an important paradox posed by Maxwell in 1872, which appeared in his \textit{Theory of Heat} \cite{maxwell1872}.  This has subsequently been discussed by generations of physicists, notably Szilard \cite{szilard29}, Brillouin\cite{Brillouin56}, Landauer \cite{landauer1961}, Bennett \cite{bennett1982} and others. 

Maxwell described his demonic setup as follows: ``Let us suppose that a vessel is divided in two 
portions, A and B, by a division in which there is a small hole, and that a being who can see individual 
molecules opens and closes this hole, so as to allow only the swifter particles to to pass from A to B, and only the slower ones to pass from B to A.   He will thus, without expenditure of work, raise the temperature of B and lower that of A, in contradiction with the second law of thermodynamics."  In attempts to save the second law from this demise, many aspects of the problem have been proposed for its resolution, including Brownian motion, quantum uncertainty and even G\"odel's Theorem.   The resolution of the paradox touches on some very fundamental issues that center on the question of how the demon might actually realize his subversive interventions. 

\begin{figure}[!t]
\begin{center}
\includegraphics[scale=0.6]{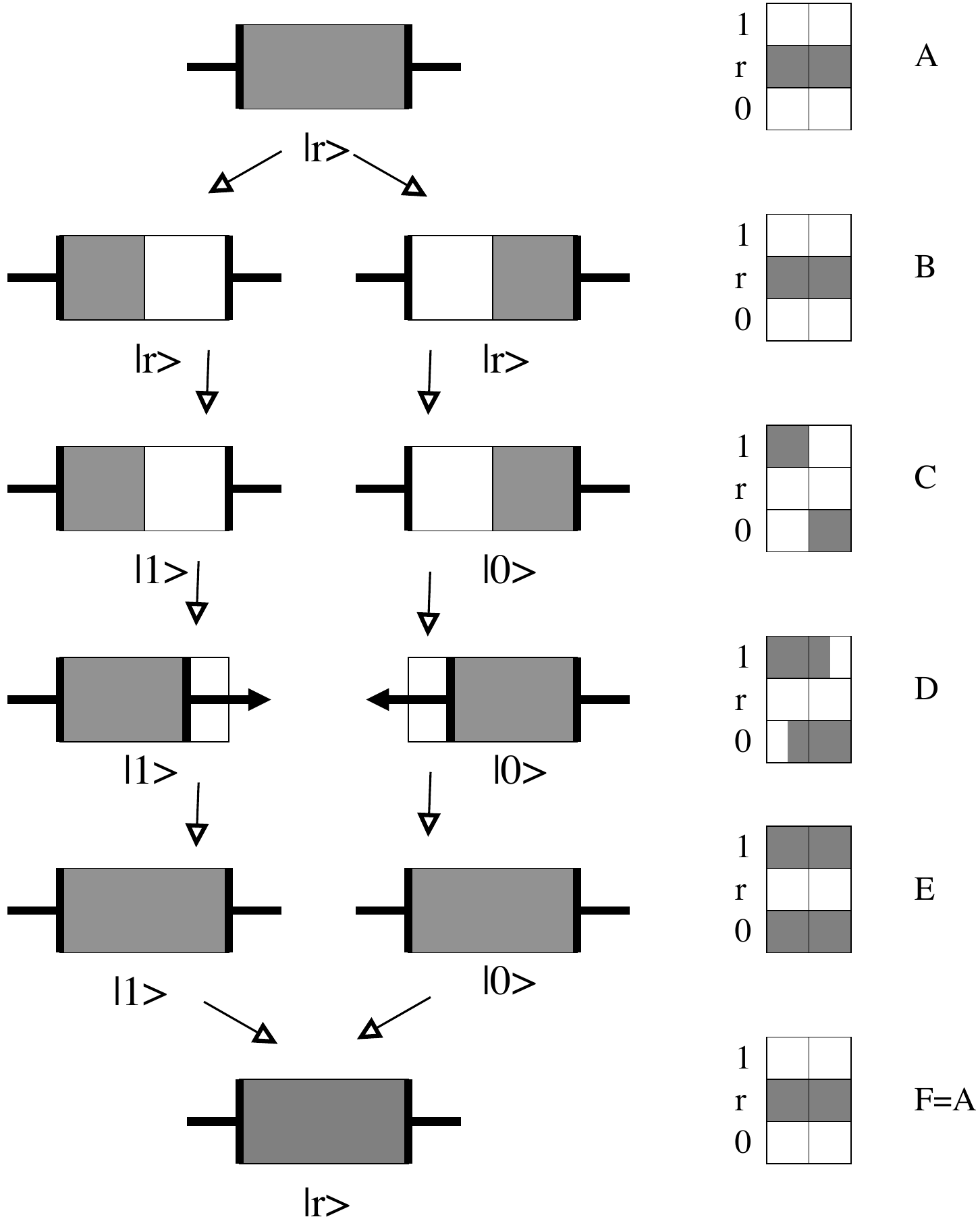}
\end{center}
\caption{The one-particle Maxwell demon apparatus as envisaged by Bennett \cite{bennett1982, bennett1987}. An explanation is given in the text.}
\label{demon}
\end{figure}

Szilard clarified the discussion by introducing an engine (or thermodynamic cycle), which is depicted in the left half of figure \ref{demon}.  He and Brillouin focused on the measurement the demon has to perform in order to find out in which half of the vessel the particle is located after the partition has been put into  place.  For the demon to ``see" the actual molecules he has to use a measurement device, such as a source of light (photons) and a photon detector.  He will in principle be able to measure whether a molecule is faster or slower then the thermal average by scattering a photon off of it.  Brillouin tried to argue that the entropy increase to the whole system once the measurement is included would always be larger or equal then the entropy gain achieved by the subsequent actions of the demon.  However, this argument didn't hold; people were able to invent devices that got around the measurement problem, so that it appeared the demon could beat the second law. 

Instead, the resolution of the paradox comes from a very different source.  In 1982 Bennett gave a completely different argument to rescue the second law.  The fundamental problem is that under Landauer's principle, production of heat is necessary for erasure of information (see section~\ref{sec:landauer}).   Bennett showed that a reversible measurement could in principle be made, so that Brillouin's original argument was wrong -- measurement does not necessarily produce any entropy.  However, to truly complete the thermodynamic cycle, the demon has to erase the information he obtained about the location of the gas molecule. As we already discussed in section~\ref{sec:landauer}, erasing that information produces entropy.  It turns out that the work that has to be done to erase the demon's memory is at least as much as was originally gained. 

Figure \ref{demon} illustrates the one-particle Maxwell demon apparatus as envisaged by Bennett \cite{bennett1982, bennett1987}, which is a generalization of the engine proposed by  Szilard \cite{szilard29}. On the left in row (A) is a gas container containing one molecule with a partition and two pistons.   On the right is a schematic representation of the phase space of the system, including the demon.  The state of mind of the demon can be in three different states:  He can know the molecule is on the right (state $0$), on the left (state $1$), or he can be in the reference or blank state $r$, where he lacks any information, and knows that he doesn't know where the particle is.  In the schematic diagram of the phase space, shown on the right, the vertical direction indicates the state of memory of the demon and the horizontal direction indicates the position of the particle.  In step (B) a thin partition is placed in the container, trapping the particle in either the left or right half.  In step (C) the demon makes a (reversible) measurement to determine the location of the particle.  This alters his state of mind as indicated -- if the particle is on the right, it goes into state $0$, if on the left, into state $1$. In step (D), depending on the outcome of the measurement, he moves either the right or left piston in and removes the partition. 
In (E) the gas freely expands, moving the piston out and thereby 
doing work. In state (E) it appears as if the system has returned 
to its original state -- it has the same volume, temperature and 
entropy -- yet work has been done. What's missing? The problem 
is that in (E) the demon's mind has not returned to its original 
blank state.  He needs to know that he doesn't know the position of the particle.  Setting the demon's memory back into its original state requires erasing a bit of information.  This is evident in the fact that to go from (E) to (F) the occupied portion of the phase space is reduced by a factor of two.  This reduction in entropy has to be accompanied by production of heat as a consequence of Landauer's principle (see figure ~\ref{landauer} and figure~\ref{erasure}) -- the work that is done to erase a bit of information is greater than or equal to the work gained by the demon.  This ensures that the full cycle of the complete system respects the second law after all.

This resolution of the paradox is remarkable, because it is not the acquisition of information (the measurement) which is irreversible and thermodynamically costly, but it is the process of erasure, which is both logically and thermodynamically irreversible, that leads to the increase of entropy required by the second law. The information comes for free, but it poses a waste disposal problem which is costly. It is gratifying to see information theory come to rescue of one of the most  cherished physical laws. 

\subsection{The Gibbs paradox}

The Gibbs paradox provides another interesting  chapter in the debate on the meaning of entropy. The basic question is to what extent entropy is a subjective notion. 
In its simplest form the paradox concerns the  mixing of
two ideal gases (kept at the same temperature and pressure) after
removing a partition. If it has been removed the gases will
mix, and if the particles of the two gases are distinguishable the entropy will increase due to this mixing.
However, if the gases are
\textit{identical}, so that their particles are indistinguishable from those on the other side, there is
no increase in the entropy.  Maxwell imagined the situation where the gases were
initially supposed to be identical, and only later recognized to be different.  This reasoning led to the
painful conclusion that the notion of irreversibility and
entropy would depend on our knowledge of physics. He concluded that the
entropy would thus depend on the state of mind of the experimenter
and therefore lacked an objective ground.  It was again Maxwell with a simple question who created  an uncomfortable situation which caused a long debate.  After the development of quantum mechanics, it became clear that particles of the same species are truly indistinguishable.  There is no such thing as labeling $N$ individual electrons, and therefore interchanging electrons doesn't change the state and this fact reduces the number of states by a relative factor of N!. Therefore the conclusion is that the 
entropy does not increase when the gases have the same constituent particles, and it does increase when they are different. 

However, the resolution of Gibbs paradox does not really depend on quantum mechanics.  Jaynes has emphasized that in the early works of Gibbs, the correct argument  was already given (well before the advent of quantum 
mechanics) \cite{jaynes96}.  Gibbs made an operational definition, saying that if ``identical" 
means anything, it means that there is no way an ``unmixing" apparatus could determine whether a particular 
molecule came from a given side of the box, short of having followed its entire trajectory.  Thus if the particles 
of the gas are identical in this sense, the entropy will not change.
We conclude that the adequate definition of entropy reflects the
objective physical constraints we put on the system, i.e. what measurements are possible or admissible. This has nothing to do with our lack of knowledge but rather with our choices. The `incompleteness of our
knowledge' is an exact and objective reflection of a particular
set of macroscopic constraints imposed on the
physical system we want to describe. The system's behavior depends
on these constraints, and so does  the entropy.

\subsection{The maximal entropy principle of Jaynes}

\begin{quote}
{\footnotesize  The statistical practice of physicists has tended
to lag about 20 years behind current developments in the field of
basic probability and statistics.
\\\mbox{}\hfill E.T.~Jaynes~(1963)}
\end{quote}

There are two equivalent sets of postulates that can be used as a
foundation to derive an equilibrium distribution in statistical
mechanics. One is to begin with the hypothesis that equilibrium
corresponds to a minimum of the free energy, and the other is that
it corresponds to a maimum of the entropy.  The latter approach
is a relatively modern development. Inspired by Shannon, Jaynes
turned the program of statistical mechanics upside down
\cite{jaynes1983}. Starting from a very general set of axioms he
showed that under the assumption of equilibrium the Gibbs
expression for the entropy is unique. Under Jaynes' approach, any
problem in equilibrium statistical mechanics is reduced to finding
the set of $p_i$ for which the entropy is maximal, under a set of
constraints that specify the macroscopic conditions, which may
come from theory or may come directly from observational data
\cite{jaynes1963}.  This variational approach removes some of the
arbitrariness that was previously present in the foundations of
statistical mechanics. The principle of maximum entropy is very
simple and has broad application.  For example if one maximizes
$S$ only under the normalization condition $\sum_i p_i=1$, then
one finds the unique solution that $p_i=1/N$ with $N$ the total
number of states.  This is the uniform probability distribution
underlying the equipartition principle.  Similarly, if we now add
the constraint that energy is conserved, i.e. $\sum_i
\varepsilon_i p_i= U$, then the unique solution is given by the
Boltzmann distribution,  equation (\ref{psubi}).  The
maximum entropy principle as a starting point clearly
separates the physical input  and purely probabilistic arguments that enter the theory. 
Let us derive the Maxwell-Boltzmann distribution to illustrate the maximal entropy principle.
We start with the  function  $L(p_i, \alpha, \beta)$ which depends on the probability distribution and two Lagrange multipliers to impose the constraints:
\begin{equation}
\label{eq: constrainedentropy }
L(p_i, \alpha, \beta) = - \sum_{i=1}^N p_i\ln p_i - \alpha (\sum_{i=1}^N p_i - 1) - \beta (\sum_{i=1}^N p_i \varepsilon_i -U)
\end{equation}
The maximum is determined by setting the partial derivatives of $L$ equal zero:
\begin{eqnarray} \label{eq:extremal}
\frac{\partial L}{\partial p_i} & = &  -\ln p_i -1-\alpha - \beta \varepsilon_i= 0\\
\frac{\partial L}{\partial \alpha} & = &  \sum_{i=1}^N p_i - 1   = 0 \\
 \frac{\partial L}{\partial \beta} & = & \sum_{i=1}^N p_i \varepsilon _i- U = 0 \label{eq:three}
 \end{eqnarray} 
From the first equation we immediately obtain that:
 \begin{equation}
\label{ eq:solutionfirst}
p_i = e^{-(1+\alpha+\beta \varepsilon_i)}  \Rightarrow p_i = \gamma e^{-\beta \varepsilon_i}
\end{equation} 
The parameters $\gamma$ and $\beta$ are determined by the constraint equations. If we first substitute the above solution in the normalisation constraint, and then use the defining equation for the partition sum (\ref{partitionfunction}), we find that $\gamma = 1/Z$. The solution for $\beta$ is most easily obtained using the following argument.
First substitute  (\ref{ eq:solutionfirst}) in the definition (\ref{S-psubi}) of $S$ to obtain the relation:
\begin{equation}
\label{U-S relation}
S= \beta U - const. \;.
\end{equation}
Next we use the  thermodynamic relation between energy and entropy (\ref{1stlaw2}), from which we obtain hat $\partial U/\partial S = T$. Combining these two relations we find that $\beta =1/T$, which yields the thermal equilibrium distribution (\ref{psubi}).
  
The \textit{maximal entropy formalism} has a much wider validity
than just statistical mechanics.   It is widely used for
statistical inference in applications such as optimizing data
transfer and statistical image improvement.   In these contexts it
provides a clean answer to the question, ``given the constraints I
know about in the problem, what is the model that is as random as
possible (i.e. minimally biased) subject to these constraints?". A
common application is missing data:  Suppose one observes a series
of points $x_i$ at regular time intervals, but some
of the observations are missing.  One can make a good guess for the missing values by solving for the distribution that maximizes the entropy, subject to the
constraints imposed by the know data points. 

One must always bear in mind, however, that in physics the maximum
entropy principle only implies to equilibrium situations, which
are only a small subset of the problems in physics.   For systems
that are not in equilibrium one must take a different approach.
Attempts to understand non-equilibrium statistical mechanics have
led some researchers to explore the use of alternative notions of
entropy, as discussed in Section~\ref{tsallis}.

\subsection{Ockham's razor}

\begin{quote}
{\footnotesize Entia non sunt multiplicanda praeter neccessitatem
\\ (\textit{Entities should not be introduced except when strictly necessary})
 \\\mbox{}\hfill William van Ockham (1285-1347)}
\end{quote}

An interesting and important application of information is to the process of modeling itself.  When developing 
a model it is always necessary to make a tradeoff between models that are too simple and fail to explain the 
data properly, and models that are too complicated and fit fluctuations in the data that are really just noise.    
The desirability of simpler models is often called ``Ockham's razor'':  If two models fit the existing data equally 
well, the simplest model is preferable, in the sense
that the simpler model is more likely to make good predictions for data that has not yet been seen.    While the 
value of using simple models seems like something we can all agree on, the tradeoff in real problems is 
typically not so obvious.  Suppose model A fits the data a little better than model B, but has one more 
parameter.  How does one trade off goodness of fit
against number of parameters?   

Using ideas from information theory Akaike \cite{Akaike74} introduced a method for making tradeoffs 
between goodness of fit and model complexity that can be applied in the context of simple linear models.  
Rissenen subsequently introduced a more general framework to think about this
problem based on a principle that he called minimum description length (MDL) \cite
{Rissanen78,Grunwald04,thisbook3}.   The basic idea is that the ability to make predictions and the ability to compress 
information are essentially two sides of the same coin.  We can only compress data if it contains regularities, 
i.e. if the structure of the data is at least partially predictable.  We can therefore find a good prediction model 
by seeking the model that gives the shortest description of the data we already have.  When we do this we 
have to take the description length of the model into account, as well as the description length of the 
deviations between the model's predictions and the actual data.   The deviations between the model and the 
data can be treated as probabilistic events.   A model that
gives a better fit has less deviation from the data, and hence
implies a tighter probability distribution, which translates into
a lower entropy for the deviations from the data.  This entropy is then added to the information
needed to specify the model and its parameters.  The best model is the one with the
lowest sum, i.e. the smallest total description length.  By characterizing the goodness of fit in terms of
bits, this approach puts the complexity of the model and the
goodness of fit on the same footing, and gives the correct
tradeoff between goodness of fit and model complexity, so that the
quality of any two models can be compared, at least in principle. 

This shows how at some level the concept of entropy underlies the whole scientific
method, and indeed, our ability to make sense out of the world.
To describe the patterns in the
world, we need to make a trade-off between overfitting
(fitting every bump even if it is a random variation, i.e. fitting noise) and
overgeneralization (identifying events that really are different). 
A similar trade-off  arises in assigning a causal mechanism  to the occurrence of an event or explaining it as random. This problem of how  to exactly make such trade-offs based on time series analysis has a rather  long history but on the other hand is still an active topic of research \cite{Kantz2006, Crutchfield1989, Still2007}.  Even if we do not do these trade-offs 
perfectly and do not think about it quantitatively, when we discover and model regularities in the world,
we are implicitly relying on a model selection process of this
type.  Any generalization makes a judgment that trades off the
information needed to specify the model and the entropy  of the
fit of the model to the world.

\subsection{Coarse graining and irreversibility}
\begin{quote}
{\footnotesize Our aim is not to `explain irreversibility' but to
describe and predict the observable facts. If one succeeds in
doing this correctly, from first principles, we will find that
philosophical questions about the 'nature of irreversibility' will
either have been answered automatically or else will be seen as
ill considered and irrelevant. \\ \mbox{} \hfill E.T.~Jaynes}
\end{quote}

The second law of thermodynamics says that for a closed system the
entropy will increase until it reaches its equilibrium value. This
corresponds to the irreversibility we all know from daily
experience.  If we put a drop of ink in a glass of water the drop
will diffuse through the water and dilute until the ink is
uniformly spread through the water.  The increase of entropy is
evident in the fact that the ink is initially in a small region,
with $p_i = 0$ except for this region, leading to a probability
distribution concentrated on a small region of space and hence a
low entropy. The system will not return to its original
configuration. Although this  is not impossible in principle, it
is so improbable that it will never be observed\footnote{``Never
say never" is a saying of unchallenged wisdom. What we mean here
by ``never", is inconceivably stronger then ``never in a
lifetime", or even ``never in the lifetime of the universe". Let's
make a rough estimate: consider a dilute inert gas, say helium,
that fills the left half of a container of volume $V$. Then we
release the gas into the full container and ask what the recurrence
time would be, i.e. how long it would take before all particles
would be in the left half again. A simple argument giving a reasonable
estimate, would be as follows:  At any given instant the
probability for a given particle to be in the left half is $1/2$,
but since the particles are independent, the probability of $N\sim
N_A$ particles to be in the left half is $ P=(1/2)^{{10}^{23}}
\approx 10^{(-10^{20})}$.  Assuming a typical time scale for completely
rearranging all the particles in the container of, say, $\tau_0 = 10^{-3}$ seconds, the typical time
that will pass before such a fluctuation occurs is $\tau = \tau_0/P = 10^{{10}^{20}} 10^{-3} \approx 10^{{10}^{20}}\; sec$.}.

Irreversibility is hard to understand from
the microscopic point of view because the microscopic laws of
nature that determine the time evolution of any physical system on
the fundamental level are all symmetric under time reversal.  That
is, the microscopic equations of physics, such as $F = ma$, are
unchanged under the substitution $t \to -t$.  How can
irreversibility arise on the macroscopic level if it has no
counterpart on the microscopic level?

In fact, if we compute the entropy at a completely microscopic
level it is conserved, which seems to violate the second law of
thermodynamics. This follows from the fact that momentum is
conserved, which implies that volumes in phase space are
conserved.  This is called Liouville's theorem.  It is easy to prove that this
implies that the entropy $S$ is conserved. This doesn't depend on
the use of continuous variables -- it only depends on applying the
laws of physics at the microscopic level.  It reflects the idea of
Laplace, which can be interpreted as a statement that statistical
mechanics wouldn't really be necessary if we could only measure
and track all the little details.
The ingenious argument that Gibbs used to clarify this, and
thereby to reconcile statistical mechanics with the second law of
thermodynamics, was to introduce the notion of \textit{coarse
graining}. This procedure corresponds to a systematic description
of what we could call ``zooming out''. As we have already
mentioned, this zooming out involves dividing phase space up in
finite regions $\delta$ according to a partition $\Pi$. Suppose,
for example, that at a microscopic level the system can be
described by discrete probabilities $p_i$ for each state.  Let us
start with a closed system in equilibrium, with a uniform
distribution over the accessible states. For the Ising system, for
example, $p_i = 1/g(N, i)$ is the probability of a particular
configuration of spins.  Now we replace in each little region
$\delta$ the values of  $p_i$ by its average value $\bar{p}_i$
over $\delta$:
\begin{equation}\label{coarse-p}
  \bar{p}_i \equiv \frac{1}{\delta}\sum_{i\in\delta} \; p_i,
\end{equation}
and consider the associated coarse grained entropy
\begin{equation}\label{coarse-S}
  \bar{S}\equiv -\sum_i\; \bar{p}_i\ln \bar{p}_i.
\end{equation}
Because we start  at time $t=0$ with  a uniform probability distribution, 
$S(0)=\bar{S}(0)$. Next we change the situation 
by removing a constraint of the system so
that it is no longer in equilibrium. In other words, we enlarge the
space of accessible states but have as an initial condition that
the probabilities are zero for the new states. For the new situation we still have that $S(0)=\bar{S}(0)$, and now  we can  compare
the evolution of the fine-grained entropy $S(t)$ and the
coarse-grained entropy $\bar{S}(t)$.  The evolution of $S(t)$ is
governed by the reversible microscopic dynamics and therefore it
stays constant, so that $S(t) = S(0)$. To study the evolution of
the coarse-grained entropy we can use a few simple mathematical
tricks. First, note that because $\bar{p}_i$ is constant over each
region with $\delta$ elements,
\begin{equation}\label{S1}
  \bar{S}(t)= -\sum_i \bar{p}_i \ln \bar{p}_i = -\sum_i p_i \ln \bar{p}_i.
\end{equation}
Then we may write
\begin{equation}\label{S-difference}
  \bar{S}(t)- \bar{S}(0)  =  \sum_i p_i (\ln p_i- \ln \bar{p}_i)
  =\sum_i p_i \ln \frac{p_i}{\bar{p}_i} = \sum_i \bar{p}_i
  (\frac{p_i}{\bar{p}_i}\ln
  \frac{p_i}{\bar{p}_i}),
\end{equation}
 which in information theory is called he Kullback-Leibler divergence.
The mathematical inequality $ x\ln x \geq (x-1)$, with $x = p_i/\bar{p_i}$, then implies the Gibbs inequality:
\begin{equation}\label{S-difference2}
  \bar{S}(t)- \bar{S}(0) \geq \sum_i p_i - \sum_i \bar{p}_i = 1-1 = 0.
\end{equation}
Equality only occurs if $p_i/\bar{p}_i =1$ throughout, so
except for the special case where this is true, this is a strict
inequality and the entropy increases. We see how the second law is
obtained as a consequence of coarse graining.

The second law describes mathematically the irreversibility we
witness when somebody blows smoke in the air. Suppose we make a
film of the developing smoke cloud. If we film the movie at an
enormous magnification, so that what we see are individual
particles whizzing back and forth, it will be impossible to tell
which way the movie is running -- from a statistical point of view
it will look the same whether we run the movie forward or
backward.  But if we film it at a normal macroscopic scale of
resolution, the answer is immediately obvious -- the direction of
increasing time is clear from the diffusion of the smoke from a
well-defined thin stream to a diffuse cloud.

From a philosophical point of view one should ask to what extent
coarse graining introduces an element of subjectivity into the
theory. One could object that the way we should coarse grain is
not decided upon by the physics but rather by the person who
performs the calculation.  The key point is that, as in so many
other situations in physics, we have to use some common sense, and
distinguish between observable and unobservable quantities.
Entropy does not increase in the highly idealized classical world
that Laplace envisioned, as long as we can observe all the
microscopic degrees of freedom and there are no chaotic dynamics.
However, as soon as we violate these conditions and observe the
world at a finite level of resolution (no matter how accurate),
chaotic dynamics ensures that we will lose information and entropy
will increase. While the coarse graining may be subjective, this
is not surprising -- measurements are inherently subjective
operations.   
In most systems one will have that the entropy 
may stabilize on plateaus corresponding to certain ranges of the fineness
of the coarseness.  In many applications the increase of entropy will therefore be constant (i.e. well defined) for a sensible choice for the scale  of coarse graining.  
The increase in (equilibrium) entropy between the microscopic scale 
and the macroscopic scale can also be seen as the amount of information 
that is lost by increasing the graining scale  from the microscopic to the macroscopic.
A relevant remark at this point is that a system is of course never perfectly closed
-- there are always small perturbations from the environment that
act as a stochastic perturbation of the system, thereby
continuously smearing out the actual distribution in phase space
and simulating the effect of coarse graining.  Coarse graining
correctly captures the fact that entropy is a measure of our
uncertainty; the fact that this uncertainty does not exist for
regular motions and perfect measurements is not relevant to most
physical problems. 

\subsection{Coarse graining and renormalization}
In a written natural language not all finite combinations of
letters are words, not all finite combinations of words are
sentences, and not all finite sequences of sentences make sense.
So by identifying what we call meaningful with accessible, what we
just said means that compared with arbitrary letter combinations,
the entropy of a language is extremely small.

Something similar is true for the structures studied in science. We are used to
thinking of the rich diversity of biological, chemical and physical
structures as being enormous, yet relative to what one might
imagine, the set of possibilities is highly constrained.
The complete hierarchy starting from
the most elementary building blocks of matter such as
\textit{leptons} and \textit{quarks}, all the way up to living
organisms, is surprisingly restricted. This has to
do with the very specific nature of the interactions between these
building blocks.  To our knowledge at the microscopic level there are only four fundamental forces
that control all interactions.
At each new structural level (quarks, protons and neutrons, nuclei, atoms,
molecules, etc) there is a more or less autonomous theory
describing the physics at that level involving only the relevant
degrees of freedom at that scale.  Thus moving
up a level corresponds to throwing out an enormous part
of the phase space available to the fundamental degrees of freedom in the
absence of interactions.  For example, at the highest, most macroscopic
levels of the hierarchy only the long range interactions
(electromagnetism and gravity) play an important role -- the structure
of quantum mechanics and the details of the other two fundamental forces
are more or less irrelevant. 
 
We may call the structural hierarchy we just described ``coarse
graining" at large.  Although this ability to leave the details of each level behind 
in moving up to the next is essential to science, there is no cut and dried procedure
that tells us how to do this.  The only exception is that in some situations it is possible to do this coarse 
graining exactly by a procedure called
 \textit{renormalization} \cite{zinnjustin1989}.  This is done by systematically
studying how a set of microscopic degrees of freedom at one level can be averaged
together to describe the degrees of freedom at the next level.  There are some situations,
such as phase transitions, where this process can then be used repeatedly to
demonstrate the existence of fixed points of the mapping from one level to the next (an example of a phase 
transition is the change from a liquid to a gas).  This procedure has provided important insights in the nature of phase 
transitions, and in many cases it has been shown that some of their properties are universal, in the sense that they do not 
depend on the details of the microscopic interactions.

\subsection{Adding the entropy of subsystems}

Entropy is an extensive quantity. Generally speaking the extensivity of entropy means that
it has to satisfy the fundamental linear scaling property
\begin{equation}\label{entropyscaling}
  S(T,qV,qN) = qS(T,V,N), \;\;\; 0 < q <  \infty.
\end{equation}
Extensivity translates into additivity of entropies:  If we combine
two noninteracting systems (labelled 1 and 2) with entropies
$S_1$ and $S_2$, then the total number of states will just be the
product of those of the individual systems.  Taking the
logarithm, the entropy of the total system $S$ becomes:
\begin{equation}\label{entropysum}
  S = S_1 + S_2.
\end{equation}
Applying this to two spin systems without an external field, the
number of states of the combined system is $w = 2^{N_1 + N_2}$, i.e. $w=w_1\; w_2$.  Taking the logarithm establishes
the additivity of entropy.

However if we allow for a nonzero magnetic field, this result is
no longer obvious. In Section
\ref{spins} we calculated the number of configurations with a
given energy $\varepsilon_k = - k \mu H$ as $g(N,k)$. If we now
allow  two systems  to exchange energy but keep the total energy
fixed, then this generates a dependence between the two systems
that lowers the total entropy.  We illustrate this with an example:

Let the number of spins pointing up in system $1$ be $k_1$ and the
number of particles be $N_1$, and similarly let this be $k_2$ and
$N_2$  for system $2$. The total energy $k = k_1 + k_2$ is
conserved, but the energy in either subsystem ($k_1$ and $k_2$) is
not conserved.   The total number of spins, $N = N_1 + N_2$ is
fixed, and so are the spins ($N_1$ and $N_2$) in either subsystem.
Because the systems only interact when the number of up spins in
one of them (and hence also the other one) changes, we can write
the total number of states for the combined system as
\begin{equation}\label{combined}
g(N,k) = \sum_{k_1} g_1(N_1,k_1)g_2(N_2,k_2),
\end{equation}
where we are taking advantage of the fact that as long as $k_1$ is
fixed, systems one and two are independent. Taking the log of the
above formula clearly does not lead to the  additivity of
entropies because we have to sum over $k_1$. This little
calculation  illustrates the remark made before:  Since we
have relaxed the constraint that each system has a fixed energy to
the condition that only the sum of their energies is fixed, the
number of accessible states for the total system is increased.  The
subsystems themselves are no longer closed and therefore the
entropy will change.

The extensivity of entropy is recovered in the thermodynamic limit in the above
example, i.e. when $N \rightarrow \infty$.  Consider the contributions to the sum in (\ref{combined})
as a function of $k_1$, and let the value of $k_1$ where $g$ reaches a maximum be
$k_1=\hat{k}_1$.   We can now write the contribution in the sum in terms
$\delta = k_1 - \hat{k}_1$ as
\begin{equation}\label{combined2}
\Delta g(N,k) = g_1(N_1,\hat{k}_1+\delta)g_2(N_2,\hat{k}_2-
\delta)=  f(\delta)g_1(N_1,\hat{k}_1)g_2(N_2,\hat{k}_2)\;,
\end{equation}
where the correction factor can be calculated by expanding the $g$
functions around their respective $\hat{k}$ values. Not surprisingly, in the limit
where $N$ is large it
turns out that $f$ is on the order of $f \sim \exp(-2\delta^2)$ so
that the contributions to $g(N,k)$ of the nonmaximal terms in the
sum (\ref{combined}) are exponentially suppressed. Thus in
the limit that the number of particles goes to infinity
the entropy becomes additive. This exercise shows
that when a system gets large we may replace the averages of
a quantity  by its value in the most probable configuration, as
our intuition would have suggested. From a mathematical point of
view this result follows from the fact that the binomial
distribution approaches a gaussian for large values of N, which becomes
ever sharper as $N \rightarrow \infty$.  This simple example shows that the extensivity of
entropy may or may not be true, depending on the context of the physical situation and in particular on the range of the inter-particle forces.

When two subsystems interact,
it is certainly possible that the entropy of one decreases at the
expense of the other. This can happen, for example, because system
one does work on system two, so the entropy of system one goes up
while that of system two goes down.  This is very important for
living systems, which collect free energy from their environment
and expel heat energy as waste. Nonetheless, the total entropy $S$
of an organism plus its environment still increases, and
so does the sum of the independent entropies of the non
interacting subsystems. That is, if at time zero
\begin{equation}\label{combined3}
S(0) = S_1(0) + S_2(0)\; ,
\end{equation}
then at time $t$ it may be true that
\begin{equation}\label{combined4}
S(t) \leq S_1(t) +S_2(t)\; ,
\end{equation}
This is due to the fact that only interactions with other parts of
the system can lower the entropy of a given subsystem. In such a
situation we are of course free to call the difference between the
entropy of the individual systems and their joint entropy a
\textit{negative} correlation entropy. However, despite this
apparent decrease of entropy, both the
total entropy and the sum of the individual entropies can only
increase, i.e.
\begin{eqnarray}\label{combined5}
  S(t) &\geq& S(0) \\ \nonumber
S_1(t) + S_2(t)&\geq& S_1(0) + S_2(0).
\end{eqnarray}
The point here is thus that equations (\ref{combined4}) and
(\ref{combined5}) are not in conflict.

\subsection{Beyond the Boltzmann, Gibbs and Shannon entropy: the
Tsallis entropy \label{tsallis}}
\begin{quote}
{\footnotesize The equation $S= k \log W\; +\; const$
appears without an elementary theory - or however one wants to say
it - devoid of any meaning from a phenomenological point of view.
\\\mbox{}\hfill A.~Einstein~(1910)}
\end{quote}

As we have already stressed, the definition of entropy as $-
\sum_i p_i \log p_i$ and the associated exponential distribution
of states apply only for systems in equilibrium.  Similarly, the
requirements for an entropy function as laid out by Shannon and
Khinchin are not the  only possibilities. By modifying these
assumptions there are other entropies that are useful.  We have
already mentioned the R\'enyi entropy, which has proved to be
valuable to describe multi-fractals.

Another context where considering an alternative definition of entropy
appears to be useful concerns power laws.  Power laws are ubiquitous in both natural and social
systems.  A {power law}\footnote{It is also possible to have a
power law at zero or any other limit, and to have $\alpha <  0$,
but for our purposes here most of the examples of interest involve
the limit $x \to \infty$ and positive $\alpha$.} is something that
behaves for large $x$ as $f(x) \sim x^{-\alpha}$, with $\alpha > 
0$.  Power law probability distributions decay much more slowly
for large values of $x$ than exponentials, and as a result have
very different statistical properties and are less well-behaved\footnote{The $m^{th}$ moment $\int x^m p(x) dx
$ of a power law distribution $p(x) \sim x^{-\alpha}$ does not exist when $m >  \alpha$.}.
Power law distributions are observed in phenomena as diverse as
the energy of cosmic rays, fluid turbulence, earthquakes, flood
levels of rivers, the size of insurance claims, price
fluctuations, the distribution of individual wealth, city size,
firm size, government project cost overruns, film sales, and word
usage frequencies \cite{Newman05,Farmer06}.   Many different models can produce
power laws, but so far there is no unifying theory, and it is not
yet clear whether any such unifying theory is even possible.   It
is clear that power laws (in energy, for instance) can't be explained by equilibrium
statistical mechanics, where the resulting distributions are
always exponential.   A common property of all the
physical systems that are known to have power laws and the models
that purport to explain them is that they are in some sense nonequilibrium
systems.  The ubiquity of power laws suggests that there might be
nonequilibrium generalizations of statistical mechanics for which
they are the standard probability distribution in the same way
that the exponential is the standard in equilibrium systems.

From simulations of model systems with long-range interactions
(such as stars in a galaxy) or systems that remain for long
periods of time at the ``edge of chaos", there is mounting
evidence that such systems can get stuck in nonequilibrium
meta-stable states with power law probability distributions for very long periods of time before they finally
relax to equilibrium.  Alternatively, power laws also occur in
many driven systems that are maintained in a steady state away from equilibrium.
Another possible area of applications is describing the behaviour of small subsystems of finite systems. 

From a purely statistical point of view it is interesting to ask
what type of entropy functions are allowed.  The natural assumption
to alter is the last of the Khinchin postulates as discussed in Section 5.2. 
The question becomes which entropy
functions satisfy the remaining  two conditions, and some sensible
alternative for the third? It turns out that there is at least one
interesting class of solutions called q-entropies introduced in
1988 by Tsallis \cite{tsallis1988,gellmann2004}. The parameter $q$
is usually referred to as the \textit{bias} or
\textit{correlation} parameter.   For $q \ne 1$ the expression for
the q-entropy $S_q$ is
\begin{equation}\label{q-entropy}
S_q[p] \equiv \frac{1- \sum_i p_i^q}{q-1}.
\end{equation}
For $q=1$, $S_q$ reduces to the  standard Gibbs entropy by taking
the limit as $q \to 1$.  Following Jaynes's approach to
statistical mechanics, one can maximize this entropy function
under suitable constraints to obtain distribution functions that
exhibit power law behavior for $q\neq 1$.  These functions are
called q-exponentials and are defined as
\begin{equation}\label{q-exp}
  e_q(x) \equiv \left\{ \begin{array}{ll}
    [1+(1-q)x]^{1/(1-q)} & (1+(1-q)x\rangle 0)\\
    0  & (1+(1-q)x\langle 0).
  \end{array}\right.
\end{equation}
An important property of the q-exponential function is that for $q >  1$ and $x  \ll -1$ it has a  power law decay.
The inverse of the q-exponential is the  $\ln_q(x)$ function
\begin{equation}\label{q-log}
  \ln_q \equiv \frac{x^{1-q} -1}{1-q}.
\end{equation}
The q-exponential can also be obtained as the solution of the equation
\begin{equation}
\frac{dx}{dt} = x^q.
\end{equation}
This is the typical behavior for  a dynamical system at the edge
of linear stability, where the first term in its Taylor series
vanishes.  This gives some alternative insight into one possible
reason why such solutions may be prevalent.  Other typical
situations involve long range interactions (such as the
gravitational interactions between stars in galaxy formation) or
nonlinear generalizations of the central limit theorem \cite{Umarov06} for variables with strong
correlations.

At first sight a problem with q-entropies is that for
$q\neq 1$ they are not additive.  In fact the following equality holds:
\begin{equation}\label{q-additivity}
S_q[p^{(1)}p^{(2)}] = S_q[p^{(1)}] + S_q[p^{(2)}] +
(1-q)S_q[p^{(1)}] S_q[p^{(2)}]
\end{equation}
with the corresponding product rule for the q-exponentials:
\begin{equation}\label{q-product}
e_q(x)e_q(y) =e_q(x +y +(1-q)xy)
\end{equation}
This is why the q-entropy is often referred to as a non-extensive
entropy.  However, this is in fact a blessing in disguise.  If the  appropriate type of scale invariant correlations
between subsystems are typical, then the q-entropies for $q \neq
1$ are strictly additive.  When there are sufficiently long-range interactions Shannon entropy is not extensive; Tsallis entropy provides a substitute by that is additive (under the right class of long-range interactions), thereby capturing an underlying regularity with a simple description. 

This alternative statistical mechanical theory involves another
convenient definition which makes the whole formalism look like
the ``old" one. Motivated by the fact that  the Tsallis entropy
weights all probabilities according to $p_i^q$, it is possible to
define an ``escort" distribution $P^{(q)}_i$ 
\begin{equation}\label{escort}
P^{(q)}_i \equiv \frac{(p_i)^q}{\sum_j (p_j)q},
\end{equation}
as introduced by Beck \cite{beck2001}.  One can then define the corresponding
expectation values of a variable $A$ in terms of the escort
distribution as
\begin{equation}\label{q-averages}
  \langle A\rangle _q =\sum_i P^{(q)}_i A_i.
\end{equation}
With these definitions the whole formalism runs parallel to the
Boltzmann-Gibbs program.

One can of course ask what the Tsallis entropy ``means".  The entropy $S_q$ is a measure of lack of information along the same lines as the Boltzmann-Gibbs-Shannon entropy is. In particular, perfect knowledge of the microscopic state of the system yields $S_q=0$, and maximal uncertainty (i.e., all $W$ possible microscopic states are equally probable) yields maximal entropy, $S_q=\ln_q W$. 
The question remains how generic such
correlations are and which physical systems exhibit them, though
at this point quite a lot of empirical evidence is accumulating to
suggest that such functions are at least a good approximation in
many situations.  In addition recent results have shown that q-exponentials
obey a central limit-like behavior for combining random variables with appropriate long-range correlations. 

A central question is what determines $q$? There is a class of natural, artificial and social systems  for which it is possible to choose a unique value of q such that the entropy is simultaneously extensive (i.e., $S_q(N)$ proportional to the number of elements $N$, $N \gg 1$) and there is finite entropy production per unit time (i.e., $S_q(t)$ proportional to time $t$, $t \gg 1$)\cite{tsallis2005a,tsallis2005b}. It is possible to acquire some intuition about the nature and meaning of the index q through the following analogy:  If we consider an idealized planar surface, it is only its $d=2$ Lebesgue measure which is finite; the measure for any $d > 2$ vanishes, and that for any $d < 2$ diverges. If we have a fractal system, only the $d=d_f$ measure is finite, where $d_f$ is the Hausdorff dimension; any $d > d_f$ measure vanishes, and any $d < d_f$ measure diverges. Analogously, only for a special value of $q$ does the entropy $S_q$ match the thermodynamical requirement of extensivity and the equally physical requirement of finite entropy production.  The value of $q$ reflects the geometry of the measure in phase space on which probability is concentrated.

Values of $q$ differing from unity are consistent with the recent $q$-generalization of the Central Limit Theorem and the alpha-stable (Levy) distributions. Indeed, if instead of adding a large number of exactly or nearly independent  random variables, we add globally correlated random variables, the attractors shift from Gaussians and Levy distributions to q-Gaussian and ($q$,$\alpha$)-stable distributions respectively \cite{moyano2006,Umarov06,umarov2006b}.

The framework described above is still in development.  It may
turn out to be relevant to `statistical mechanics' not only in
nonequilibrium physics, but  also in quite different arenas, such
as economics.

\section{Quantum information}
\begin{quotation} {\footnotesize
Until recently, most people thought of quantum mechanics in terms of the uncertainty principle and unavoidable limitations on measurement. Einstein and Schr\"odinger understood early on the importance of entanglement, but most people failed to notice, thinking of the EPR paradox as a question for philosophers. The appreciation of the positive application of quantum effects to information processing grew slowly.\\
\mbox{}\hfill Nicolas Gisin
}
\end{quotation} 

Quantum mechanics provides a fundamentally different means of computing, and potentially makes it possible to solve problems that would be intractable on classical computers.  For example, with a classical computer the typical time it takes to 
factor a number grows exponentially with the size of the number, but using quantum computation Shor has shown that this can be done in polynomial time \cite{shor1994}.  Factorization is one of the main tools in cryptography, so this is not just a matter of academic interest.  To see the huge importance of exponential vs. polynomial scaling, suppose an elementary computational step takes $\Delta t$ seconds.  If the number of steps increases exponentially, factorizing a number with $N$ digits will take $\Delta t \exp(aN)$ seconds, where $a$ is a constant that depends on the details of the algorithm.  For example, if $\Delta t = 10^{-6}$ and $a = 10^{-2}$, factoring a number with $N = 10,000$ digits will take $10^{37}$ seconds, which is much, much longer than the lifetime of the universe (which is a mere $4.6 \times 10^{17}$ seconds).  In contrast, if the number of steps scales as the third power of the number of digits, the same computation takes $a' \Delta t N^3$ seconds, which with $a' = 10^{-2}$, is $10^4$ seconds or a little under three hours.  Of course the constants $a$, $a'$ and $\Delta t$ are implementation dependent, but because of the dramatic difference between exponential vs. polynomial scaling, for sufficiently large $N$ there is always a fundamental difference in speed. 
In fact for the factoring problem as such, the situation is more subtle: at present the best available classical algorithm  requires $\exp(O(n^{1/3} \log^{2/3} n))$ operations, whereas the best available quantum algorithm would require $O(n^2 \log n \log\log n)$ operations.  Factorization is only one of several problems that could potentially benefit from quantum computing.  The implications go beyond quantum computing, and include diverse applications such as quantum cryptography and quantum communication \cite{nielsen1990, kaye2007,mermin2007,lloyd2008}.

The possibility for such huge speed-ups comes from the intrinsically parallel nature of quantum systems.  The reasons for this are sufficiently subtle that it took many decades after the discovery of quantum mechanics before anyone realized that its computational properties are fundamentally different.  The huge interest in quantum computation in recent years has caused a re-examination of the concept of information in physical systems, spawning a field that is sometimes referred to as ``quantum information theory".

Before entering the specifics of quantum information and computing, we give a brief introduction to the basic setting  of quantum theory and contrast it with its classical counterpart. We describe the physical states of a quantum systems, the definition of quantum observables, and time evolution according to the Schr{\"o}dinger equation. Then we briefly explain the measurement process, the basics of quantum teleportation and quantum computation. To connect to classical statistical physics we describe the density matrix and the von Neumann entropy.  Quantum computation in practice involves sophisticated and highly specialized subfields of experimental physics which are beyond the scope of this brief review -- we have tried to limit the discussion to the essential principles.  

\subsection{Quantum states and the definition of a qubit}
In classical physics we describe the state of a system by specifying the values of dynamical variables, for example, the position and velocity of a particle at a given instant in time. The time evolution is then described by Newton's laws, and any uncertainty in its evolution is driven by the accuracy of the measurements.  As we described in Section~\ref{chaos}, uncertainties can be amplified by chaotic dynamics, but within classical physics there is no fundamental limit on the accuracy of measurements -- by measuring more and more carefully, we can predict the time evolution of a system more and more accurately.  At a fundamental level, however, all of physics behaves according to the laws of quantum mechanics, which are very different from the laws of classical physics.  At the macroscopic scales of space, time and energy where classical physics is a good approximation, the predictions of classical and quantum theories have to be roughly the same, a statement that is called the {\it correspondence principle}.   Nonetheless, understanding the emergence of classical physics from an underlying quantum description is not always easy.
  
The scale of the quantum regime is set by Planck's constant, which has dimensions of $energy \times time$ (or equivalently  $momentum \times length$).  It is extremely small in ordinary units\footnote{We are using the reduced Planck's constant, $\hbar = h/2 \pi$.}: $\hbar=1.05 \times 10^{-34}$ Joule-seconds. 
This is why quantum properties only manifest themselves at very small scales or very low temperatures.  One has to keep in mind however, that  radically different properties at a microscopic scale (say at the level of atomic and molecular structure) will also lead to fundamentally different collective behavior on a macroscopic scale.  Most phases of condensed matter realized in nature, such as crystals, super, ordinary or semi-conductors or magnetic materials,   can only be understood from the quantum mechanical perspective.   The stability and structure of matter is to a large extent a consequence of the quantum behavior of its fundamental constituents.    

To explain the basic ideas of quantum information theory we will restrict our attention to systems of \textit{qubits}, which can be viewed as the  basic building blocks of quantum information systems.  The physical state of a quantum system is described by a wavefunction  that can be thought of a vector in an abstract multidimensional space, called a {\it Hilbert space}.   For our purposes here, this is just a finite dimensional vector space where the vectors have complex rather than real coefficients, and where the length of a vector is the usual length in such a space, i.e. the square root of the sum of the square amplitudes of its components\footnote{More generally it is necessary to allow for the possibility of infinite dimensions, which introduces complications about the convergence of series that we do not need to worry about here.}.  Hilbert space replaces the concept of phase space in classical mechanics.   
Orthogonal basis vectors defining the axes of the space correspond to different values of measurable quantities, also called observables, such as spin, position, or momentum.  

As we will see, an important difference from classical mechanics is that many quantum mechanical quantities, such as position and momentum or spin along the $x$-axis and spin along the $y$-axis, cannot be measured simultaneously.  Another essential difference from classical physics is that the dimensionality of the state space of the quantum system is huge compared to that of the classical phase space.  To illustrate this drastic difference think of a particle that can move along an infinite line with an arbitrary momentum.  From the classical perspective it has a phase space that  is  two dimensional and real (a position $x$ and a momentum $p$ ), but from the quantum point of view it it is given by a wavefunction  $\Psi$ of one variable (typically the position $x$ or the momentum $p$). This wave function corresponds to an element in an infinite dimensional Hilbert space.

We discussed the classical Ising spin in section \ref{spins}.  It is a system with only two states, denoted by $s= \pm 1$, called spin up or spin down, which can be thought of as representing a classical bit with two possible states, ``0" and ``1".  The quantum analog of the Ising spin is a very different kind of animal.  Where the Ising spin corresponds to a classical bit, the quantum spin corresponds to what is called a \textit{qubit}. As we will make clear in a moment, the state space of a qubit is much larger then that of its classical counterpart, making it possible to store much more information.
This is only true in a certain sense, as one has to take into account to what extent the state is truly observable and whether it can be precisely prepared, questions we will return to later.  

Any well-defined two level quantum system can be thought of as representing a qubit.  Examples of two state quantum systems are a photon, which possesses two polarization states, an electron, which possesses two possible spin states, or a particle in one of two possible energy states.  In the first two examples the physical quantities in the Hilbert space are literally spins, corresponding to angular momentum, but in the last example this is not the case.  This doesn't matter -- even if the underlying quantities have nothing to do with angular momentum, as long as it is a two state quantum system we can refer to it as a ``spin".  We can arbitrarily designate one quantum state as ``spin up", represented by the symbol $| 1 \rangle$, and the other ``spin down", represented by the symbol $| 0 \rangle$.   

The state of a qubit is described by a wavefunction or state vector $|\psi\rangle $, which can be written as
\begin{equation}
|\psi\rangle  = \alpha |1\rangle  + \beta |0\rangle   \mbox{ with }  |\alpha|^2 +|\beta|^2= 1. 
\label{twoSpinQuantum}
\end{equation}
Here $\alpha$ and $\beta$ are complex numbers\footnote{A complex number $\alpha$ has a real and imaginary part $\alpha= a_1+i a_2$, where $a_1$ and $a_2$ are both real, and $i$ is the imaginary unit with the property $i^2=-1$.  Note that a complex number can therefore also be thought of as a vector in a two dimensional real space.The complex conjugate is defined as $\alpha^*= a_1-i a_2$ and the square of the modulus, or absolute value, as $|\alpha|^2=\alpha^* \alpha= a_1^2+a_2^2.$  }, and thus we can think of $|\psi\rangle $ as a vector in the  2-dimensional complex vector space, denoted $\mathbf{C}^2$, and we can represent the state as a column vector $\begin{pmatrix}
  \alpha  \\
  \beta 
\end{pmatrix}$.  We can also define a dual vector space in $\mathbf{ C}^2$ with dual vectors that can either be represented as row vectors or alternatively be written 
\begin{equation}
\langle \psi| = \langle 0|\alpha^*  + \langle 1|\beta^* \;. 
\label{dualspin}
\end{equation}
This allows us to define the inner product between two state vectors  $|\psi\rangle $ and $|\phi\rangle = \gamma|1\rangle  + \delta |0\rangle $ as
\begin{equation}
\label{ innerproduct}
\langle \phi|\psi\rangle =\langle \psi|\phi\rangle ^*= \gamma^*\alpha +\delta^*\beta\;. 
\end{equation}
Each additional state (or configuration) in the classical system yields  an additional orthogonal dimension (complex parameter) in the quantum system. Hence a finite state classical system will lead to a finite dimensional complex vector space for the corresponding quantum system. 

Let us describe the geometry of the quantum configuration space of a single qubit in more detail.
The constraint   $|\alpha|^2 +|\beta|^2= 1$ says that the state vector has unit length, which defines the complex unit circle in $\mathbf{ C^2}$, but if we write the complex numbers in terms of their real and imaginary parts as  $\alpha= a_1 + i a_2$ and $\beta= b_1+i b_2$, then we obtain $|a_1 + a_2 i|^2 + |b_1+b_2 i|^2= a_1^2 + a_2^2 + b_1^2 + b_2^2 =  1$. The geometry of the space described by the latter equation is just the three dimensional unit sphere $S^3$ embedded in a four dimensional Euclidean space, $\mathbf{ R}^4$.  

To do any nontrivial quantum computation we need  to consider a system with multiple qubits.  Physically it is easiest to imagine a system of $n$ particles, each with its own spin.  (As before, the formalism does not depend on this, and it is possible to have examples in which the individual qubits might correspond to other physical properties).  The mathematical space in which the $n$ qubits live is the tensor product of the individual qubit spaces, which we may write as $\mathbf{ C}^2\otimes \mathbf{ C}^2\otimes...\otimes\mathbf{ C}^2= \mathbf{ C}^{2^n}$.   For example, the Hilbert space for two qubits is $\mathbf{ C}^2\otimes \mathbf{ C}^2$.  This is a four dimensional complex vector space spanned by the vectors $| 1 \rangle \otimes | 1 \rangle$, $| 0 \rangle \otimes | 1 \rangle$, $| 1 \rangle \otimes | 0 \rangle$, and $| 0 \rangle \otimes | 0 \rangle$.   For convenience we will often abbreviate the tensor product by omitting the tensor product symbols, or by simply listing the spins.  For example
\[
|1 \rangle \otimes |0 \rangle = |1 \rangle | 0 \rangle = |10 \rangle.
\]
The tensor product of two qubits with wave functions $|\psi\rangle  = \alpha |1\rangle  + \beta |0\rangle$ and $|\phi\rangle  = \gamma |1\rangle  + \delta |0\rangle$ is 
\[
|\psi\rangle \otimes |\phi\rangle = | \psi \rangle | \phi \rangle =  \alpha \gamma |11\rangle +\gamma \delta  |10 \rangle  +\beta \gamma  |01 \rangle +\beta \delta |00 \rangle.
\]
The most important feature of the tensor product is that it is multi-linear, i.e. $(\alpha |0 \rangle +  \beta |1 \rangle) \otimes |\psi \rangle = \alpha |0 \rangle \otimes | \psi \rangle + \beta |1 \rangle \otimes |\psi \rangle$.     
Again we emphasize that whereas the classical $n-$bit system has $2^n$ states, the $n-$qubit system corresponds to a  vector of unit length in a $2^n$ dimensional complex space, with twice as many degrees of freedom.  For example a three-qubit can be expanded as:
\begin{eqnarray*} |\psi\rangle = \alpha_1 |000\rangle &+&\alpha_2 |001\rangle +\alpha_3 |010\rangle  + \alpha_4 |011\rangle  \\ \nonumber
&+&\alpha_5 |100\rangle +\alpha_6|101\rangle +\alpha_7 |110\rangle + \alpha_8 |111\rangle 
\label{eq:psitensor}
\end{eqnarray*}
Sometimes it is convenient to denote the state vector by the column vector of its components $\alpha_1,\alpha_2,..., \alpha_{2^n}$.

\subsection{Observables}
How are ordinary physical variables such as energy, position, velocity, and spin retrieved from the state vector? In the quantum formalism observables are defined as \textit{hermitian} operators acting on the state space.  In quantum mechanics an {\it operator} is a linear transformation that maps one state into another, which providing the state space is finite dimensional, can be represented by a matrix. A hermitian operator or matrix satisfies the condition $A=A^\dagger$, where $A^\dagger= (A^{tr})^*$ is the complex conjugate of the transpose of $A$. The fact that observables are represented by operators reflects the property  that measurements may alter the state and that outcomes of different measurements may depend on the order in which the measurements are performed. In general observables in quantummechanics do not necessarily \textit{commute}, by which we mean that for the product of two observables $A$ and $B$ one may have that $AB\neq BA$. The reason that observables have to be hermitian is because the outcome of measurements are the eigenvalues of observables, and hermitian operators are guaranteed to have real eigenvalues.

For example consider a single qubit.  The  physical observables are the components of the spin along the $x$, $y$ or $z$ directions, which are by convention written $s_x=\frac{1}{2}\sigma_x$, $s_y=\frac{1}{2}\sigma_y$, etc.  The operators $\sigma$ are the Pauli matrices
\begin{equation}
\label{spinmatrices}
\sigma_x = \begin{pmatrix}
    0  & 1   \\
     1 &  0
\end{pmatrix} \;,\;
\sigma_y =  \begin{pmatrix}
     0 & -i   \\
    i  & 0 
\end{pmatrix}\;,\;
\sigma_z =  \begin{pmatrix}
    1  &  0  \\
     0 & -1 
\end{pmatrix},
\end{equation}
which obviously do not commute.
In writing the spin operators this way we have arbitrarily chosen\footnote{We can rotate into a different representation that makes either of the other two axes diagonal, and in which the $z$-axis is no longer diagonal -- it is only possible to make one of the three axes diagonal at a time.  Experimental set-ups often have conditions that break symmetry, such as an applied magnetic fields, in which case it is most convenient to let the symmetry breaking direction be the $z$-axis.}  the $z$-axis to have a diagonal representation, so that the {\it eigenstates}\footnote{The eigenstates $|\chi_k\rangle $ of a linear operator $A$ are defined by  the equation $A |\chi_k\rangle = \lambda_k|\chi_k\rangle$.  If $A$ is hermitian the eigenvalue $\lambda_k$ is a real number.    It is generally possible to choose the eigenstates as orthonormal, so that $\langle \chi_j|\chi_k\rangle =\delta_{jk}$, where $\delta_{ij} = 1$ when $i = j$ and $\delta_{ij} = 0$ otherwise.} for spin along the $z$ axis are the column matrices 
\[
| 1 \rangle = \begin{pmatrix} 1\\ 0 \end{pmatrix}, ~~~~~~~ | 0 \rangle = \begin{pmatrix} 0\\ 1 \end{pmatrix}.
\]

\subsection{Quantum evolution: the Schr\"odinger equation}
The wave function of a quantum system evolves in time according to the famous Schr\"odinger equation. Dynamical changes in a physical system are induced by the underlying forces acting on the system and between its constituent parts, and their effect can be represented in terms of what is called the  energy or Hamiltonian operator $H$.   For a single qubit system the operators can be represented as $2 \times 2$ matrices, for a two qubit system they are $4 \times 4$ matrices, etc.  The Schr\"odinger equation can be written
\begin{equation}
\label{ Schrodinger}
i\hbar\frac{d|\psi(t)\rangle }{dt}= H|\psi(t)\rangle. 
\end{equation}
This is a linear differential equation expressing the property that the time evolution of a quantum system is generated by its energy operator.  Assuming that H is constant, given an initial state $|\psi(0)\rangle $ the solution is  simply
\begin{equation}
\label{timeevolution}
|\psi(t)\rangle =U(t)|\psi(0)\rangle  \mbox{ with } U(t)= e^{-iHt/\hbar}.
\end{equation}
The time evolution is {\it unitary}, meaning that the operator  $U(t)$ satisfies $UU^\dagger=1$.  
\begin{equation}
\label{ unitarity}
U^\dagger = \exp(-iHt/\hbar)^\dagger= \exp(iH^\dagger t/\hbar) = \exp(iHt/\hbar= U^{-1}. 
\end{equation}
Unitary time evolution means that the length of the state vector remains invariant, which is necessary to  preserve the total probability for the system to be in any of its possible states.  The unitary nature of the the time evolution operator $U$ follows directly from the fact that $H$ is hermitian.  Any hermitean $2 \times 2$ matrix can be written
\begin{equation}
A= \begin{pmatrix}
  a    & b+i c    \\
    b - i c  & -a  
\end{pmatrix},
\end{equation}
where $a$, $b$ and $c$ are real numbers\footnote{ We omitted a component proportional to the unit matrix as it acts trivially on any state.}. 

For the simple example of a single qubit, suppose the initial state is
 \[|\psi(0)\rangle =
\sqrt{\frac{1}{2}}(|1\rangle +|0\rangle ) \equiv \sqrt{\frac{1}{2}}\begin{pmatrix}
     1  \\ 1
       \end{pmatrix}. \]  
On the right, for the sake of convenience, we have written the state as a column vector.   Consider the energy of a spin in a magnetic field B directed along the positive z-axis\footnote{Quantum spins necessarily have a magnetic moment, so in addition to carrying angular momentum they also interact with a magnetic field.}.   In this case $H$ is given by $H=B s_z$.  From
 (\ref{spinmatrices})
 \begin{equation}
\label{U(t) }
U(t)= \exp(\frac{-iBt}{2\hbar}\sigma_z) =  
\begin{pmatrix}
    \exp(-iBt/2\hbar)  & 0   \\
     0 & \exp (iBt/2\hbar) 
\end{pmatrix}.
\end{equation}
Using (\ref{timeevolution}) we obtain an oscillatory time dependence for the state, i.e.
\begin{equation}
\label{ psi(t)}
|\psi(t)\rangle  = \sqrt{\frac{1}{2}}\begin{pmatrix}
   e^{-iBt/2\hbar}    \\
    e^{iBt/2\hbar} 

\end{pmatrix} =
   \sqrt{\frac{1}{2}} \left[ \cos \frac{Bt}{2\hbar} \begin{pmatrix}
      1  \\
      1  
\end{pmatrix} + i  \sin \frac{Bt}{2\hbar}  \begin{pmatrix} -1 \\ 1        
\end{pmatrix} \right].
\end{equation}

We thus see that, in contrast to classical mechanics, time evolution in quantum mechanics is always linear.  It is in this sense much simpler than classical mechanics.  The complication is that when we consider more complicated examples, for example corresponding to a macroscopic object such as a planet, the dimension of the space in which the quantum dynamics takes place becomes extremely high.

\subsection{Quantum measurements}
Measurement in classical physics is conceptually trivial:  One simply estimates the value of the classical state at finite precision and approximates the state as a real number with a finite number of digits.  The accuracy of measurements is limited only by background noise and the precision of the measuring instrument.  The measurement process in quantum mechanics, in contrast, is not at all trivial.  One difference with classical mechanics is that in many instances the set of measurable states is discrete, with quantized values for the observables.   It is this property that has given the theory of quantum mechanics its name.  But perhaps an even more profound difference is that quantum measurement typically causes a radical alteration of the wavefunction.  Before the measurement of an observable we can only describe the possible outcomes in terms of probabilities, whereas after the measurement the outcome is known with certainty, and the wavefunction is irrevocably altered to reflect this.  In the conventional Copenhagen interpretation of quantum mechanics the wave function is said to ``collapse" when a measurement is made.  In spite of the fact that quantum mechanics makes spectacularly successful predictions, the fact that quantum measurements are inherently probabilistic and can "instantly" 
alter the state of the system has caused a great deal of controversy.  In fact, one can argue that historically the field of quantum computation emerged from thinking carefully about the measurement problem \cite{Deutsch85}. 

In the formalism of quantum mechanics the possible outcomes of an observable quantity $A$ are given by the eigenvalues of the matrix $A$.  For example, the three spin operators defined in Eq.~\ref{spinmatrices} all have the same two eigenvalues $\lambda_{\pm}=\pm 1/2$. This means that the possible outcomes of a measurement of the spin in any direction can only be plus or minus one half. This is completely different than a spinning object in classical physics, which can spin at any possible rate in any direction.  This is why quantum mechanics is so nonintuitive! 

If a quantum system is in an eigenstate then the outcome of measurements in the corresponding direction is certain.  For example, imagine we have a qubit in the state with $\alpha=1$ and $\beta=0$ so $|\psi\rangle = |1\rangle  $.  It is then in the eigenstate of $s_z$ with eigenvalue $+\frac{1}{2}$, so the measurement of $s_z$ will always yield that value.  This is reflected in the mathematical machinery of quantum mechanics by the fact that for the spin operator in the $z-$direction, $A = s_z$, the eigenvector with eigenvalue $\lambda_+= +1/2$ is $|1\rangle=\begin{pmatrix}  1\\0   \end{pmatrix} $ and the eigenvector with $\lambda_-=-1/2$ is $|0\rangle =\begin{pmatrix} 0\\1  \end{pmatrix}$. 
In contrast, if we make measurements in the orthogonal directions to the eigenstate, e.g. $A=s_x$, the outcomes become probabilistic.  In the example above the eigenvectors of $s_x$ are $|\chi_+\rangle=\sqrt{\frac{1}{2}}(|1\rangle +|0\rangle )$ and   $|\chi_-\rangle=\sqrt{\frac{1}{2}}(|1\rangle -|0\rangle )$.
In general the probability of finding the system in a given state in a measurement is computed by first expanding the given state $|\psi\rangle $ into the eigenstates $ |\chi_k\rangle $ of the matrix $A$ corresponding to the observable, i.e.
\begin{equation}
\label{eigenexpansion}
|\psi\rangle = \sum_k  \alpha_k |\chi_k\rangle    \; \mbox{where} \; \alpha_k= \langle \chi_k|\psi\rangle.  
\end{equation} 
The probability of measuring the system in the state corresponding to eigenvalue $\lambda_k$ is $p_k = |\alpha_k|^2$. 
The predictions of quantum mechanics are therefore probabilistic but the theory is essentially  different from classical probability theory. On the one hand it is clear that a given operator defines a probability measure on Hilbert space, however as the operators are non-commuting (like matrices) one is dealing with a non-commutative probability theory \cite{Holevo1982}. It is the non-commutativity of observables that gives rise to the intricacies  in the quantum theory of measurement.

Let us discuss an example for clarification. Consider the spin in the $x$-direction, $A=s_x$, and $|\psi\rangle =|1\rangle $, i.e. spin up in the $z$-direction.  Expanding in eigenstates of $\sigma_x$ we get $|\psi\rangle  = |1\rangle = \sqrt{\frac{1}{2}}|\chi_+\rangle  + \sqrt{\frac{1}{2}}|\chi_-\rangle $.  The probability of measuring spin up along the $x$-direction is $|\alpha_+|^2 = 1/2$, and the probability of measuring spin down along the $x$-direction is $|\alpha_-|^2 = 1/2$.  We see how probability enters quantum mechanics at a fundamental level.  The average of an observable is its {\it expectation value}, which is the weighted sum
\begin{equation}
\label{expectation value }
\langle \psi| A|\psi \rangle  = \sum_k  |\alpha_k|^2 \lambda_k = \sum_k  p_k \lambda_k.
\end{equation} 
In the example at hand $\langle\sigma_x \rangle=0$.

The act of measurement influences the state of the system.  If we measure $s_x=+\frac{1}{2}$ and then measure it again immediately afterward, we will get the same value with certainty. Stated differently, doing the measurement somehow forces the system into the eigenstate $|\chi_+\rangle$, and once it is there, in the absence of further interactions, it stays there.  This strange property of measurement, in which the wavefunction collapses onto the observed eigenstate,  was originally added to the theory in an ad hoc manner, and is called the {\it projection postulate}.    This postulate introduces a rather arbitrary element into the theory that appears to be inconsistent:  The system evolves under quantum mechanics according to the Schr\"odinger equation until a measurement is made, at which point some kind of magic associated with the classical measurement apparatus takes place, which lies completely outside the rest of the theory. 

To understand the measurement process better it is necessary to discuss the coupling of a quantum system and a classical measurement apparatus in more detail.   A measurement apparatus, such as a pointer on a dial or the conditional emission of a light pulse, is also a quantum mechanical system.  If we treat the measurement device quantum mechanically as well, it should be possible to regard the apparent ``collapse" of the wavefunction as the outcome of the quantum evolution of the combined system of the measurement device and the original quantum system under study, without invoking the projection postulate.  We return to this when we discuss decoherence in Section~\ref{decoherence} .

Note that a measurement does not allow one to completely determine the state. A complete measurement of the two-qubit system yields at most two classical bits of information, whereas determining the full quantum state requires knowing seven real numbers ( four complex numbers subject to a normalization condition). In this sense one cannot just say that a quantum states ``contains" much more information that its classical counterpart.  In fact, due to the non-commutativity of the observables, with simultaneous measurements one is able to extract less information than from the corresponding classical system.  

There are two ways to talk about quantum theory:  If one insists it is a theory of a single system, then one has to live with the fact that it only predicts the probability of things to happen and as such is a retrenchment from the ideal of classical physics. Alternatively one may take the view that quantum theory is a theory that only applies to ensembles of particles. To actually measure probability distributions one has to make many measurements on ``identically prepared" quantum systems.  From this perspective the dimensionality of Hilbert space should be compared to that of classical distributions defined over a classical phase space, which makes the difference between classical and quantum theories far less dramatic. This raises the quest for a theory underlying quantum mechanics which applies for a single system. So far nobody has succeeded in producing such a theory, and on the contrary, attempts to build such theories based on ``hidden variables" have failed.  The Bell inequalities suggest that such a theory is probably impossible \cite{omnes1999}. 

\subsection{Multi qubit states and entanglement\label{entanglement}}
When we have more than one qubit an important practical question is when and how measurements of a given qubit depend on measurements of other qubits.  Because of the deep properties of quantum mechanics, qubits can be coupled in subtle ways that produce consequences for measurement that are very different from classical bits.   Understanding this has proved to be important for the problems of computation and information transmission.  To explain this we need to introduce the opposing concepts of separability and entanglement, which describe whether measurements on different qubits are statistically independent or statistically dependent.
 
An $n$-qubit state is {\it separable} if it can be factored into n-single qubit states\footnote{Strictly speaking this is only true for pure states, which we define in the next section.}, i.e. if it can be written as $n-1$ tensor products of sums of qubits, with each factor depending only on a single qubit.  
An example of a separable two-qubit is
\begin{equation}
\label{eq:factorizable }
|\psi\rangle =\frac{1}{2}(|00\rangle +|01\rangle  +|10\rangle  +|11\rangle )= \frac{1}{2}(|0\rangle +|1\rangle )\otimes (|0\rangle +|1\rangle ).
\end{equation} 
If an $n$-qubit state is separable then measurements on individual qubits are statistically independent, i.e. the probability of making a series of measurements on each qubit can be written as a product of probabilities of the measurements for each qubit.

An $n$-qubit state is {\it entangled} if it is not separable.   An example of an entangled two-qubit state is  
\begin{equation}
\label{eq:entangled}
|\psi\rangle  = \frac{1}{\sqrt{2}}(|00\rangle  + |11\rangle ),
\end{equation}
which cannot be factored into a single product.  For entangled states measurements on individual qubits depend on each other.

We now illustrate this for the two examples above.  Suppose we do an experiment in which we measure the spin of the first qubit and then measure the spin of the second qubit.  For both the separable and entangled examples, there is a $50\%$ chance of observing either spin up or spin down on the first measurement.  Suppose it gives spin up.   For the separable state this transforms the wave function as
 \[
\frac{1}{2}(|0\rangle +|1\rangle )\otimes (|0\rangle +|1\rangle ) \rightarrow \frac{1}{\sqrt{2}}(|1\rangle )\otimes (|0\rangle +|1\rangle ) = \frac{1}{\sqrt{2}}(|10\rangle + |11\rangle ).
\]
If we now measure the spin of the second qubit, the probability of measuring spin up or spin down is still $50\%$.  The first measurement has no effect on the second measurement.

In contrast, suppose we do a similar experiment on the entangled state of equation~\ref{eq:entangled} and observe spin up in the first measurement.  This transforms the wave function as
\begin{equation}
\frac{1}{\sqrt{2}}(|00\rangle +|11\rangle ) \longrightarrow  |11\rangle.
\end{equation}
(Note the disappearance of the factor $1/\sqrt{2}$ due to the necessity that the wave function remains normalized).  If we now measure the spin of the second qubit we are certain to observe spin up!  Similarly, if we observe spin down in the first measurement, we will also observe it in the second.  For the entangled example above the measurements are completely coupled -- the outcome of the first determines the second.  This property of entangled states was originally pointed out by Einstein, Podolski and Rosen \cite{Einstein1935}, who expressed concern about the possible consequences of this when the qubits are widely separated in space.  This line of thinking did not point out a fundamental problem with quantum mechanics as they perhaps originally hoped, but rather led to a deeper understanding  of the quantum measurement problem and to the practical application of quantum teleportation as discussed in Section~\ref{teleportation}.
 
The degree of entanglement of a system of qubits is a reflection of their past history.  By applying the right time evolution operator, i.e. by introducing appropriate interactions,  we can begin with a separable state and entangle it, or begin with an entangled state and separate it.  Separation can be achieved, for example, by applying the inverse of the operator that brought about the entanglement in the first place -- quantum dynamics is reversible.  Alternatively separation can be achieved by transferring the entanglement to something else, such as the external environment.  (In the latter case there will still be entanglement, but it will be between one of the qubits and the environment, rather than between the two original qubits).

\subsection{Entanglement and entropy}

So far we have assumed that we are able to study a single particle or a few particles with perfect knowledge of the state.  This is called a statistically pure state, or often more simply, a {\it pure state}.  In experiments it can be difficult to prepare a system in a pure state.  More typically there is an ensemble of particles that might be in different states, or we might have incomplete knowledge of the states.  Such a situation, in which there is a nonzero probability for the particle to be in more than one state, is called a {\it mixed state}.   As we explain below, von Neumann developed an alternative formalism for quantum mechanics in terms of what is called a density matrix, which replaces the wavefunction as the elementary level of description.  The density matrix representation very simply handles mixed states, and leads to a natural way to measure the entropy of a quantum mechanical system and measure entanglement.

Consider a mixed state in which there is a probability $p_i$ for the system to have wavefunction $\psi_i$ and an observable characterized by operator $A$.  The average value measured for the observable (also called its expectation value) is
\begin{equation}
\langle A \rangle = \sum_i p_i \langle \psi_i | A | \psi_i \rangle.
\label{expectation}
\end{equation}
We can expand each wavefunction $\psi_i$ in terms of a basis $| \chi_j \rangle$ in the form
\[
| \psi_i \rangle = \sum_j  \langle \chi_j | \psi_i \rangle | \chi_j \rangle,
\]
where in our earlier notation $\langle \chi_j | \psi_i \rangle = \alpha_{j}^{(i)}$.  Performing this expansion for the dual vector $\langle \psi_i |$ as well, substituting into (\ref{expectation}) and interchanging the order of summation gives
\begin{eqnarray*}
\langle A \rangle & = & \sum_{j,k} \left( \sum_i p_i \langle \chi_j | \psi_i \rangle \langle \psi_i  | \chi_k \rangle \right) \langle \chi_k | A | \chi_j \rangle\\
& = & \sum_{j,k} \langle \chi_j | \rho | \chi_k \rangle \langle \chi_k | A | \chi_j \rangle\\
& = & \mbox{tr}(\rho A),\\
\end{eqnarray*}
where
\begin{equation}
\rho = \sum_i p_i | \psi_i \rangle \langle \psi_i  | 
\end{equation}
is called the {\it density matrix}\footnote{The density matrix provides an alternative representation for quantum mechanics --  the Schr\"odinger equation can be rewritten in terms of the density matrix so that we never need to use wavefunctions at all.}.  Because the trace $\mbox{tr}(\rho A)$ is independent of the representation this can be evaluated in any convenient basis, and so provides an easy way to compute expectations.  Note that $\mbox{tr}(\rho) = 1$.  For a pure state $p_i = 1$ for some value of $i$ and $p_i = 0$ otherwise.  In this case the density matrix has rank one.  This is obvious if we write it in a basis in which it is diagonal -- there will only be one nonzero element.  When there is more than one nonzero value of $p_i$ it is a mixed state and the rank is greater than one.

To get a better feel for how this works, consider the very simple example of a single qubit, and let $\psi_1 = |1 \rangle$.  If this is a pure state then the density matrix is just 
\[
\rho = |1 \rangle \langle 1 | = \begin{pmatrix}
    1  & 0   \\
     0 &  0
\end{pmatrix}.
\]
The expectation of the spin along the $z$-axis is $\mbox{tr} (\rho s_z) = 1/2$.  If, however, the system is in a mixed state with $50\%$ of the population spin up and $50\%$ spin down, this becomes
\[
\rho = \frac{1}{2} \left( |1 \rangle \langle 1 |  + |0 \rangle \langle 0 | \right) = \frac{1}{2} \begin{pmatrix}
    1  & 0   \\
     0 &  1
\end{pmatrix}.
\]
In this case the expectation of the spin along the $z$-axis is $\mbox{tr} (\rho s_z) = 0$. 

This led von Neumann to define the entropy of a quantum state in analogy with the Gibbs entropy for a classical ensemble as
\begin{equation}
\label{ eq:Neumannentropy}
S(\rho)= -\mbox{tr}\; \rho \log \rho = -\sum_i p_i \log p_i \;.
\end{equation}
The entropy of a quantum state provides a quantitative measure of ``how mixed" a system is.
The entropy of a pure state is equal to zero, whereas the entropy of a mixed state is greater than zero. 

In some situations there is a close relationship between entangled and mixed states.
An entangled but pure state in a high dimensional multi-qubit space can appear to be a mixed state when viewed from the point of view of a lower dimensional state space.   The view of the wavefunction from a lower dimensional subspace is formally taken using a \textit{partial} trace.  This is done by summing over all the coordinates associated with the subspaces we want to ignore.   This corresponds to leaving some subsystems out of consideration, for example, because we can only measure a certain qubit and can't measure the qubits on which we perform the partial trace.  As an example consider the entangled state of equation (\ref{eq:entangled}), and trace it with respect to the second qubit.  To do this we make use of the fact that $\mbox{tr} (| \psi \rangle \langle \phi |) = \langle \psi | \phi \rangle$.  Using labels $A$ and $B$ to keep the qubits straight, and remembering that because we are using orthogonal coordinates terms of the form $\langle 0 | 1 \rangle = 0$, the calculation can be written
\begin{eqnarray*}
\mbox{tr} \left( | \psi_{AB} \rangle \langle \psi_{AB} | \right) & = & \frac{1}{2} \mbox{tr} \left( |1 \rangle_A \langle 1 |_B  + |0 \rangle_A \langle 0 |_B \right) \left( \langle 0 |_B \langle 0 |_A + \langle 1 |_B \langle 1 |_A \right)\\
& = & \frac{1}{2} \left( |1 \rangle_A \langle 1 |_A  \langle 1 | 1 \rangle_B + |0 \rangle_A  \langle 0 |_A \langle 0 | 0 \rangle_B \right)\\
& = & \frac{1}{2} \left( |1 \rangle_A \langle 1 |_A + |0 \rangle_A \langle 0 |_A \right)
\end{eqnarray*}
This is a mixed state with probability $1/2$ to be either spin up or spin down.  The corresponding entropy is also higher:  In base two $S= - \log(1/2) =1$ bit, while for the original pure state $S=\log 1 = 0$.  In general if we begin with a statistically pure separable state and perform a partial trace we will still have a pure state, but if we begin with an entangled state, when we perform a partial trace we will get a mixed state.  In the former case the entropy remains zero, but in the latter case it increases.  Thus the von Neumann entropy yields a useful measure of entanglement. 

\subsection{Measurement and Decoherence}
\label{decoherence}
In this section we return to the measurement problem and the  complications that arise if one wants to couple a classical measurement device to a quantum system.  A classical system is by definition described in terms of macro-states, and one macro-state can easily correspond to $10^{40}$ micro-states.  A classical measurement apparatus like a Geiger counter or a photo multiplier tube is prepared in a meta-stable state in which an interaction with the quantum system can produce a decay into a more stable state indicating the outcome of the measurement.  For example, imagine that we want to detect the presence of an electron.  We can do so by creating a detector consisting of a meta-stable atom.  If the electron passes by its interaction with the meta-stable atom via its electromagnetic field can cause the decay of the meta-stable atom, and we observe the emission of a photon.  If it doesn't pass by we observe nothing.  There are very many possible final states for the system, corresponding to different micro-states of the electron and the photon, but we aren't interested in that -- all we want to know is whether or not a photon was emitted.  Thus we have to sum over all possible combined photon-electron configurations.  This amounts to tracing the density matrix of the complete system consisting of the electron and the measurement apparatus over all states in which a photon is present in the final state. This leads to a reduced density matrix describing the electron after the measurement, with the electron in a mixed state, corresponding to the many possible photon states.  Thus even though we started with a zero entropy pure state in the combined system of the electron and photon, we end up with a positive entropy mixed state in the space of the electron alone.    The state of the electron is reduced to a classical probability distribution, and due to the huge number of microstates that are averaged over, the process of measurement is thermodynamically irreversible. Even if we do not observe the outcoming photon with our own eyes, it is clear whether or not the metastable atom decayed, and thus whether or not the electron passed by.

The description of the measurement process above is an example of {\it decoherence}, i.e. of a process whereby quantum mechanical systems come to behave as if they were governed by classical probabilities.  A common way for this to happen is for a quantum system to interact with its environment, or for that matter any other quantum system, in such a way that the reduced density matrix for the system of interest becomes diagonal in a particular basis.  The phases are randomized, so that after the measurement the system is found to be in a mixed state.  
According to this view, the wavefunction does not actually collapse, there is just the appearance of a collapse due to quantum decoherence.  The details of how this happen remain controversial, and is a subject of active research \cite{Zurek91,Zurek03,Schlosshauer04,omnes1999}.   In Section~\ref{quantumGates} we will give an example of how decoherence can be generated even by interactions between simple systems.

\subsection{The no-cloning theorem}
We have seen that  by doing a measurement we may destroy the original state. One important consequence  connected to this destructive property of the act of measurement is that a quantum state cannot be cloned; one may be able to transfer a state from one register to another but one cannot make a  xerox copy of a given quantum state.  This is expressed by the no-cloning theorem \cite{wootters1982, dieks1982}.
Worded differently, the no-cloning theorem states that for an arbitrary state $|\psi_1\rangle $ on one qubit and some particular state $|\phi\rangle $  on another, there is no quantum device $[A]$ that transforms $ |\psi_1\rangle \otimes  |\phi\rangle  \rightarrow |\psi_1\rangle  \otimes |\psi_1\rangle $, i.e. that transforms $| \phi \rangle$ into $\psi_1 \rangle$.   Letting $U_A$ be the unitary operator representing $A$, this can be rewritten $ |\psi_1\rangle  |\psi_1\rangle = U_A|\psi_1 \rangle  |\phi\rangle $.  For a true cloning device this property has to hold for any other state $|\psi_2 \rangle $, i.e. we must also have $ |\psi_2 \rangle  |\psi_2\rangle = U_A|\psi_2 \rangle  |\phi\rangle $.   We now show the existence of such a device leads to a contradiction.   Since $\langle \phi | \phi \rangle = 1$ and $U_A^\dagger U_A = 1$, and $U_A|\psi_i \rangle  |\phi\rangle = U_A  |\phi\rangle |\psi_i \rangle$, the existence of a device that can clone both $\psi_1$ and $\psi_2$ would imply that
\begin{eqnarray*}
\langle \psi_1 | \psi_2 \rangle & = &\left( \langle \psi_1| \langle \phi | \right) \left( | \phi \rangle | \psi_2 \rangle \right) = ( \langle \psi_1| \langle \phi |U_A^\dagger ) \left( U_A|\phi\rangle |\psi_2\rangle \right) = (\langle \psi_1 | \langle \psi_1 |)(| \psi_2 \rangle | \psi_2 \rangle)\\
& = & \langle \psi_1 | \psi_2 \rangle^2.
\end{eqnarray*}
The property $\langle \psi_1 | \psi_2 \rangle = \langle \psi_1 | \psi_2 \rangle^2$ only holds if $\psi_1$ and $\psi_2$ are either orthogonal or equal, i.e. it does not hold for arbitrary values of $\psi_1$ and $\psi_2$, so there can be no such general purpose cloning device.  In fact, in view of the uncertainty of quantum measurements, the no-cloning theorem does not come as a surprise:  If it were possible to clone wavefunctions, it would be possible to circumvent the uncertainty of quantum measurements by making a very large number of copies of a wavefunction, measuring different properties of each copy, and reconstructing the exact state of the original wavefunction.  

\subsection{Quantum teleportation\label{teleportation}}

Quantum teleportation provides a method for privately sending messages in a way that ensures that the receiver will know if anyone eavesdrops.  This is possible because a quantum state is literally teleported, in the sense of StarTrek:  A quantum state is destroyed in one place and recreated in another.  Because of the no-cloning theorem, it is impossible to make more than one copy of this quantum state, and as a result when the new teleported state appears, the original state must be destroyed.  Furthermore, it is impossible for both the intended receiver and an eavesdropper to have the state at the same time, which helps make the communication secure.

\begin{figure}[!t]
\begin{center}
\includegraphics[scale=0.6]{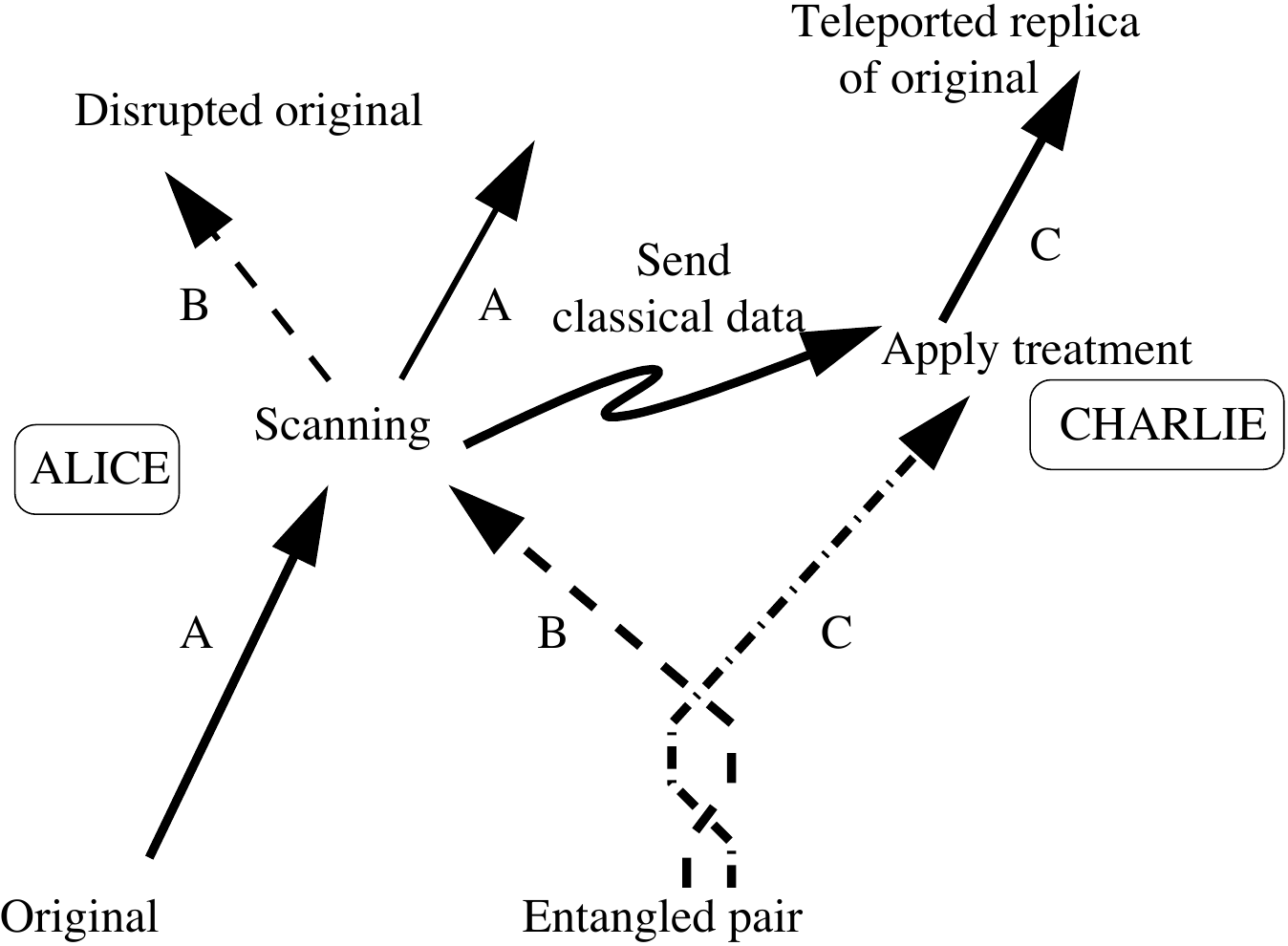}
\end{center}
\caption{Quantum teleportation of a quantum state as proposed by Bennett et al.\cite{bennett1993}, using an entangled pair. An explanation is given in the text.}
\label{fig:teleportation}
\end{figure}

Quantum teleportation takes advantage of the correlation between entangled states as discussed in Section~\ref{entanglement}.   
Suppose Alice wants to send a secure message to Charlie at a (possibly distant) location.   The process of teleportation depends on Alice and Charlie sharing different qubits of an entangled state.  Alice makes a measurement of her part of the entangled state, which is coupled to the state she wants to teleport to Charlie, and sends him some classical information about the entangled state.  With the classical information plus his half of the entangled state, Charlie can reconstruct the teleported state. We have indicated the process in figure \ref{fig:teleportation}.   We follow the method proposed by Bennett et al. \cite{bennett1993}, and first realized in an experimental setup by the group of Zeilinger in 1997 \cite{bouwmeester1997}.
In realistic cases the needed qubit states are typically implemented as left and right handed polarized light quanta (i.e. photons). 

The simplest example of quantum teleportation can be implemented with three qubits.  The (A) qubit  is the unknown state to be teleported,
\begin{equation}
\label{ original}
|\psi_A\rangle = \alpha|1\rangle  + \beta|0\rangle. 
\end{equation}
This state is literally teleported from one place to another.  If Charlie likes, once he has the teleported state he can make a quantum measurement and extract the same information about $\alpha$ and $\beta$ that he would have been able to extract had he made the measurement on the original state.

The teleportation of this state is enabled by an auxiliary two-qubit entangled state.   We label these two qubits $B$ and $C$.  For technical reasons it is convenient to represent this in a special basis consisting of four states, called Bell states, which are written
\begin{eqnarray}
\label{ bellbasis}
|\Psi_{BC}^{(\pm)}\rangle  &=& \sqrt{\frac{1}{2}}(|1_B\rangle |0_C\rangle  \pm |0_B\rangle |1_C\rangle ) \nonumber \\
|\Phi_{BC}^{(\pm)}\rangle  &=& \sqrt{\frac{1}{2}}(|1_B\rangle |1_C\rangle  \pm |0_B\rangle |0_C\rangle ).
\end{eqnarray}

The process of teleportation can be outlined as follows (please refer to Figure~\ref{fig:teleportation}).
\begin{enumerate}
\item
Someone prepares an entangled two qubit state $BC$ (the {\it Entangled pair} in the diagram).
\item
Qubit $B$ is sent to Alice and qubit $C$ is sent to Charlie.
\item
In the {\it Scanning} step, Alice measures in the Bell states basis the combined wavefunction of qubits $A$ (the {\it original} in the diagram) and the entangled state $B$, leaving behind the {\it Disrupted original}.
\item
Alice sends two bits of classical data to Charlie telling him the outcome of her measurements ({\it Send classical data}).
\item
Based on the classical information received from Alice, Charlie applies one of four possible operators to qubit $C$ ({\it Apply treament}), and thereby reconstructs $A$, getting a {\it teleported replica of the original}.  If he likes, he can now make a measurement on $A$ to recover the message Alice has sent him.
\end{enumerate}

We now explain this process in more detail.  In step (1) an entangled two qubit state $\psi_{BC}$ such as that of (\ref{eq:entangled}) is prepared.  In step (2) qubit $B$ is transmitted to Alice and qubit $C$ is transmitted to Charlie.  This can be done, for example, by sending two entangled photons, one to each of them.  In step (3)  Alice measures the joint state of qubit $A$ and $B$ in the Bell states basis, getting two classical bits of information, and projecting the joint wavefunction $\psi_{AB}$ onto one of the Bell states.  The Bell states basis has the nice property that the four possible outcomes of the measurement have equal probability.  To see how this works, for convenience suppose the entangled state $BC$ was prepared in state $|\Psi_{BC}^{(-)}\rangle$.  In this case the combined wavefunction of the three qubit state is
\begin{eqnarray}
\label{threestate}
|\psi_{ABC}\rangle &=& |\psi_A\rangle |\Psi_{BC}^{(-)}\rangle \\ 
&=&{\frac{\alpha}{\sqrt{2}}}(|1_{A}\rangle  |1_B\rangle |0_C\rangle -|1_A\rangle |0_B\rangle |1_C\rangle ) +{\frac{\beta}{\sqrt{2}}}(0_{A}\rangle  |1_B\rangle |0_C\rangle -|0_A\rangle |0_B\rangle |1_C\rangle ).\nonumber
\end{eqnarray}
If this is expanded in the Bell states basis for the pair AB, it can be written in the form 
\begin{eqnarray}
\label{threestate2}
|\psi_{ABC}\rangle  = &\frac{1}{2} \left[ |\Psi_{AB}^{(-)}\rangle  ( -\alpha |1_C\rangle - \beta |0_C\rangle)   + |\Psi_{AB}^{(+)}\rangle ( -\alpha |1_C\rangle + \beta |0_C\rangle) \right.\nonumber \\
&\left. |\Phi_{AB}^{(-)}\rangle (\beta |1_C\rangle +  \alpha |0_C\rangle)  +   |\Phi_{AB}^{(+)}\rangle ( - \beta |1_C\rangle + \alpha |0_C\rangle )  \right]\;.
\end{eqnarray}  
We see that the two qubit $AB$ has equal probability to be in the four possible states $|\Psi_{AB}^{(-)}\rangle$, $|\Psi_{AB}^{(+)}\rangle$, $|\Phi_{AB}^{(-)}\rangle$ and $|\Phi_{AB}^{(+)}\rangle$.  

In step (4), Alice transmits two classical bits to Charlie, telling him which of the four basis functions she observed.  Charlie now makes use of the fact that in the Bell basis there are four possible states for the entangled qubit that he has, and his qubit C was entangled with Alice's qubit $B$ before she made the measurement.   In particular, let $|\phi_C\rangle$ be the state of the $C$ qubit, which from (\ref{threestate2}) is one of the four states:
\begin{equation}
\label{ }
|\phi_C\rangle =
\begin{pmatrix} \alpha \\  \beta \end{pmatrix}; \begin{pmatrix} -\alpha \\  \beta \end{pmatrix}; \begin{pmatrix} \beta \\  \alpha \end{pmatrix}; \mbox{ and } \begin{pmatrix} -\beta \\  \alpha \end{pmatrix}\;.
\end{equation}   In step (5), based on the information that he receives from Alice, Charlie selects one of four possible operators $F_i$ and uses it to measure the $C$ qubit.  There is one operator $F_i$ for each of the four possible Bell states, which are respectively:
\begin{equation}
\label{treatmentoperators }
 F = -\begin{pmatrix}
     1 & 0   \\
     0 & 1 
\end{pmatrix}\; ;\; \begin{pmatrix}
   -1   &  0  \\
     0 &  1
\end{pmatrix} \;;\; \begin{pmatrix}
   0   & 1   \\
  1    &  0
\end{pmatrix} \;;\; \mbox{ and } \begin{pmatrix}
   0   &  -1  \\
    1  &  0
\end{pmatrix}\;.
\end{equation} 
 Providing Charlie has the correct classical information and an intact entangled state he can reconstruct the original $A$ qubit by measuring $| \phi_C \rangle$ with the appropriate operator $F_i$. 
\begin{equation}
\label{ treatment}
|\psi_A\rangle = \alpha | 1 \rangle + \beta | 0\rangle =  F_i |\phi_C\rangle\;.
\end{equation}
By simply multiplying each of the four possibilities it is easy to verify that as long as his information is correct, he will correctly reconstruct the $A$ qubit $\alpha | 1_A \rangle + \beta | 0_A \rangle$.

We stress that Charlie needs the classical measurement information from Alice.   If he could do without it the teleportation process would violate causality, since information could be transferred instantaneously from Alice to Charlie.  That is, when Alice measures the $B$ qubit, naively it might seem that because the $B$ and $C$ qubits are entangled, this instantaneously collapses the $C$ qubit, sending Charlie the information about Alice's measurement, no matter how far away he is.   To understand why such instantaneous communication is not possible, suppose Charlie just randomly guesses the outcome and randomly selects one of the four operators $F_i$.   Then the original state will be reconstructed as a random mixture of the four possible incoming states  $|\phi_C\rangle$.  This mixture does not give any information about the original state $|\psi_A\rangle$. 

The same reasoning also applies to a possible eavesdropper, conveniently named Eve.   If she manages to intercept qubit (C) and measures it before Charlie does, without the two bits of classical information she will not be able to recover the original state.   Furthermore she will have affected that state.   If Charlie somehow gets the mutilated state he will not be able to reconstruct the original state $A$. Security can be achieved if Alice first sends a sequence of known states which can be checked by Charlie after reconstruction. If the original and reconstructed sequence are perfectly correlated then that guarantees that Eve is not interfering.  Note that the cloning theorem is satisfied, since when Alice makes her measurement she alters the state $\psi_A$ as well as her qubit $B$.  Once she has done that, the only hope to reconstruct the original $\psi_A$ is for her to send her measurement to Charlie, who can apply the appropriate operator to his entangled qubit $C$.  

The quantum security mechanism of teleportation is based on strongly correlated, highly non-local entangled states. While a strength, the non-locality of the correlations is also a weakness.  Quantum correlations are extremely fragile and can be corrupted by random interactions with the environment, i.e. by decoherence. As we discussed before, this is a  process in which the quantum correlations are destroyed and information gets lost. The problem of decoherence is the main stumbling block in making progress towards large scale development and application of quantum technologies.  Nevertheless, in 2006 the research group of Gisin at the University of Geneva succeeded in demonstrating  teleportation over a distance of 550 meters using the optical fiber network of Swisscom \cite{gisin2007}.

\subsection{Quantum computation}

Quantum computation is performed by setting up controlled interactions with non-trivial dynamics that successively couple individual qubits together and alter the time evolution of the wavefunction in a predetermined manner.  A multi-qubit system is first prepared in a known initial state, representing the input to the program.  Then interactions are switched on by applying forces, such as magnetic fields, that determine the direction in which the wavefunction rotates in its state space. Thus a quantum program is just a sequence of unitary operations that are externally applied to the initial state.  This is achieved in practice by a corresponding sequence of quantum gates.  When the computation is done measurements are made  to read out the final state.

Quantum computation is essentially a form of analog computation.  A physical system is used to simulate a mathematical problem, taking advantage of the fact that they both obey the same equations.   The mathematical problem is mapped onto the physical system by finding an appropriate arrangement of magnets or other fields that will generate the proper equation of motion.  One then prepares the initial state, lets the system evolve, and reads out the answer.  Analog computers are nothing new.  For example, Leibnitz built a mechanical calculator for performing multiplication in 1694, and in the middle of the twentieth century, because of their vastly superior speed in comparison with digital computers, electronic analog computers were often used to solve differential equations. 

Then why is quantum computation special?  The key to its exceptional power is the massive parallelism at intermediate stages of the computation.  Any operation on a given state works exactly the same on all basis vectors.  The physical process that defines the quantum computation for an $n$ qubit system thus acts in parallel on a set of $2^n$ complex numbers, and the phases of these numbers (which would not exist in a classical computation) are important in determining the time evolution of the state.  When the measurement is made to read out the answer at the end of the computation we are left with the n-bit output and the phase information is lost.

Because quantum measurements are generically probabilistic, it is possible for the `same' computation to yield different ``answers", e.g. because the measurement process projects the system onto different eigenstates.  This can require the need for error correction mechanisms, though for some problems, such as factoring large numbers, it is possible to test for correctness by simply checking the answer to be sure it works.  It is also possible for quantum computers to make mistakes due to decoherence, i.e. because of essentially random interactions between the quantum state used to perform the computation and the environment.  This also necessitates error correction mechanisms.

The problems caused by decoherence are perhaps {\it the} central difficulty in creating physical implementations of quantum computation.  These can potentially be overcome by constructing systems where the quantum state is not encoded locally, but rather globally, in terms of topological properties of the system that cannot be disrupted by external (local) noise.  This is called \textit{topological quantum computing}.  This interesting possibility arises in certain two-dimensional physical media which exhibit  \textit{topological order}, referring to states of matter in which the essential quantum degrees of freedom and their interactions are topological \cite{kitaev2003, dassarma2007}.

\subsection{Quantum gates and circuits\label{quantumGates}} 
\begin{figure}[h!]
\begin{center}
\includegraphics[scale=0.8]{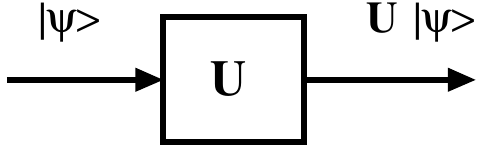}
\caption{The diagram representing the action of a unitary matrix  $U$ corresponding to a quantum gate on a qubit in a state $|\psi\rangle $. }
\label{fig:su2action}
\end{center}
\end{figure}

In the same way that classical gates are the building blocks of classical computers, quantum gates are the basic building blocks of quantum computers.   A gate used for a classical computation implements binary operations on binary inputs, changing zeros into ones and vice versa.  
For example, the only nontrivial single bit logic operation is $NOT$, which takes $0$ to $1$ and $1$ to $0$. 
In a quantum computation the situation is quite different, because qubits can exist in superpositions of $0$ and $1$.    
The set of allowable single qubit operations consists of unitary transformations corresponding to $2 \times 2$ complex matrices $U$ such that $U^\dagger U = 1$.  
The corresponding action on a single qubit is represented in a circuit as illustrated in figure \ref{fig:su2action}.
Some quantum gates have classical analogues, but many do not.  For example, the operator $X = \begin{pmatrix}
    0  &   1 \\
     1 &  0
\end{pmatrix}$ is the quantum equivalent of the classical $NOT$ gate, and serves the function of interchanging spin up and spin down.  In contrast, the operation $\begin{pmatrix}
    1  &  0  \\
     0 &  -1
\end{pmatrix}$  rotates the phase of the wavefunction by 180 degrees and has no classical equivalent.  

A general purpose quantum computer has to be able to transform an arbitrary $n$-qubit input into an $n$-qubit output corresponding to the result of the computation.  In principle implementing such a computation might be extremely complicated, and might require constructing quantum gates of arbitrary order and complexity.  

Fortunately, it is possible to prove that the transformations needed to implement a universal quantum computer can be generated by a simple -- so-called universal -- set of elementary quantum gates, for example involving a well chosen pair of a one-qubit and a two-qubit gate.  Single qubit gates are unitary matrices with three real degrees of freedom. If we allow ourselves to work with finite precision, the set of all gates can be arbitrary well approximated by a small well chosen set. There are many possibilities -- the optimal choice depends on the physical implementation of the qubits.   
Typical  one-qubit logical gates are for example the following:
\begin{eqnarray}
X &=& \begin{pmatrix}
    0  &   1 \\
     1 &  0
\end{pmatrix}\\
P(\theta) &=& \begin{pmatrix}
    1  &  0  \\
     0 &  \exp^{i\theta}
\end{pmatrix}\\
H &=& \sqrt{\frac{1}{2}}\begin{pmatrix}
  1    &  1  \\
   1   &  -1
\end{pmatrix}
\end{eqnarray}
$X$ is the quantum equivalent of the classical $NOT$ gate, serving the function of interchanging $|1\rangle$ and $|0\rangle$.  The two other ones have no classical equivalent.   The $P(\theta)$ operation corresponds to the phase gate, it  changes  the relative phase by  $\theta$ degrees, typically with $\theta$ an irrational multiple of  $\pi$.  For the third gate we can choose  the so-called Hadamard gate $H$ which creates a superposition of the basis states: $|1\rangle \Rightarrow \frac{1}{2}(|1\rangle + |0\rangle )$. 

\begin{figure}[h!]
\begin{center}
\includegraphics[scale=0.5]{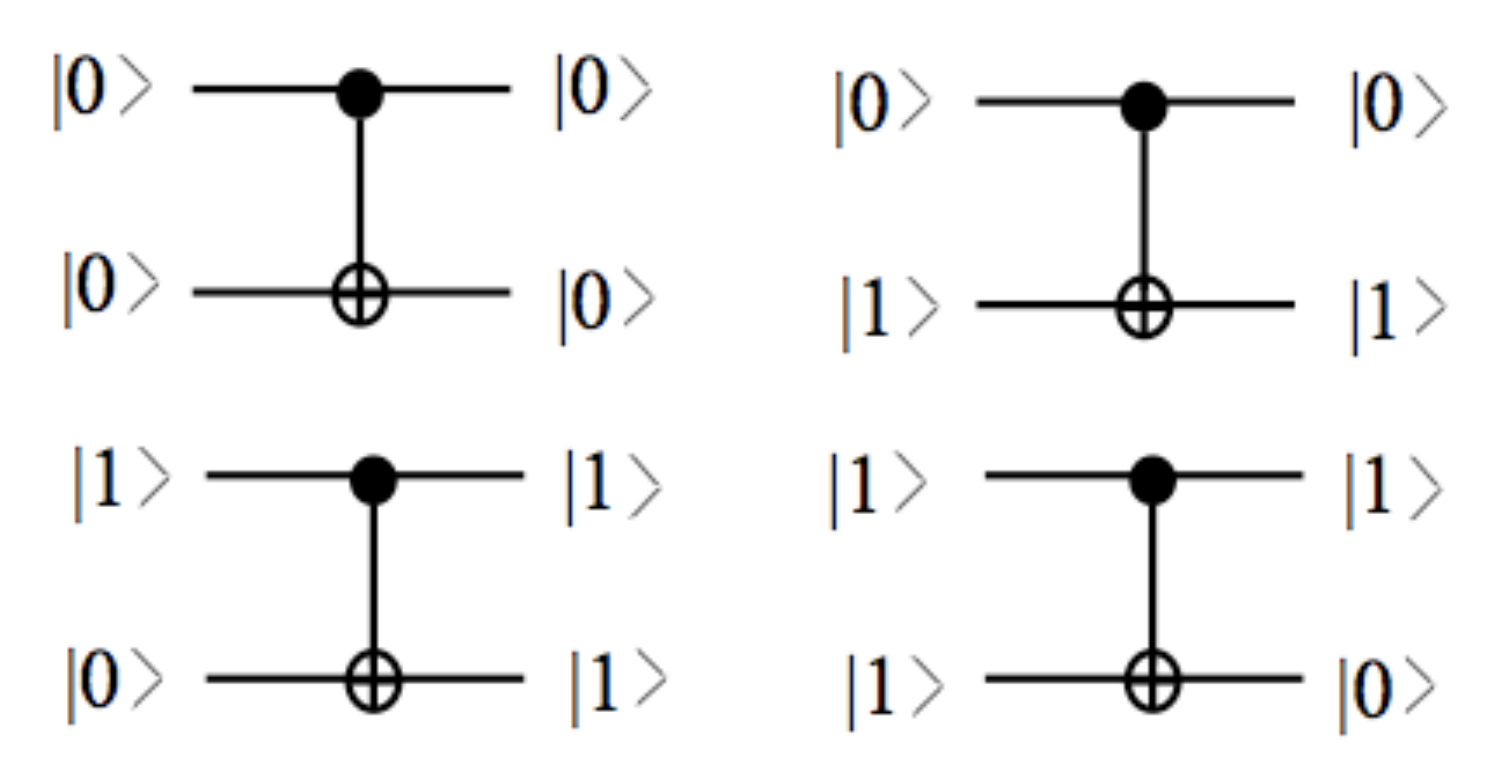}
\caption{The circuit diagram representing the action of the $CNOT$ gate defined in (\ref{CNOTmatrix}) on the four possible two-qubit basis states. The filled dot on the upper qubit denotes the control and the cross is the symbol for the conditional one qubit $NOT$ gate.} 
\label{fig:CNOT}
\end{center}
\end{figure}

From the perspective of experimental implementation, a convenient two-qubit gate  is the $CNOT$ gate. It has been shown that the $CNOT$ in combination with the Hadamard gate forms a universal set \cite{barenco1995}.
The $CNOT$ gate acts as follows on the state $|A\rangle \otimes|B\rangle $:
\begin{equation}
CNOT:  |A\rangle \otimes |B\rangle  \Rightarrow 
 |A\rangle \otimes  |[A+B] \mathrm{mod}\; 2\rangle   
\end{equation}
In words, the $CNOT$ gate flips the state of $B$ if $A = 1$, and does nothing if $A = 0$.  In matrix form one may write the $CNOT$ gate  as
\begin{equation}
\label{CNOTmatrix} CNOT: \;
\begin{pmatrix}
     1 &0 &0&0    \\
      0&1 &0 &0   \\  
      0&0 & 0& 1  \\
     0 &0 &1 & 0     \end{pmatrix}.
\end{equation}
We have fully specified its action on the basis states in figure \ref{fig:CNOT}. 

With the CNOT gate one can generate an entangled state from a separable one, as follows:
\begin{equation}
\label{qcomputation}
\mathrm{CNOT}: \frac{1}{\sqrt{2}} (|0\rangle +|1\rangle )\otimes |0\rangle  \Rightarrow \frac{1}{\sqrt{2}}(|00\rangle +|11\rangle )\;.
\end{equation}
In fact, from an intuitive point of view the ability to generate substantial speed-ups using a quantum computer vs. a classical computer is related to the ability to operate on the high dimensional state space including the  entangled states. 
To describe a separable $n$-qubit state with $k$ bits of accuracy we only need to describe each of the individual qubits separately, which only requires the order of $n k$ bits.  In contrast, to describe an $n$-qubit entangled state we need the order of $k$ bits for each dimension in the Hilbert space, i.e. we need the order of $k 2^n$ bits.  If we were to simulate the evolution of an entangled state on a classical computer we would have to process all these bits of information and the computation would be extremely slow.  Quantum computation, in contrast, acts on all this information at once -- a quantum computation acting on an entangled state is just as fast as one acting on a separable state.  Thus, if we can find situations where the evolution of an entangled state can be mapped into a hard mathematical problem, we can sometimes get substantial speedups.

The CNOT gate can also be used to illustrate how decoherence comes about.  Through the same action that allows it to generate an entangled state from a separable state, when viewed from the perspective of a single qubit, the resulting state becomes decoherent.  That is, suppose we look at (\ref{qcomputation}) in the density matrix representation.   Looking at the first qubit only, the wavefunction of the separable state is $| \psi \rangle = 1/\sqrt{2} (| 1 \rangle + | 0 \rangle)$, or in the density matrix representation
\begin{eqnarray*}
| \psi \rangle \langle \psi | & = & \frac{1}{2} ( |1 \rangle \langle 1 | + |1 \rangle \langle 0 |  + |0 \rangle \langle 1 |  + |0 \rangle \langle 0 | )\\
& = & \frac{1}{2} \begin{pmatrix}
     1 & 1    \\
      1 & 1  \\    \end{pmatrix}.
 \end{eqnarray*}
 Under the action of CNOT this becomes $\frac{1}{2} \begin{pmatrix}
     1 & 0    \\
      0 & 1  \\    \end{pmatrix}$, i.e. it becomes diagonal. 

\subsection{Applications.}

At the present point in time there are many different efforts in progress to implement quantum computing.  In principle all that is needed is a simple two level quantum system that can easily be manipulated and scaled up to a large number of qubits.   The first requirement is not very restrictive, and many different physical implementations of systems with a single or a few qubits have been achieved, including NMR, spin lattices, linear optics with single photons, quantum dots, Josephson junction networks, ion traps and atoms and polar molecules in optical lattices \cite{divincenzo1990}. The much harder problem that has so far limited progress toward practical computation is to couple the individual qubits in a controllable way and to achieve a sufficiently low level of decoherence.  With the great efforts now taking place, future developments could be surprisingly fast\footnote{A first 16-qubit quantum computer has been announced by D-Wave Systems Inc. in California, but at the time of writing this product is not available yet.}. 
If we had quantum computers at our disposal, what miracles would they perform?  As we said in the introduction to this section, there are many problems where the intrinsic massive parallelism of quantum evolution might yield dramatic speedups. The point is not that a classical computer would not be able to do the same computation -- after all, one can always simulate a quantum computer on a classical one -- but rather the time that is needed.  As we mentioned already, the most spectacular speedup is the Shor algorithm (1994) for factorizing large numbers into their prime factors \cite{shor1994}.  Because many security keys are based on the inability to factor large numbers into prime factors, 
the reduction from an exponentially hard to a polynomial hard problem has many practical applications for code breaking.  Another important application is the quadratic speedup by Grover's algorithm (1996) \cite{grover1996} for problems such as the traveling salesman, in which large spaces need to be searched.  Finally, an important application is the simulation of quantum systems themselves \cite{Aspuru:05a}. 
Having a quantum computer naturally provides an exponential speed-up, which in turn feeds back directly into the development of new quantum technologies.
 
Quantum computation and security are another challenging instance of the surprising and important interplay between the basic concepts of physics and information theory. If physicists and engineers succeed in mastering quantum technologies it will mark an important turning point in information science.

 \section{Black Holes: a space time information
paradox}\label{sectionblackholes}

In this section we make a modest excursion into the realm of curved
space-time as described by Einstein's theory of general relativity.  As was realized only in the 1970's, this theory poses an interesting and still not fully resolved information paradox for fundamental physics.  In general relativity gravity is understood as a manifestation of the curvature of space-time:   the curvature of space-time determines how matter and radiation propagate, while at the same time matter and radiation determine how space-time is curved.  Particles follow geodesics in curved space-time to produce the curvilinear motion that we observe.

An unexpected and long-ignored prediction of general relativity was the
existence of mysterious objects called \textit{black holes} that correspond to solutions with a curvature singularity at their center.  Black holes can be created when a very massive star burns all of its nuclear fuel and subsequently collapses into an ultra-compact object under its own gravitational pull.  The space-time curvature at the surface of a black hole is so strong that even light cannot escape -- hence the
term ``black hole". The fact that the escape velocity from a black hole is
larger then the speed of light implies, at least
classically, that no information from inside the black hole can ever
reach  far away observers. The physical size of a black hole of mass $M$ is defined by its
\textit{event horizon}, which is an imaginary sphere centered on the
hole with a radius (called the \textit{Schwarzchild radius})
\begin{equation}\label{schwarzchild}
R_S = \frac{2 G_N M}{c^2} \;,
\end{equation}
where $G_N$ is Newton's gravitational constant and $c$ is the velocity of light.  For a black hole with the mass of the sun this yields
 $R_S = 3 km$, and for the earth only $R_S = 1 cm$!
The only measurable quantities of a black hole for an observer far away
are  its mass, its charge and its angular momentum. 

But what about the second law of thermodynamics?  If we throw an object with non-zero entropy into black hole, it naively seems that the entropy would disappear for ever and thus the total entropy of the universe would decrease, causing a blunt violation of the second law of thermodynamics.  In the early 1970's, however, Bekenstein \cite{bekenstein1973} and Hawking \cite{bardeen1973}  showed that it is possible to assign an
entropy to a black hole. This entropy is proportional to the area $A = 4\pi (R_S)^2$ of the event horizon, 
\begin{equation}\label{bhentropy}
  S = \frac{Ac^3}{4G_N \hbar}.
\end{equation}
A striking analogy with the laws  thermodynamics became evident:  The change of mass (or energy) as we throw things in leads  according to classical general relativity to a change of horizon area, as the Schwarzchild radius also increases.   For an electrically neutral,  spherically symmetric black hole, it is possible to show that the incremental change of mass $dM$ of the black hole is related to the change of area $dA$ as
\begin{equation}
dM= \frac{\kappa}{2\pi}dA 
\label{eq:firstbhlaw}
\end{equation}
where $\kappa= \hbar c/2R_s $ is the \textit{surface gravity} at the horizon.  One can make an analogy with thermodynamics, where $dA$ plays the role of ``entropy", $dM$ the role of ``heat", and the $\kappa$ the role of ``temperature".  Since no energy can leave the black hole, $dM$ is positive and  therefore $dA \geq 0$,  analogous to the second law of thermodynamics.  At this point the correspondence between black hole dynamics and thermodynamics is a mere analogy, because we know that a classical black hole does not radiate and therefore has zero temperature.
One can still argue that the information is not necessarily be lost, it is only somewhere else and unretrievable for certain observers.  

What happens to this picture if we take quantum physics into account?  Steven Hawking was the first to investigate the quantum behavior of black holes and his results  radically changed their physical interpretation. He showed \cite{hawking1974, hawking1975} that if we  apply quantum theory to the spacetime region close to the horizon  then black holes aren't black at all!  Using arguments based on the spontaneous creation of particle-antiparticle pairs in the strong gravitational field near the horizon he showed that a black hole behaves like a stove, emitting black body thermal radiation of a characteristic temperature, called the \textit{Hawking temperature},
given by\footnote{We recall that we adopted units where Boltzmann's constant k is equal to one.} 
\begin{equation}\label{hawking}
T_H \equiv \frac{\hbar c}{4\pi R_S}=\frac{\hbar c^3}{8\pi G_N M}\;\;,
\end{equation}
fully consistent with the first law (\ref{eq:firstbhlaw}). We see that the black hole temperature is inversely proportional to
its mass, which means that a black hole becomes hotter and radiates more energy as it becomes lighter.  In other words, a black hole will radiate and lose mass at an ever-increasing rate until it finally explodes\footnote{The type of blackholes that are most commonly considered are very massive objects like collapsed stars. The lifetime of a black hole is given by $\tau\simeq G_N^2M^3/\hbar c^4$ which implies that the lifetime of such a massive black hole is on the order of $\tau \geq 10^{50}$ years (much larger than the lifetime of the universe $\tau_0 \simeq 10^{10}$ y). Theoretical physicists have also considered microscopic black
holes, where the information paradox we are discussing leads to a problem of principle.}. 

We conclude that quantum mechanics indeed radically changes the picture of a black hole.  Black holes will eventually evaporate, presumably leaving nothing behind except thermal radiation, which has a nonzero entropy.  However, as we discussed in the previous section, if we start with a physical system in a pure state that develops into a black hole, which subsequently evaporates, then at the level of quantum mechanics the information about the wavefunction should be rigorously preserved -- the quantum mechanical entropy should not change.

It may be helpful to  compare the complete black hole formation and evaporation process with a similar, more familiar situation (proposed by Sidney Coleman) where we know that quantum processes conserve entropy.  Imagine a piece of coal at zero temperature (where by definition $S=0$) that gets  irradiated with a given amount of high entropy radiation, which we assume gets absorbed completely.  It brings the coal into an excited state of finite temperature. As a consequence the piece of coal starts radiating, but since there is no more incoming radiation, it eventually returns to the zero temperature state, with zero entropy.  As the quantum process of absorbing the initial radiation and emitting the outgoing radiation is unitary, it follows that the outcoming radiation should have exactly the same entropy as the incoming radiation.  

Thus, if we view the complete process of black hole formation and subsequent evaporation from a quantum mechanical point of view there should be no loss of information.  So if the initial state is  a pure state than a pure state should come out.
But how can this be compatible with the observation that only thermal radiation comes out, independent of what we throw in?  Thermal radiation is produced by entropy generating processes, is maximally uncorrelated and random, and has maximal entropy.
If we throw the Encyclopedia Brittanica into the black hole and only get radiation out, its highly correlated initial state would seem to have been completely lost.  This suggests that Hawking's quantum calculation is in some way incomplete. These conflicting views on the process of black hole formation and evaporation are  referred to as the \textit{black hole information paradox}.  It has given rise to a fundamental debate in physics between the two principle theories of nature: the theory of relativity describing space-time and gravity on one hand and the theory of quantum mechanics describing matter and radiation on the other.  Does the geometry of Einstein's theory of relativity prevail over quantum theory, or visa versa?  

If quantum theory is to survive one has to explain how the incoming information gets transferred to the outgoing radiation coming from  the horizon\footnote{It has been speculated by a number of authors that there is the logical possibility that the black hole does not disappear altogether, but leaves some remnant behind just in order to preserve the information. The final state of the remnant should then somehow contain the information of the matter thrown in.}, so that a clever quantum detective making extremely careful measurements with very fancy equipment could  recover it. If such a mechanism is not operative the incoming information is completely lost, and the laws of quantum mechanics are violated.  The question is, what cherished principles must be given up? 

There is a generic way to think about this problem along the lines of quantum teleportation and a so-called \textit{final state projection} \cite{Horowitz2004, Lloyd2006}. We mentioned that Hawking radiation can be considered as a consequence of virtual particle-antiparticle pair production near the horizon of the black hole. The pairs that are created and separated at the horizon are in a highly entangled state, leading to highly correlated in-falling and outgoing radiation.  It is then possible, at least in principle, that the interaction between the in-falling radiation and the in-falling matter (making the black hole) would lead to a projection in a given quantum state. 
 Knowing that final state - for example by proving that only a unique state is possible - one would instantaneously have teleported the information from the incoming mass ( qubit A) to the outgoing radiation (qubit C) by using the entangled pair (qubit pair BC) in analogy with the process of teleportation we discussed in section \ref{teleportation} . 
 The parallel with quantum teleportation is only partial, because in that situation  the sender Alice (inside the black hole) has to send some classical information on the outcome of her measurements to the receiver Charlie (outside the black hole) before he is able decode the information in the outcoming radiation.
But sending classical information out of a black hole is impossible.  So this mechanism to rescue the information from the interior can only work if there is a projection onto an a priori known unique final state, so that it is as if Alice made a measurement yielding this state and sent the information to Charlie.
But how this assumption could be justified is still a mystery.  

A more ambitious way to attack this problem is to attempt to construct a quantum theory of gravity, where one assumes the existence of microscopic degrees of freedom so that  the thermodynamic properties of black holes could be explained by the statistical mechanics of these underlying degrees of freedom. Giving the quantum description of these new fundamental degrees of freedom would then  allow for a unitary description. Before we explain what these degrees of freedom might be, let us first consider another remarkable property of black holes.  As we explained before, the entropy of systems that are not strongly coupled is an extensive property, i.e. proportional to volume.
The entropy of a black hole, in contrast, is proportional to the area of the event horizon rather than the volume.  This dimensional reduction of the number of degrees of freedom is highly suggestive that all the physics of a black hole takes place at its horizon, an idea introduced by 't Hooft and Susskind \cite{Susskind04}, that is called the \textit{holographic principle}\footnote{A hologram is a two dimensional image that appears to be a three dimensional image; in a similar vein, a black hole is a massive object for which everything appears to take place on the surface.}.

Resolving the clash between the quantum theory of matter and
general relativity of space-time is one of the main motivations
for the great effort to search for a theory that overarches all of
fundamental physics. At this moment the main line of attack is based on 
\textit{superstring theory}, which is a quantum theory in which both matter
and space-time are a manifestation of extremely tiny strings ($l =
10^{-35}m$). This theory incorporates microscopic degrees of
freedom that might provide a statistical  mechanical account of
the entropy of black holes.  In 1996 Strominger and Vafa\cite{Strominger1996}
managed to calculate the Bekenstein-Hawking entropy for
(extremal) black holes in terms of microscopic strings using a property of string theory 
called \textit{duality}, which allowed them to count the number of accessible quantum
states. The answer they found implied that for the exterior observer
information is preserved on the surface of the horizon, basically realizing the holographic principle. 

There are indeed situations (so-called Anti-de Sitter/Conformal Field Theory dualities or AdS/CFT models) in string theory describing space-times with a boundary where the  holographic principle is realized explicitly. One hopes that in such models   the complete process of formation and evaporation of a black hole can be described by the time evolution of its holographic image on the boundary, which in this case  is  a super-symmetric gauge theory, a well behaved quantum conformal field theory (CFT).
A caveat is that in this particular Anti-de Sitter (AdS) classical setting so far only  a static ``eternal" black hole solution has been found, so interesting as that situation may be, it doesn't yet allow for a decisive answer to a completely realistic process of black hole formation and evaporation.  Nevertheless, the  communis opinion - at least for the moment - is that the principles of quantum theory have successfully passed a severe test\footnote{Indicative is that a long standing bet between Hawking and Presskil of Caltech was settled in 2004 when Hawking officially declared defeat.   In doing so he recognized the fact that information is not lost when we throw something into a black hole --  quantum correlations between the in-falling matter and the out-coming radiation should in principle make it possible to retrieve the original information.} \cite{Susskind04}.

\section{Conclusion}

The basic method of scientific investigation is to acquire information about nature by doing measurements and then to make models which optimally compress that information.  Therefore information theoretic questions arise naturally at all levels of scientific enterprise: in the analysis of measurements, in performing computer simulations, and in evaluating  the quality of mathematical models and theories. 

The notion of entropy started in thermodynamics as a rather abstract mathematical property.  With the development of statistical mechanics it emerged as a measure of disorder, though the notion of disorder referred to a very restricted context.  With the passage of time the generality and the power of the notion of entropy became clearer, so that now the line of reasoning is easily reversed -- following Jaynes, statistical mechanics is reduced to an application of the maximum entropy principle, using constraints that are determined by the physical system.  Forecasting is a process whose effectiveness can be understood in terms of the information contained in measurements, and the rate at which the geometry of the underlying dynamical system, used to make the forecast, causes this information to be lost.   And following Rissanen, the whole scientific enterprise is reduced to the principle of minimum description length, which essentially amounts to finding the optimal compromise between the information contained  in a model and the information contained in the discrepancies between the model and the data.

Questions related to the philosophy of information have lead us naturally back to some of the profound  debates in physics on the nature of the concept of entropy as it appears in the description of systems about which we have \textit{a priori} only limited information.  The Gibbs paradox, for example, centers around the question of whether entropy is subjective or objective.  We have seen that while the description might have subjective components, whenever we use the concept of entropy to ask concrete physical questions, we always get objective physical answers.  Similarly, when we inject intelligent actors into the story, as for Maxwell's demon, we see that the second law remains valid -- it applies equally well in a universe with sentient beings.

Fundamental turning points in physics have always left important traces in information theory.  A particularly interesting example is the development of quantum information theory, with its envisaged applications to quantum security, quantum teleportation and quantum computation.  Another interesting example is the  black hole information paradox, where the notions of entropy and information continue to be central players in our attempts to resolve some of the principal debates of modern theoretical physics.  In a sense, our ability to construct a proper statistical mechanics is a good test of our theories.  If we could only formulate an underlying  statistical mechanics of black holes, we might be able to resolve fundamental questions about the interface between gravity and quantum mechanics.

Finally, as we enter the realm of nonequilbrium statistical mechanics, we see that the question of what information means and how it can be used remains vital.  New entropies are being defined, and their usefulness and theoretical consistency are topics that are actively debated.  The physics of information is an emerging field, one that is still very much in progress.\\[5mm]
\textbf{Acknowledgements:}\\
The authors would like to thank Seth Lloyd, Peter Harremo\"{e}s, David Bacon, Jim Crutchfield, Cris Moore, Constantino Tsallis, Bill Wooters and Erik Verlinde for illuminating conversations and comments. 
Doyne Farmer appreciates support from Barclays Bank and National Science Foundation grant 0624351. Sander Bais thanks the Santa Fe Institute for its hospitality, which allowed this work to take shape. Any opinions, findings, and conclusions or recommendations expressed in this material are those of the authors and do not necessarily reflect the views of the National Science Foundation. 

\vspace*{1cm}
\bibliographystyle{amsplain}
\bibliography{physinf}

\end{document}